\documentclass[a4paper,11pt]{article} 

\usepackage{xcolor}

\usepackage{amsmath, amssymb, theorem, latexsym}

\usepackage{rotating}
\usepackage{multirow}

\usepackage{epstopdf}
\usepackage{enumerate}
\usepackage{graphicx}

\def \ar{\rightarrow}

\def\R{{\mathbb R}}

\def\CD{{\mathcal D}}

\def\CS{{\mathcal S}}

\def\Ai{{\rm Ai}}

\newcommand{\sca}[2]{\langle#1,#2\rangle}
\newcommand{\barr}{\overline}

\newcommand{\beq}{\begin{equation}}
\newcommand{\eeq}{\end{equation}}

\newcommand{\Supp}{\textrm{Supp~}}
\renewcommand{\Re}{{\rm Re\,}}
\renewcommand{\Im}{{\rm Im\,}}
\newtheorem{theorem}{Theorem}[section]
\newtheorem{lemma}[theorem]{Lemma}

\newtheorem{proposition}[theorem]{Proposition}

\newtheorem{remark}[theorem]{Remark}

\def\B{{\mathcal B}}
\def\ctan{{\rm ctan}}

\makeatletter
\@addtoreset{equation}{section}

\makeatother

\title{The complex Airy operator \\ with a semi-permeable barrier}

\author{D. S. Grebenkov,\\
Laboratoire de Physique de la Mati\`ere Condens\'ee, \\
CNRS--Ecole Polytechnique, 91128
Palaiseau, France\\ ~\\
 B. Helffer\\
 Laboratoire de Math\'ematiques Jean Leray, Universit\'e de Nantes\\
 2 rue de la Houssini\`ere 44322 Nantes, France\\
 and\\
 Laboratoire de Math\'ematiques, \\ Universit\'e Paris-Sud, CNRS, Univ. Paris Saclay, France\\~\\
 R. Henry\\
   Laboratoire de Math\'ematiques, \\
   Universit\'e Paris-Sud, CNRS, Univ. Paris Saclay, France}

\date{}

\begin{document}
\maketitle

\begin{abstract}
We consider a suitable extension of the complex Airy operator,
$-d^2/dx^2 + ix$, on the real line with a transmission boundary
condition at the origin.  We provide a rigorous definition of this
operator and study its spectral properties.  In particular, we show
that the spectrum is discrete, the space generated by the generalized
eigenfunctions is dense in $L^2$ (completeness), and we analyze the
decay of the associated semi-group.  We also present explicit formulas
for the integral kernel of the resolvent in terms of Airy functions,
investigate its poles, and derive the resolvent estimates.
\end{abstract}

\section{Introduction}\label{s1}
The transmission boundary condition which is considered in this
article appears in various exchange problems such as molecular
diffusion across semi-permeable membranes
\cite{Tanner78,Powles92,Novikov98}, heat transfer between two
materials \cite{Carslaw,Gilkey,Bardos}, or transverse magnetization
evolution in nuclear magnetic resonance (NMR) experiments
\cite{Grebenkov10}.  In the simplest setting of the latter case, one
considers the local transverse magnetization $G(x,y\,;t)$ produced by
the nuclei that started from a fixed initial point $y$ and diffused in
a constant magnetic field gradient $g$ up to time $t$.  This
magnetization is also called the propagator or the Green function of
the Bloch-Torrey equation \cite{Torrey56}:
\begin{equation}
\label{eq:BT_G}
\frac{\partial}{\partial t} G(x,y\,;t) = \left(D \Delta - i \gamma g x_1 \right) G(x,y\,;t)\, ,
\end{equation}
with the initial condition 
\begin{equation}
\label{eq:G_initial}
G(x,y\,;t=0) = \delta(x-y), 
\end{equation}
where $\delta(x)$ is the Dirac distribution, $D$ the intrinsic
diffusion coefficient, $\Delta = \partial^2/\partial x_1^2 + \ldots +
\partial^2/\partial x_d^2$ the Laplace operator in $\R^d$, $\gamma$ the
gyromagnetic ratio, and $x_1$ the coordinate in a prescribed direction.

Throughout this paper, we focus on the one-dimensional situation
\break ($d = 1$), in which the operator
\begin{equation*}
D_x^2 + ix = -\frac{d^2}{dx^2}+ix
\end{equation*}
is called the complex Airy operator and appears in many contexts:
mathematical physics, fluid dynamics, time dependent Ginzburg-Landau
problems and also as an interesting toy model in spectral theory (see
\cite{Alm}).  We will consider a suitable extension $\mathcal A_1^+$ of this
differential operator and its associated evolution operator $e^{-t
\mathcal A_1^+}$.  The Green function $G(x,y\,;t)$ is the distribution
kernel of $e^{-t \mathcal A_1^+}$.  A separate article will address
this operator in higher dimensions \cite{GrHel}. \\

For the problem on the line $\R$, an intriguing property is that this
non self-adjoint operator, which has compact resolvent, has empty
spectrum (see Section \ref{Airy_line}).  However, the situation is
completely different on the half-line $\mathbb R_+$.   The eigenvalue
problem
\begin{equation*}
(D_x^2 + ix) u = \lambda u ,
\end{equation*}
for a spectral pair $(u,\lambda)$ with $u \in H^2(\mathbb R_+)$ and
$xu\in L^2(\mathbb R^+)$ has been thoroughly analyzed for both
Dirichlet ($u(0)=0$) and Neumann ($u'(0)=0$) boundary conditions.  The
spectrum consists of an infinite sequence of eigenvalues of
multiplicity one explicitly related to the zeros of the Airy function
(see \cite{Stoller91,Hel1}).  The space generated by the
eigenfunctions is dense in $L^2(\mathbb R_+)$ (completeness property)
but there is no Riesz basis of eigenfunctions (we recall that a
collection of vectors $(x_k)$ in a Hilbert space $\mathcal H$ is
called Riesz basis if it is an image of an orthonormal basis in
$\mathcal H$ under some isomorphism).  Finally, the decay of the
associated semi-group has been analyzed in detail.  The physical
consequences of these spectral properties for NMR experiments have
been first revealed by Stoller, Happer and Dyson
\cite{Stoller91} and then thoroughly discussed in
\cite{deSwiet94,Grebenkov07,Grebenkov14b}. \\

In this article, we consider another problem for the complex Airy
operator on the line but with a transmission property at $0$ which
reads \cite{Grebenkov14b}: 
\begin{equation}
\label{eq:transmission}
 \left\{\begin{array}{lll}
         u'(0_+) & = & u'(0_-)\,, \\
         u'(0) & = & \kappa\, \big(u(0_+)-u(0_-)\big)\,,
        \end{array}\right.
\end{equation}
where $\kappa \geq 0$ is a real parameter (in physical terms, $\kappa$
accounts for the diffusive exchange between two media $\R_-$ and
$\R_+$ across the barrier at $0$ and is defined as the ratio between
the barrier permeability and the bulk diffusion coefficient).  The
case $\kappa=0$ corresponds to two independent Neumann problems on
$\mathbb R_-$ and $\mathbb R_+$ for the complex Airy operator.  When
$\kappa$ tends to $+\infty$,  the second relation in
(\ref{eq:transmission}) becomes the continuity condition,
$u(0_+) = u(0_-)$, and the barrier disappears.  As a consequence, the
problem tends (at least formally) to the standard problem for the
complex Airy operator on the line.\\
The main purpose of this paper is to define the complex Airy operator
with transmission (Section \ref{s3}) and then to analyze its spectral
properties.  Before starting the analysis of the complex Airy operator
with transmission, we first recall in Section \ref{s2} the spectral
properties of the one-dimensional Laplacian with the transmission
condition, and summarize in Section \ref{s2a} the known properties of
the complex Airy operator.  New properties are also established
concerning the Robin boundary condition and the behavior of the
resolvent for real $\lambda$ going to $+\infty$.  In Section \ref{s3}
we will show that the complex Airy operator ${\mathcal A}^+_1 = D_x^2
+ ix$ on the line $\R$ with a transmission property
(\ref{eq:transmission}) is well defined by an appropriate sesquilinear
form and an extension of the Lax-Milgram theorem.  Section
\ref{s5} focuses on the exponential decay of the associated
semi-group.  In Section \ref{s6}, we present explicit formulas for the
integral kernel of the resolvent and investigate its poles.  In
Section \ref{s7}, the resolvent estimates as $|\Im \lambda|\to 0$ are
discussed.  Finally, the proof of completeness is reported in Section
\ref{s8}.  In five Appendices, we recall the basic properties of Airy
functions (Appendix \ref{AppA}), determine the asymptotic behavior of
the resolvent as $\lambda\to +\infty$ for extensions of the complex
Airy operator on the line (Appendix \ref{AppB}) and in the semi-axis
(Appendix \ref{AppC}), give the statement of the needed
Phragmen-Lindel\"of theorem (Appendix \ref{AppD}) and finally describe
the numerical method for computing the eigenvalues (Appendix
\ref{sec:numerics}).

We summarize our main results in the following:
\begin{theorem}
The semigroup $\exp(-t{\mathcal A}^+_1)$ is contracting.  The operator
${\mathcal A}^+_1$ has a discrete spectrum $\{\lambda_n(\kappa)\}$.
The eigenvalues $\lambda_n(\kappa)$ are determined as (complex-valued)
solutions of the equation
\begin{equation}
2\pi \Ai'(e^{2\pi i/3} \lambda) \Ai'(e^{-2\pi i/3} \lambda) + \kappa = 0,
\end{equation}
where $\Ai'(z)$ is the derivative of the Airy function.\\
For all $\kappa\geq 0$, there exists $N$ such that, for all $n\geq N$,
there exists a unique eigenvalue of $\mathcal A^+_1$ in the ball
$B(\lambda_n^\pm, 2 \kappa |\lambda_n^\pm|^{-1})$, where
$\lambda_n^\pm = e^{\pm 2\pi i/3} a_n'$, and $a'_n$ are the zeros of
$\Ai'(z)$. \\
Finally, for any $\kappa \geq 0$ the space generated by the
generalized eigenfunctions of the complex Airy operator with
transmission is dense in  $L^2(\R_-)\times L^2(\R_+)$.
\end{theorem}

Note that due to the possible presence of eigenvalues with Jordan
blocks, we do not prove in full generality that the eigenfunctions of
${\mathcal A}^+_1$ span a dense set in $L^2(\R_-)\times L^2(\R_+)\,$.
Numerical computations suggest actually that all the spectral
projections have rank one (no Jordan block) but we shall only prove in
Proposition \ref{prop:finite_number} that there are at most a finite
number of eigenvalues with nontrivial Jordan blocks.

\section{The free Laplacian with a semi-permeable barrier}\label{s2}

As an enlighting exercise, let us consider in this section the case of
the free one-dimensional Laplacian $-\frac{d^2}{dx^2}$ on
$\mathbb{R}\setminus\{0\}$ with the transmission condition
(\ref{eq:transmission}) at $x=0$.  We work in the Hilbert space
\[
 \mathcal H := L_-^2\times L_+^2\,,
\]
where $L_-^2:=L^2(\mathbb R_-)$ and $L_+^2:=L^2(\mathbb R_+)\,$.\\ An
element $u\in L_-^2\times L_+^2$ will be denoted by $u = (u_-,u_+)$
and we shall use the notation $H_-^s = H^s(\mathbb R_-)\,$, $H_+^s =
H^s(\mathbb R_+)\,$ for $s\geq0\,$.\\ So \eqref{eq:transmission} reads
\beq\label{Condbis}
 \left\{\begin{array}{lll}
         u_+'(0) & = & u_-'(0)\,, \\
         u_+'(0) & = & \kappa\big(u_+(0)-u_-(0)\big)\,.
        \end{array}\right.
\eeq

In order to define appropriately the corresponding operator, we start
by considering a sesquilinear form defined on the domain
\[
 V = H^1_- \times H^1_+\,.
\]
The space $V$ is endowed with the Hilbertian norm $\|\cdot\|_V$
defined for all $u=(u_-,u_+)$ in $V$ by
\[
 \|u\|_{V}^2 = \|u_-\|_{H_-^1}^2+\|u_+\|_{H_+^1}^2\,.
\]

We then define a Hermitian sesquilinear form $ a_\nu$ acting on
$V\times V$ by the formula
\begin{eqnarray}
a_\nu (u,v) &=& \int_{-\infty}^0\Big(u_-'(x)\bar v_-'(x) +\nu\, u_-(x)\bar v_-(x)\Big)\,dx \nonumber\\
& & + \int_0^{+\infty}\Big(u_+'(x)\bar v_+'(x)+\nu\, u_+(x)\bar v_+(x)\Big)\,dx \nonumber \\
& & + \, \kappa \, \big(u_+(0)-u_-(0)\big)
 \big(\barr{v_+(0)-v_-(0)}\big)\,, \label{defForma0}
\end{eqnarray}
for all pairs $u = (u_-,u_+)$ and $v=(v_-,v_+)$ in $V\,$.  For $z\in
\mathbb C$, $\bar z$ denotes the complex conjugate of $z$.  The
parameter $ \nu\geq 0$ will be determined later to ensure the
coercivity of $ a_\nu \,$.

\begin{lemma}\label{lemConta0}
The sesquilinear form $a_\nu $ is continuous on $V\,$.
\end{lemma}
\textbf{Proof:}\\
We want to show that, for any $\nu \geq 0\,$, there exists a
positive constant $c$ such that, for all $(u,v)\in V\times V\,$,
\beq\label{estaCont} 
|a_\nu (u,v)|\leq c\, \|u\|_V\|v\|_V\,.
\eeq
We have, for some $c_0>0\,$,
\begin{eqnarray*}
 && \left|\int_{-\infty}^0\Big(u_-'(x)\bar v_-'(x)+\nu\,u_-(x)\bar v_-(x)\Big)\,dx\right.  \\
&&  ~~~~~~~~~~~~ \left.+\int_0^{+\infty}\Big(u_+'(x)\bar v_+'(x)+\nu\,u_+(x)\bar v_+(x)\Big)\,dx\right|\leq c_0\, \|u\|_V\|v\|_V\,.
\end{eqnarray*}
On the other hand, 
\beq\label{estU0}
 |u_+(0)|^2 = -  \int_0^{+\infty}(u_+\bar u_+)'(x)\,dx\leq 2\, \|u\|_{L^2}\|u'\|_{L^2}\,,
\eeq
and similarly for $|u_-(0)|^2\,$, $|v_+(0)|^2$ and
$|v_-(0)|^2\,$. Thus there exists $c_1>0$ such that, for all $(u,v)\in
V\times V\,$,
\[
 \Big|\kappa\big(u_-(0)-u_+(0)\big) \big(\barr{v_-(0)-v_+(0)}\big)\Big|\leq c_1\|u\|_V\|v\|_V\,,
\]
and (\ref{estaCont}) follows with $c=c_0+c_1$.
\hfill $\square$\\

The coercivity of the sesquilinear form $a_\nu$ for $\nu$ large enough
is proved in the following lemma.  It allows us to define a closed
operator associated with $a_\nu $ by using the Lax-Milgram theorem.

\begin{lemma}\label{lemCoerca0}
There exist $\nu_0>0$ and $\alpha > 0$ such that, for all $\nu\geq\nu_0\,$,
\beq\label{coerca0} 
\forall u\in V\,,~~~~a_\nu(u,u) \geq \alpha \, \|u\|_V^2\,.
\eeq
\end{lemma}
\textbf{Proof:}
The proof is elementary for $\kappa \geq 0$.  For completeness,
we also treat the case $\kappa < 0$, in which an additional difficulty
occurs.  Except for this lemma, we keep considering the physically
relevant case $\kappa \geq 0$.\\
Using the estimate (\ref{estU0}) as well as the Young inequality
\[
 \forall e,f,\delta > 0\,,~~~ef\leq \frac{1}{2}\Big(\delta e^2+\delta^{-1}f^2\Big)\,,
\]
we get that, for all $\varepsilon>0\,$, there exists
$C(\varepsilon)>0$ such that, for all $u\in V\,$,
\beq\label{Young}
 \big|u_-(0)-u_+(0)\big|^2\leq \varepsilon\left(\int_{-\infty}^0|u_-'(x)|^2\,dx + \int_0^{+\infty}|u_+'(x)|^2\,dx\right)+C(\varepsilon)\|u\|_{L^2}^2\,.
\eeq
Thus for all $u\in V$ we have
\begin{eqnarray} 
 a_\nu (u,u)
 & \geq & (1+\kappa\varepsilon)\left(\int_{-\infty}^0|u_-'(x)|^2\,dx + \int_0^{+\infty}|u_+'(x)|^2\,dx\right) \nonumber \\
 && +\big(\nu +\kappa C(\varepsilon)\big)\|u\|_{L^2}^2\,. \label{estCoerca0}
\end{eqnarray}
Choosing $\varepsilon<|\kappa|^{-1}$ and $\nu > -\kappa C(\varepsilon)\,$, we get (\ref{coerca0}).
\hfill $\square$\\

The sesquilinear form $a_\nu $ being symmetric, continuous and
coercive in the sense of (\ref{coerca0}) on $V\times V\,$, we can use
the Lax-Milgram theorem \cite{Hel1} to define a closed, densely
defined selfadjoint operator $ S_\nu$ associated with $
a_\nu\,$. Then we set $\mathcal T_0 = S_\nu -\nu\,$. By
construction, the domain of $ S_\nu $ and $\mathcal T_0$ is
\begin{eqnarray}
 \CD(\mathcal T_0) &=& \big\{u\in V : v\mapsto a_\nu (u,v)~\textrm{ can be extended continuously} \nonumber\\
 & &\qquad\qquad  \textrm{ on }~L_-^2\times L_+^2\,\big\}\,, \label{defDomA0}
\end{eqnarray}
and the operator $\mathcal T_0$ satisfies, for all $(u,v)\in\CD(\mathcal T_0)\times V\,$,
\[
 a_\nu (u,v) = \sca{\mathcal T_0u}{v}+\nu\sca{u}{v}\,.
\]
Now we look for an explicit description of the domain
(\ref{defDomA0}). The antilinear form $a(u,\cdot)$ can be extended
continuously on $L_-^2\times L_+^2$ if and only if there exists $w_u =
(w_u^-,w_u^+)\in L_-^2\times L_+^2$ such that
\[
 \forall v\in V\,,~~~a_\nu (u,v) = \sca{w_u}{v}\,.
\]
According to the expression (\ref{defForma0}), we have necessarily
\[
 w_u = \big(-u_-''+\nu \, u_- , -u_+''+\nu \, u_+\big)\in L_-^2\times L_+^2\,,
\]
where $u_-''$ and $u_+''$ are \emph{a priori} defined in the sense of
distributions respectively in $\mathcal D'(\mathbb R_-)$ and $\mathcal
D'(\mathbb R_+)$.  Moreover $(u_-,u_+)$ has to satisfy conditions
(\ref{eq:transmission}).  Consequently we have
\begin{eqnarray*}
 \CD(\mathcal T_0)& =& \Big\{u=(u_-,u_+)\in H_-^1\times H_+^1 : (u_-'',u_+'')\in L_-^2\times L_+^2 \quad \\
&&\qquad \qquad\qquad\qquad   ~\textrm{ and }u\textrm{ satisfies conditions }(\ref{eq:transmission})\Big\}\,.
\end{eqnarray*}
Finally we have introduced a closed, densely defined selfadjoint
operator $\mathcal T_0$ acting by
\begin{equation*}
\mathcal T_0\, u = -u'' 
\end{equation*}
on $(-\infty,0)\cup(0,+\infty)\,$, with domain
\[
 \CD(\mathcal T_0) = \big\{u\in H_-^2\times H_+^2 : u \textrm{ satisfies conditions }(\ref{eq:transmission})\big\}\,.
\]
Note that at the end $\mathcal T_0$ is independent of the $\nu$
chosen for its construction.\\ 
We observe also that because of the transmission conditions
(\ref{eq:transmission}), the operator $\mathcal T_0$ might not be
positive when $\kappa <0$, hence there can be negative spectrum in the
interval $[-\nu,0)\,$, as can be seen in the following statement.
\begin{proposition}
For all $\kappa\in\mathbb{R}\,$, the essential spectrum of $\mathcal
T_0$ is
\beq
\sigma_{ess}(\mathcal T_0) = [0,+\infty)\,.
\eeq
Moreover, if $\kappa\geq0$ the operator $\mathcal T_0$ has empty
discrete spectrum and
\beq
\sigma(\mathcal T_0) = \sigma_{ess}(\mathcal T_0) = [0,+\infty)\,.
\eeq
On the other hand, if $\kappa < 0$ there exists a unique negative
eigenvalue $-4\kappa^2\,$, which is simple, and 
\beq
\sigma(\mathcal T_0) = \big\{-4\kappa^2\big\}\cup[0,+\infty)\,.
\eeq
\end{proposition}
\textbf{Proof:}
Let us first prove that $[0,+\infty)\subset\sigma_{ess}(\mathcal
T_0)\,$.  This can be achieved by a standard singular sequence
construction.\\ 
Let $(a_j)_{j\in \mathbb N}$ be a positive increasing sequence such that, for
all $j\in \mathbb N\,$, $a_{j+1}-a_j> 2j+1\,$.  Let
$\chi_j\in\mathcal{C}_0^\infty(\mathbb{R})$ ($j\in \mathbb N$) such that  $$\Supp
\chi_j\subset(a_j-j,a_j+j)\,, \,||\chi_j ||_{L^2_+}=1 \mbox{  and }
 \sup|\chi_j^{(p)}|\leq \frac{C}{j^p}\,,~~p=1,2\,,
$$
for some $C\,$ independent of $j$.
Then, for all $r\geq0\,$, the sequence
$
u_{j,r}(x) = \big(0,\, \chi_j(x)\, e^{irx}\big)\,,
$
is a singular sequence for $\mathcal T_0$ corresponding to $z = r^2$ in the
sense of \cite[Definition IX.$1.2$]{EdEv}. Hence according to
\cite[Theorem IX.$1.3$]{EdEv}, we have
$[0,+\infty)\subset\sigma_{ess}(\mathcal T_0)\,$.\\

Now let us prove that $(\mathcal T_0-\mu)$ is invertible for all
$\mu\in(-\infty,0)$ if $\kappa\geq0\,$, and for all
$\mu\in(-\infty,0)\setminus\{-4\kappa^2\}$ if $\kappa< 0\,$.\\ Let
$\mu<0$ and $f = (f_-,f_+)\in L_-^2\times L_+^2\,$.  We are going to
determine explicitly the solutions $u=(u_-,u_+)$ to the equation
\beq\label{eqNonHomA0}
\mathcal T_0\, u = \mu\, u + f\,.
\eeq
Any solution of the equation $-u_\pm'' = \mu \, u_\pm+f_\pm$ has the form
\beq\label{formSolA0}
 u_\pm(x) = \frac{1}{2\sqrt{-\mu}}\int_0^xf_\pm(y)\big(e^{-\sqrt{-\mu}\,(x-y)}-e^{\sqrt{-\mu}\,(x-y)}\big)\,dy + A_\pm e^{\sqrt{-\mu}\,x}+B_\pm e^{-\sqrt{-\mu}\,x}\,,
\eeq
for some $A_\pm,B_\pm\in\mathbb{R}\,$.\\ 
We shall now determine $A_+\,$, $A_-\,$, $B_+$ and $B_-$ such that
$(u_-,u_+)$ belongs to the domain $\CD(\mathcal T_0)\,$.  The
conditions (\ref{eq:transmission}) yield
\[
\left\{
\begin{array}{rcl}
  A_+-B_+ & = & A_--B_-\,,\\
  \sqrt{-\mu}\, \big(A_+-B_+\big) & = &- \kappa\, \big(A_-+B_--A_+-B_+\big)\,.
\end{array}
\right.
\]
Moreover, the decay conditions at $\pm \infty$  imposed by $u_\pm\in H_\pm^2$
lead to the following values for $A_+$ and $B_-\,$:
\beq\label{exprA+B-}
 A_+ = \frac{1}{2\sqrt{-\mu}}\int_0^{+\infty}f_+(y)e^{-\sqrt{-\mu}\,y}\,dy\,,~~~~B_- = \frac{1}{2\sqrt{-\mu}}\int_{-\infty}^0f_-(y)e^{\sqrt{-\mu}\,y}\,dy\,.
\eeq
The remaining constants $A_-$ and $B_+$ have to satisfy the system
\beq\label{systA-B+}
\left\{
\begin{array}{rcl}
  A_-+B_+ & = & A_++B_-\,,\\
 - \kappa A_- + \big(\sqrt{-\mu}+\kappa\big)B_+ & = & \big(\sqrt{-\mu}- \kappa\big)A_+ +\kappa B_-\,.
\end{array}
\right.
\eeq
We then notice that the equation (\ref{eqNonHomA0}) has a unique
solution $u=(u_-,u_+)$ if and only if $\kappa\geq0$ or $\mu\neq
-4\kappa^2\,$.\\Finally in the case $\kappa< 0$ and $\mu =
-4\kappa^2\,$, the homogeneous equation associated with
(\ref{eqNonHomA0}) (\emph{i.e} with $f\equiv0$) has a one-dimensional
space of solutions, namely $u(x) = \big(A_-e^{-2\kappa x} ,
B_+a^{2\kappa x}\big)$ with $A_- = B_+\,$, or equivalently $u(x) =
Ke^{2\kappa|x|}\,$, $K\in\mathbb{R}\,$.  Consequently if $\kappa<
0\,$, the eigenvalue $\mu = -4\kappa^2$ is simple, and the desired
statement is proved.
\hfill $\square$\\

The expression (\ref{exprA+B-}) along with the system (\ref{systA-B+})
yield the values of $A_-$ and $B_+$ when $\mu\notin\sigma(\mathcal T_0)\,$:
\[
\begin{array}{ll}
 A_- &= \frac{2\kappa}{2\sqrt{-\mu}\big(\sqrt{-\mu}+2\kappa\big)}\int_0^{+\infty}f_+(y)e^{-\sqrt{-\mu}\,y}\,dy \\ &\quad   +\frac{1}{2\big(\sqrt{-\mu}+2\kappa\big)}
 \int_{-\infty}^0f_-(y)e^{\sqrt{-\mu}\,y}\,dy
 \end{array}
\]
and
\[
\begin{array}{ll}
 B_+ & = \frac{1}{2\big(\sqrt{-\mu}+ 2\kappa\big)}\int_0^{+\infty}f_+(y)e^{-\sqrt{-\mu}y\,}\,dy \\ 
	& \quad + \frac{2\kappa}{2\sqrt{-\mu}\big(\sqrt{-\mu}  + 2\kappa\big)}
 \int_{-\infty}^0f_-(y)e^{\sqrt{-\mu}\,y}\,dy\,.
 \end{array} 
\]
Using (\ref{formSolA0}), we are then able to obtain the expression of
the integral kernel of $(\mathcal T_0-\mu)^{-1}\,$.  More precisely we
have, for all $f=(f_-,f_+)\in L_-^2\times L_+^2\,$,
\[
 (\mathcal T_0-\mu)^{-1}= 
 \left(
 \begin{array}{cc}
  \mathcal{R}_\mu^{--} & \mathcal{R}_\mu^{-+} \\
  \mathcal{R}_\mu^{+-} & \mathcal{R}_\mu^{++}
 \end{array}
\right)\,,
\]
where for $\varepsilon,\sigma\in\{-,+\}$ the operator
$\mathcal{R}_\mu^{\varepsilon\,\sigma} :\mathbb{R}^\sigma
\rightarrow\mathbb{R}^\varepsilon$ is an integral operator whose
kernel (still denoted $\mathcal{R}_\mu^{\varepsilon\,\sigma}\,$) is
given for all $(x,y)\in\mathbb{R}^\varepsilon\times\mathbb{R}^\sigma$
by
\beq\label{KerA0}
\mathcal{R}_\mu^{\varepsilon,\sigma}(x,y) =
 \frac{1}{2\sqrt{-\mu}}e^{-\sqrt{-\mu}\,|x-y|} + \varepsilon\sigma\,\frac{1}{2\big(\sqrt{-\mu} + 2\kappa\big)}e^{-\sqrt{-\mu}(|x|+|y|)}\,.
\eeq
Noticing that the first term in the right-hand side of (\ref{KerA0})
is the integral kernel of the Laplacian on $\mathbb{R}\,$, and that
the second term is the kernel of a rank one operator, we finally get
the following expression of $(\mathcal T_0-\mu)^{-1}$ as a rank one
perturbation of the Laplacian:
\[
\begin{array}{ll}
 (\mathcal T_0-\mu)^{-1} &  = (-\Delta-\mu)^{-1} \\ & \quad + \frac{1}{2\big(\sqrt{-\mu}+ 2\kappa\big)}
 \left(
 \begin{array}{cc}
 \sca{\,\cdot\,}{\ell_\mu}_- (\ell_\mu)_- &  -\sca{\,\cdot\,}{\ell_\mu}_+\ (\ell_\mu)_- \\
 -\sca{\,\cdot\,}{\ell_\mu}_- (\ell_\mu)_+ & \sca{\,\cdot\,}{\ell_\mu}_+ (\ell_\mu)_+
 \end{array}
\right)\,,
\end{array}
\]
where $\ell_\mu(x) = e^{-\sqrt{-\mu}\,|x|}$ and
$\sca{\,\cdot\,}{\,\cdot\,}_\pm$ denotes the $L^2$ scalar product on
$\mathbb{R}^\pm\,$.\\ Here the operator $(-\Delta-\mu)^{-1}$ denotes
the operator acting on $L_-^2\times L_+^2$ like the resolvent of the
Laplacian on $L^2(\mathbb{R})$:
\[
 (-\Delta-\mu)^{-1}(u-,u_+) := (-\Delta-\mu)^{-1}(u_-\, \textbf{1}_{(-\infty,0)}+u_+\, \textbf{1}_{(0,+\infty)})\,,
\]
composed with the map $L^2(\mathbb R)\ni v \mapsto (v_{/\mathbb R_-},
v_{/ \mathbb R_+}) \in L^2_-\times L^2_+\,.$\\

\section{Reminder on the complex Airy operator}\label{s2a}

Here we recall relatively basic facts coming from
\cite{Mar,Alm,BM,Hel1,Hel2,Hen,Hen1} and discuss new questions
concerning estimates on the resolvent  and the Robin boundary
condition. Complements will also be given in Appendices \ref{AppA}, \ref{AppB} and  \ref{AppC}.

\subsection{The complex Airy operator on the line}
\label{Airy_line}

The complex Airy operator on the line can be defined as the closed
extension $\mathcal A^+$ of the differential operator $\mathcal A_0^+
:= D_x^2 + i\,x\,$ on $C_0^\infty (\mathbb R)$.  We observe that
$\mathcal A^+ = (\mathcal A_0^-)^*$ with $\mathcal A_0^- := D_x^2 -
i\,x\,$ and that its domain is
\begin{equation*}
D(\mathcal A^+)= \{u\in H^2(\mathbb R)\,,\, x\,u\in L^2(\mathbb R)\}\,.
\end{equation*}
In particular, $ \mathcal A^+$ has a compact resolvent.  It is also
easy to see that $ - \mathcal A^+$ is the generator of a semi-group $
S_t$ of contraction,
\begin{equation}\label{a5} 
S_t =\exp( - t \mathcal A^+)\,.
\end{equation}
Hence the results of the theory of semi-groups can be applied (see for
example \cite{Dav}).
\\ In particular, we have, for $ \Re
\lambda < 0\,$,
\begin{equation}\label{a6}
||(\mathcal A^+-\lambda)^{-1}|| \leq \frac{1}{|\Re \lambda|}\,.
\end{equation}
A very special property of this operator is that, for any $ a\in
\mathbb R$,
\begin{equation} 
T_a\, \mathcal A^+ = (\mathcal A^+ - i a)\,T_a\,,
\end{equation}
where $ T_a$ is the translation operator $ (T_a u)(x) = u (x-a)\,$.\\
As an immediate consequence, we obtain that the spectrum is empty and
that the resolvent of $ \mathcal A^+$,
\begin{equation*}
\mathcal G_0^+(\lambda) = (\mathcal A^+-\lambda)^{-1}
\end{equation*}
which is defined for any $ \lambda\in \mathbb C$, satisfies
\begin{equation} 
||(\mathcal A^+ -\lambda)^{-1}|| = ||(\mathcal A^+ - \Re\lambda)^{-1}||\,.
\end{equation}
The most interesting property is the control of the resolvent for 
$\Re \lambda \geq 0$.
\begin{proposition}[W. Bordeaux-Montrieux \cite{BM}]\label{PropBM}~\\
As $ \Re \lambda \ar +\infty$, we have
\begin{equation} \label{eq:BM}
||\mathcal G_0^+ (\lambda) || \sim \sqrt{\frac \pi 2}  (\Re \lambda)^{-\frac 14} \exp \left(\frac 43 (\Re \lambda)^{\frac 32}\right)\,,
\end{equation}
where $f(\lambda) \sim g(\lambda)$ means that the ratio
$f(\lambda)/g(\lambda)$ tends to $1$ in the limit $\lambda\to
+\infty$.
\end{proposition}
This improves a previous result (see Appendix \ref{AppB}) by
J. Martinet \cite{Mar} (see also in \cite{Hel1,Hel2}) who also
proved\footnote{The coefficient was wrong in \cite{Mar} and is
corrected here, see Appendix \ref{AppB}. }
\begin{proposition}\label{PropMart}
\begin{equation}
\label{eq:Martinet}
 \| \mathcal G_0^+ (\lambda)\|_{HS} =  \| \mathcal G_0^+ (\Re \lambda)\|_{HS}\,,
 \end{equation}
 and 
\begin{equation}\label{BMM}  
 \| \mathcal G_0^+(\lambda)\|_{HS} \sim \sqrt{ \pi/2 }  \, (\Re \lambda)^{-\frac 14}  \exp \left(\frac 43  (\Re \lambda)^\frac 32\right)\,,
\end{equation}
\end{proposition}
where $\| \cdot \|_{HS}$ is the Hilbert-Schmidt norm.  This is
consistent with the well-known translation invariance properties of
the operator $\mathcal A^+\,$, see \cite{Hel1}.  The comparison
between the $HS$-norm and the norm in $\mathcal L (L^2(\mathbb R))$,
immediately implies that Proposition \ref{PropMart} gives the upper
bound in Proposition \ref{PropBM}.

\subsection{The complex Airy operator on the half-line: Dirichlet case}

It is not difficult to define the Dirichlet realization $\mathcal
A^{\pm,D}$ of $D_x^2 \pm ix$ on $\mathbb R_+$ (the analysis on the
negative semi-axis is similar).  One can use for example the
Lax-Milgram theorem and take as form domain
\begin{equation*}
V^D :=\{ u \in H_0^1(\mathbb R_+)\,,\, x^\frac 12 u \in L^2_+\}\,.
\end{equation*}
It can also be shown that the domain is
\begin{equation*}
\mathcal D^D:= \{u\in V^D\,,\, u\in H^2_+\}\,.
\end{equation*}
This implies
\begin{proposition}\label{SchD}
The resolvent $\mathcal G^{\pm,D}(\lambda) := ({\mathcal A}^{\pm,D} -
\lambda)^{-1}$ is in the Schatten class $C^p$ for any $p>\frac 32$
(see \cite{DS} for definition), where ${\mathcal A}^{\pm,D} =
D_x^2 \pm ix$ and the superscript $D$ refers to the Dirichlet case.
\end{proposition}
More precisely we provide the distribution kernel ${\mathcal
G}^{-,D}(x,y\,;\lambda)$ of the resolvent for the complex Airy operator
$D_x^2-ix$ on the positive semi-axis with Dirichlet boundary condition
at the origin (the results for ${\mathcal G}^{+,D}(x,y\,;\lambda)$
are similar).  Matching the boundary conditions, one gets
\begin{equation}
\mathcal G ^{-,D} (x,y\,;\lambda) = \begin{cases}
2\pi \frac{\Ai(e^{-i\alpha} w_y)}{\Ai(e^{-i\alpha} w_0)} \bigl[\Ai(e^{i\alpha} w_x) \Ai(e^{-i\alpha} w_0) \cr
\hskip 30mm - \Ai(e^{-i\alpha} w_x) \Ai(e^{i\alpha} w_0)\bigr]  \quad (0 < x < y)\,,  \cr 
2\pi \frac{\Ai(e^{-i\alpha} w_x)}{\Ai(e^{-i\alpha} w_0)} \bigl[\Ai(e^{i\alpha} w_y) \Ai(e^{-i\alpha} w_0) \cr
\hskip 30mm - \Ai(e^{-i\alpha} w_y) \Ai(e^{i\alpha} w_0)\bigr]  \quad (x > y)\,,   \end{cases}
\end{equation}
where $\Ai(z)$ is the Airy function, $w_x = ix + \lambda$, and
$\alpha = 2\pi/3\,$. \\ The above expression can also be written as
\begin{equation}
\mathcal G ^{-,D}(x,y\,;\lambda) =  \mathcal G _0 ^-(x,y\,;\lambda) + \mathcal G ^{-,D}_1(x,y\,;\lambda),
\end{equation}
where $ \mathcal G _0 ^-(x,y\,;\lambda)$ is the resolvent for the
complex Airy operator $D_x^2 - ix$ on the whole line,
\begin{equation}
\label{eq:G0_free}
\mathcal G _0 ^-(x,y\,;\lambda) = \begin{cases} 2\pi \Ai(e^{i\alpha} w_x) \Ai(e^{-i\alpha} w_y)  \quad (x<y), \cr
2\pi \Ai(e^{-i\alpha} w_x) \Ai(e^{i\alpha} w_y)  \quad (x>y), \end{cases}
\end{equation}
and
\begin{equation}
\label{eq:Resolvent_Dirichlet}
\mathcal G ^{-,D}_1(x,y\,;\lambda) = - 2\pi \frac{\Ai(e^{i\alpha} \lambda)}{\Ai(e^{-i\alpha} \lambda)} 
\Ai\bigl(e^{-i\alpha} (ix+\lambda)\bigr) ~\Ai\bigl(e^{-i\alpha} (iy+\lambda)\bigr)  .
\end{equation}
The resolvent is compact.  The poles of the resolvent are determined
by the zeros of $\Ai(e^{-i\alpha} \lambda)$, i.e., $\lambda_n =
e^{i\alpha} a_n\,$, where the  $a_n$ are zeros of the Airy function:
$\Ai(a_n) = 0\,$.  The eigenvalues have multiplicity $1$ (no Jordan
block).  See Appendix \ref{AppA}.\\

As a consequence of the analysis of the numerical range of the
operator, we have
\begin{proposition}
  \begin{equation}
  || \mathcal G^{\pm,D}(\lambda)|| \leq \frac{1}{|\Re \lambda|}\,,\, \mbox{ if } \Re \lambda < 0\,;
  \end{equation}
  and
  \begin{equation}
  || \mathcal G^{\pm ,D}(\lambda)|| \leq \frac{1}{|\Im \lambda|}\,,\, \mbox{ if } \mp \Im \lambda > 0\,.
  \end{equation}
\end{proposition}
This proposition together with the Phragmen-Lindel\"of principle
(Theorem~\ref{thPL}) and Proposition~\ref{SchD} implies (see
\cite{Agm} or \cite{DS})
\begin{proposition}\label{CompD}
The space generated by the eigenfunctions of the Dirichlet realization
$\mathcal A^{\pm,D}$ of $D_x^2\pm i x$ is dense in $L^2_+$.
\end{proposition}
It is proven in \cite{Hen} that there is no  Riesz basis of
eigenfunctions. 

At the boundary of the numerical range of the operator, it is
interesting to analyze the behavior of the resolvent.  Numerical
computations lead to the observation that 
\begin{equation}
\label{eq:auxil_8}
\lim_{\lambda \ar +\infty} || \mathcal G^{\pm ,D}(\lambda) ||_{\mathcal L(L^2_+)} =0 \,.
\end{equation}
As a new result, we will prove
\begin{proposition}\label{conja}
When $\lambda$  tends to $+\infty$, we have
\begin{equation}
\label{eq:GD_HS}
 || \mathcal G^{\pm ,D}(\lambda) ||_{HS} \approx  \lambda^{-\frac 14} (\log \lambda)^\frac 12 \,.
\end{equation}
\end{proposition}
Here we use the convention that ``$A(\lambda) \approx B(\lambda)$ as
$\lambda \ar +\infty$'' means that there exist $C$ and $\lambda_0$
such that
\begin{equation*}
\frac 1C \leq \frac{|A(\lambda)|}{|B(\lambda)|} \leq C\,,\quad \forall \lambda \geq \lambda_0\,,
\end{equation*}
or, in other words, $A = \mathcal O (|B|)$ and $B=\mathcal O
(|A|)$. \\ 
The proof of this proposition will be given in Appendix
\ref{AppC}.\\ Note that, as $|| \mathcal G^{\pm ,D}(\lambda)
||_{\mathcal L (L^2)} \leq || \mathcal G^{\pm ,D}(\lambda) ||_{HS} $,
the estimate \eqref{eq:GD_HS} implies \eqref{eq:auxil_8}.\\

\subsection{The complex Airy operator on the half-line: Neumann case}

Similarly, we can look at the Neumann realization $\mathcal A^{\pm,N}$
of $D_x^2 \pm ix$ on $\R_+$ (the analysis on the negative semi-axis is
similar).

One can use for example the Lax-Milgram theorem and take as form
domain
\begin{equation*}
V^N =\{ u \in H^1_+ \,,\, x^\frac 12 u \in L^2_+\}.
\end{equation*}
We recall that the Neumann condition appears when writing the domain
of the operator ${\mathcal A}^{\pm,N}$.\\ 
As in the Dirichlet case (Proposition \ref{SchD}), this implies
\begin{proposition}\label{SchN}
The resolvent $\mathcal G^{\pm,N}(\lambda) := ({\mathcal A}^{\pm,N} -
\lambda)^{-1}$ is in the Schatten class $C^p$ for any $p>\frac 32$\,.
\end{proposition}
More explicitly, the resolvent of ${\mathcal A}^{-,N}$ is obtained as
\begin{equation*}
\mathcal G^{-,N} (x,y\,;\lambda) =  \mathcal G _0 ^-(x,y\,;\lambda) + \mathcal G _1^{-,N}(x,y\,;\lambda)\,  \quad \textrm{for}~(x,y)\in \mathbb R_+^2,
\end{equation*}
where  $\mathcal G_0^-(x,y\,;\lambda)$ is given by
(\ref{eq:G0_free}) and $\mathcal G ^{-,N}_1(x,y\,;\lambda)$ is
\begin{equation}
\label{eq:Resolvent_N}
\mathcal G ^{-,N}_1(x,y\,;\lambda) = - 2\pi \frac{e^{i\alpha} \Ai'(e^{i\alpha} \lambda)}{e^{-i\alpha} \Ai'(e^{-i\alpha} \lambda)} 
\Ai\bigl(e^{-i\alpha} (ix+\lambda)\bigr)~ \Ai\bigl(e^{-i\alpha} (iy+\lambda)\bigr)\,. 
\end{equation}
The poles of the resolvent are determined by zeros of
$\Ai'(e^{-i\alpha} \lambda)$, i.e., $\lambda_n = e^{i\alpha} a'_n$,
where $a'_n$ are zeros of the derivative of the Airy function:
$\Ai'(a'_n) = 0$.  The eigenvalues have multiplicity $1$ (no Jordan
block).  See Appendix \ref{AppA}.\\
As a consequence of the analysis of the numerical range of the
operator, we have
\begin{proposition}
\begin{equation}
  || \mathcal G^{\pm,N}(\lambda)|| \leq \frac{1}{|\Re \lambda|}\,,\, \mbox{ if } \Re \lambda < 0\,;
  \end{equation}
  and
  \begin{equation}
  || \mathcal G^{\pm,N}(\lambda)|| \leq \frac{1}{|\Im \lambda| }\,,\, \mbox{ if } \mp  \Im \lambda  > 0\,.
  \end{equation}
\end{proposition}
This proposition together with Proposition \ref{SchN} and the
Phragmen-Lindel\"of principle implies the completeness of the
eigenfunctions:
\\
\begin{proposition}\label{CompN}
The space generated by the eigenfunctions of the Neumann realization
$\mathcal A^{\pm,N}$ of $D_x^2 \pm ix$ is dense in $L^2_+$.
\end{proposition}

At the boundary of the numerical range of the operator, we have
\begin{proposition}\label{conjb}
When $\lambda$  tends to $+\infty$, we have
\begin{equation}
 || \mathcal G^{\pm ,N}(\lambda) ||_{HS} \approx \lambda^{-\frac 14} (\log \lambda)^\frac 12 \,.
\end{equation}
\end{proposition}
{\bf Proof}\\
Using the Wronskian \eqref{eq:Wronskian} for Airy functions, we have
\begin{equation}
\mathcal G^{-,D} (x,y\,;\lambda) - \mathcal G^{-,N} (x,y\,;\lambda) = - i e^{i\alpha} \frac{\Ai(e^{-i\alpha} w_x) 
\Ai(e^{-i\alpha} w_y) }{ \Ai (e^{-i\alpha} \lambda) \Ai'(e^{-i\alpha} \lambda)}\,.
\end{equation}
Hence
\begin{equation*}
|| \mathcal G^{-,D} (x,y\,;\lambda) - \mathcal G^{-,N} (x,y\,;\lambda) ||_{HS}^2 = 
\frac{(\int_0^{+\infty} |\Ai(e^{-i\alpha} w_x) |^2\,dx) ^2}{ | \Ai (e^{-i\alpha} \lambda) |^2\,  |\Ai'(e^{-i\alpha} \lambda)|^2}\,.
\end{equation*}
We will show in \eqref{estI0} that
\begin{equation*}
\int_0^{+\infty} |\Ai(e^{-i\alpha} w_x) |^2\,dx \leq C \lambda^{-\frac 12} \exp \left(\frac 43 \lambda^\frac 32\right)\,.
\end{equation*}
On the other hand, using \eqref{5} and \eqref{5a}, we obtain, for
$\lambda \geq \lambda_0$
\begin{equation*}
|\Ai (e^{-i\alpha} \lambda) \Ai'(e^{-i\alpha} \lambda)| \geq \frac{1}{4\pi} \exp\left( \frac43 \lambda^\frac 32 \right)
\end{equation*}
(this argument will also be used in the proof of \eqref{eq:est_f}).
We have consequently obtained that there exist $C>0$ and $\lambda_0
>0$ such that, for $\lambda \geq \lambda_0$,
\begin{equation}
|| \mathcal G^{-,D} (\lambda) - \mathcal G^{-,N} (\lambda) ||_{HS} \leq C\, |\lambda|^{-\frac 14}\,.
\end{equation}
The proof of the proposition follows from Proposition \ref{conja}.

\subsection{The complex Airy operator on the half-line: Robin case}

For completeness, we  provide new results for the complex Airy
operator on the half-line with the Robin boundary condition that
naturally extends both Dirichlet and Neumann cases:
\begin{equation}
\left[\frac{\partial}{\partial x} \mathcal G^{-,R} (x,y\,;\kappa, \lambda) - \kappa \,  \mathcal G^{-,R} (x,y\,;\kappa, \lambda)\right]_{x=0} = 0\, ,
\end{equation}
with a positive parameter $\kappa$.  The operator is associated with
the sesquilinear form defined on $H^1_+ \times H^1_+$ by
\begin{equation}
a^{-,R} (u,v) = \int_0^{+\infty} u'(x)\bar v '(x) \,dx - i \int_0^{+\infty} x u(x) \bar v (x)\, dx + \kappa \, u(0) \bar v (0)\,.
\end{equation}
The distribution kernel of the resolvent is obtained as
\begin{equation*}
\mathcal G^{-,R} (x,y\,;\lambda) =  \mathcal G _0 ^-(x,y\,;\lambda) + \mathcal G _1^{-,R}(x,y\,;\kappa, \lambda)\,  \quad \textrm{for}~ (x,y)\in \mathbb R_+^2,
\end{equation*}
where 
\begin{equation}
\begin{split}
\mathcal G ^{-,R}_1(x,y\,;\kappa, \lambda) & = - 2\pi \frac{i e^{i\alpha} \Ai'(e^{i\alpha} \lambda) - \kappa \Ai(e^{i\alpha}\lambda)}
{i e^{-i\alpha} \Ai'(e^{-i\alpha} \lambda) - \kappa \Ai(e^{-i\alpha} \lambda)}  \\
& \times  \Ai\bigl(e^{-i\alpha} (ix+\lambda)\bigr)~ \Ai\bigl(e^{-i\alpha} (iy+\lambda)\bigr)\,.   \\
\end{split}
\end{equation}
Setting $\kappa = 0$, one retrieves Eq. (\ref{eq:Resolvent_N}) for the
Neumann case, while the limit $\kappa\to +\infty$ yields
Eq. (\ref{eq:Resolvent_Dirichlet}) for the Dirichlet case, as
expected.  As previously, the resolvent is compact and actually in the
Schatten class $\mathcal C^p$ for any $p>\frac 32$ (see Proposition
\ref{SchD}).  Its poles are determined as (complex-valued)
solutions of the equation
\begin{equation}\label{eigenvalueR}
f^R(\kappa, \lambda):=i e^{-i\alpha} \Ai'(e^{-i\alpha} \lambda) - \kappa \Ai(e^{-i\alpha} \lambda) = 0\,.
\end{equation}
Except for the case of small $\kappa$, in which the eigenvalues might
be localized close to the eigenvalues of the Neumann problem (see
Section \ref{s3} for an analogous case), it does not seem easy to
localize all the solutions of \eqref{eigenvalueR} in general.
Nevertheless one can prove that the zeros of $f^R(\kappa, \cdot)$ are
simple.  If indeed $\lambda$ is a common zero of $f^R$ and $(f^R)'$,
then either $\lambda +\kappa^2=0$, or $e^{-i \alpha} \lambda$ is a
common zero of $\Ai$ and $\Ai'$.  The second option is excluded by the
properties of the Airy function, whereas the first option is excluded
for $\kappa \geq 0$ because the spectrum is contained in the positive
half-plane.

As a consequence of the analysis of the numerical range of the
operator, we have
\begin{proposition}
  \begin{equation}
  || \mathcal G^{\pm,R}(\kappa, \lambda)|| \leq \frac{1}{|\Re \lambda|}\,,\, \mbox{ if } \Re \lambda < 0\,;
  \end{equation}
  and
  \begin{equation}
  || \mathcal G^{\pm ,R}(\kappa,\lambda)|| \leq \frac{1}{|\Im \lambda|}\,,\, \mbox{ if } \mp \Im \lambda > 0\,.
  \end{equation}
\end{proposition}
This proposition together with the Phragmen-Lindel\"of principle
(Theorem~\ref{thPL}) and the fact that the resolvent is in the
Schatten class $\mathcal C^p$, for any $p>\frac 32$, implies
\begin{proposition}\label{CompR}
For any $\kappa \geq 0$, the space generated by the eigenfunctions of
the Robin realization $\mathcal A^{\pm,R}$ of $D_x^2\pm i x$ is dense
in $L^2_+$.
\end{proposition}

At the boundary of the numerical range of the operator, it is
interesting to analyze the behavior of the resolvent.  Equivalently to
Propositions \ref{conja} or \ref{conjb}, we have
\begin{proposition}\label{conjc}
When $\lambda$  tends to $+\infty$, we have
\begin{equation}
 || \mathcal G^{\pm ,R}(\kappa, \lambda) ||_{HS} \approx \lambda^{-\frac 14} (\log \lambda)^\frac 12 \,.
\end{equation}
\end{proposition}
{\bf Proof}.\\ The proof is obtained by using Proposition \ref{conjb}
and computing, using \eqref{eq:Wronskian},
\begin{eqnarray*}
|| \mathcal G^{-,N} (\lambda) &-& \mathcal G^{-, R} (\kappa, \lambda)||_{HS}^2 =  \left(\int_0^{+\infty} |\Ai (e^{-i\alpha} w_x )|^2 dx\right)^2  \\
&\times& \frac{\kappa}{2 \pi}\,  \frac{1}{ | i e^{-i \alpha}\, \Ai'(e^{-i\alpha} \lambda) -
\kappa \Ai(e^{-i\alpha} \lambda)|^2\, | \Ai'(e^{-i\alpha} \lambda) |^2 } .
\end{eqnarray*}
As in the proof of Proposition \ref{conjb}, we show the existence, for
any $\kappa_0 > 0$, of $C>0$ and $\lambda_0$ such that, for $\lambda
\geq \lambda_0$ and $\kappa \in [0,\kappa_0]$,
\begin{equation*}
|| \mathcal G^{-,N} (\lambda) - \mathcal G^{-, R}
(\kappa, \lambda)||_{HS} \leq C |\kappa|  \lambda^{-\frac 34}\,.
\end{equation*}

\section{The complex Airy operator with a semi-permeable barrier: definition and properties}\label{s3}

In comparison with Section \ref{s2}, we now replace the differential
operator $-\frac{d^2}{dx^2}$ by $\mathcal A_1^+ = -\frac{d^2}{dx^2} +
i x $ but keep the same transmission condition.  To give a precise
mathematical definition of the associated closed operator, we consider
the sesquilinear form $a_\nu$ defined for $u=(u_-,u_+)$ and
$v=(v_-,v_+)$ by
\begin{eqnarray}
a_\nu (u,v) &=& \int_{-\infty}^0\Big(u_-'(x)\bar v_-'(x) +i \,xu_-(x)\bar v_-(x)+\nu\, u_-(x)\bar v_-(x)\Big)\,dx \nonumber \\
& & + \int_0^{+\infty}\Big(u_+'(x)\bar v_+'(x) +i \,xu_+(x)\bar v_+(x)+\nu\, u_+(x)\bar v_+(x)\Big)\,dx \nonumber \\
& &  + \kappa \big(u_+(0)-u_-(0)\big)
 \big(\barr{v_+(0)-v_-(0)}\big)\,, \label{defForm1d}
\end{eqnarray}
where the form domain $\mathcal V$ is
\[ 
 \mathcal V := \Big\{u=(u_-,u_+)\in H_-^1\times H_+^1 : |x|^\frac 12 u\in L_-^2\times L_+^2\Big\}\,.
\]
The space $\mathcal V$ is endowed with the  Hilbertian norm
\[
 \|u \|_{\mathcal V} := \sqrt{ { \|u_-\|_{H_-^1}^2+\|u_+\|_{H_+^1}^2 } +\||x|^{1/2}u\|_{L^2}^2}\,.
\]
We first observe
\begin{lemma}
For any $\nu \geq 0$, the sesquilinear form $a_\nu $ is
continuous on $\mathcal V\,$.
\end{lemma}
\textbf{Proof:}
The proof is similar to that of Lemma \ref{lemConta0}, the additional
term $i \Big(\int_{-\infty}^0 x\, u_-(x)\,\bar v_-(x)\,dx+
\int_0^{+\infty}x\, u_+(x)\, \bar v_+(x)\,dx\Big)$ being obviously bounded by
$\, \|u\|_V\|v\|_V\,$.
\hfill $\square$\\

Let us notice that, if $u$ and $v$ belong to $H_-^2\times H_+^2$ and
satisfy the boundary conditions (\ref{eq:transmission}), then an
integration by parts yields
\begin{eqnarray*}
 a_\nu (u,v) & = & \int_{-\infty}^0\big(-u_-''(x)+i xu_-(x)+\nu\, u_-(x)\big)\bar v_-(x)\,dx\\
 & & + \int_0^{+\infty}\big(-u_+''(x)+i xu_+(x)+\nu\, u_+(x)\big)\bar v_+(x)\,dx \\
 & & +\big(u_+'(0) + \kappa(u_-(0)-u_+(0))\big)\big(\barr{v_-(0)-v_+(0)}\big)\\
 & = & \sca{\left(-\frac{d^2}{dx^2}+i x+\nu\right)u}{v}_{L_-^2\times L_+^2}\,.
\end{eqnarray*}
Hence the operator associated with the form $a_\nu\,$, once
defined appropriately, will act as $-\frac{d^2}{dx^2}+ix+\nu$ on
$C_0^\infty(\mathbb{R}\setminus\{0\}) \,$.\\

As the imaginary part of the potential $ix$ changes sign, it is not
straightforward to determine whether the sesquilinear form $a_\nu$
is coercive, i.e., whether there exists $\nu_0$ such that for
$\nu \geq \nu_0$ the following estimate
\beq\label{coercive}
 \exists\alpha>0\,,~\forall u\in \mathcal V\,,~~~|a_\nu (u,u)|\geq\alpha\|u\|_{\mathcal V}^2
\eeq
holds. \\
Let us show that it is indeed not true. Consider for instance the sequence 
\begin{equation*}
 u_n(x) = (\chi(x+n),\chi(x-n))\,,~~~~n\geq1\,,
\end{equation*}
where $\chi\in\mathcal{C}_0^\infty(-1,1)$ is an even function such
that $\chi(x) = 1$ for $x\in[-1/2,1/2]\,$. \\ 
Then $\|u_n'\|_{L^2(-\infty,0)}$ and $\|u_n'\|_{L^2(0,+\infty)}$ are
bounded, and
\[
 \int_\mathbb{R}x\, |u_n(x)|^2\,dx = 0\,,
\]
since $x\mapsto x|u_n(x)|^2$ is odd, whereas
$\||x|^{1/2}u_n\|_{L^2}\longrightarrow+\infty$ as
$n\rightarrow+\infty\,$.  Consequently
\begin{equation*}
 \frac{|a_\nu (u_n,u_n)|}{\|u_n\|_{\mathcal V}^2}\longrightarrow0
\mbox{ as }n\rightarrow+\infty\,,
\end{equation*}
and (\ref{coercive}) does not hold.\\

Due to the lack of coercivity, the standard version of the Lax-Milgram
theorem does not apply.  We shall instead use the following
generalization introduced in \cite{AlmHel}.

\begin{theorem}\label{thmLaxMilgramAH}
Let $\mathcal{V}\subset\mathcal{H}$ be two Hilbert spaces such that
$\mathcal{V}$ is continuously embedded in $\mathcal{H}\,$ and
$\mathcal{V}$ is dense in $\mathcal{H}\,$.  Let $a$ be a continuous
sesquilinear form on $\mathcal{V}\times\mathcal{V}\,$, and assume that
there exists $\alpha>0$ and two bounded linear operators $\Phi_1$ and
$\Phi_2$ on $\mathcal{V}$ such that, for all $u\in\mathcal{V}\,$,
\beq\label{PseudoCoerc}
   \left\{
   \begin{array}{ccc}
    |a(u,u)|+|a(u,\Phi_1u)|& \geq& \alpha\, \|u\|_\mathcal{V}^2\,,\\
    |a(u,u)|+|a(\Phi_2u,u)|&\geq&\alpha \, \|u\|_\mathcal{V}^2\,.
   \end{array}
\right.
\eeq
Assume further that $\Phi_1$ extends to a bounded linear operator on
$\mathcal{H}\,$.\\ Then there exists a closed, densely-defined
operator $S$ on $\mathcal{H}$ with domain
\[
 \CD(S) = \big\{u\in\mathcal{V} : v\mapsto a(u,v)~\textrm{ can be extended continuously on }~\mathcal{H}\,\big\}\,,
\]
such that, for all $u\in\CD(S)$ and $v\in \mathcal V\,$,
\[
 a(u,v) = \sca{Su}{v}_\mathcal{H}\,.
\]
\end{theorem}

Now we want to find two operators $\Phi_1$ and $\Phi_2$ on $\mathcal
V$ such that the estimates (\ref{PseudoCoerc}) hold for the form
$a_\nu $ defined by (\ref{defForm1d}).\\ 
First we have, as in
(\ref{estCoerca0}),
\begin{eqnarray*}
\Re a_\nu (u,u) & \geq & (1-|\kappa| \varepsilon)\left(\int_{-\infty}^0|u_-'(x)|^2\,dx + \int_0^{+\infty}|u_+'(x)|^2\,dx\right)  \\
 && +\big(\nu-|\kappa| C(\varepsilon)\big)\|u\|_{L^2}^2\,.
\end{eqnarray*}
Thus by choosing $\varepsilon$ and $\nu$ appropriately we get, for
some $\alpha_1> 0\,$,
\beq\label{estpoids1}
|a_\nu (u,u)|\geq\alpha_1\left(\int_{-\infty}^0|u_-'(x)|^2\,dx+\int_0^{+\infty}|u_+'(x)|^2\,dx + \|u\|_{L^2}^2\right)\,.
\eeq
It remains to estimate the term $\||x|^{1/2}u\|_{L^2}$ appearing in
the norm $\|u\|_{\mathcal{V}}\,$.  For this purpose, we introduce the operator
\[
 \rho :(u_-,u_+)\longmapsto (-u_-,u_+)\,,
\]
which corresponds to the multiplication operator by the function
$\textrm{sign}\,x\,$.\\ It is clear that $\rho$ maps $\mathcal
H$ onto $\mathcal H$ and $\mathcal V$ onto $\mathcal V$.  Then we have
\beq
\Im a_\nu (u,\rho u) = \, \||x|^{1/2}u\|_{L^2}^2\,.
\eeq
Thus using (\ref{estpoids1}), there exists $\alpha_0$ such that, for
all $u\in \mathcal{V}\,$,
\[
 |a_\nu (u,u)|+|a_\nu (u,\rho u)|\geq\alpha\|u\|_{\mathcal V}^2\,.
\]
Similarly, for all $u\in \mathcal{V}\,$,
\[
|a_\nu (u,u)|+|a_\nu (\rho u,u)|\geq\alpha\|u\|_{\mathcal V}^2\,.
\]
In other words, the estimate (\ref{PseudoCoerc}) holds, with 
$\Phi_1=\Phi_2=\rho\,$.  Hence the assumptions of Theorem
\ref{thmLaxMilgramAH} are satisfied, and we can define a closed
operator $\mathcal A^+_1 :=S-\nu$, which is given by the identity
\[
 \forall u\in\CD(\mathcal A^+_1)\,,~\forall v\in \mathcal V\,,~~~a_\nu (u,v) = \sca{\mathcal A^+_1u + \nu u}{v}_{L_-^2\times L_+^2}
\]
on the domain
\begin{eqnarray*}
 \CD(\mathcal A^+_1) =\CD(S) &= &\big\{u\in\mathcal{V} : v\mapsto a_\nu (u,v)~\textrm{ can be extended continuously }\\
 & &~~~~~~~~~~~~~~~~~ \textrm{ on }~L_-^2\times L_+^2\,\big\}\,.
\end{eqnarray*}
Now we shall determine explicitly the domain $\CD(\mathcal A^+_1)\,$.

Let $u\in \mathcal V\,$. The map $v\mapsto a_\nu (u,v)$ can be extended
continuously on $L_-^2\times L_+^2$ if and only if there exists some
$w_u=(w_u^-,w_u^+)\in L_-^2\times L_+^2$ such that, for all $v\in
\mathcal V\,$, $a_\nu (u,v) = \sca{w_u}{v}_{L^2}\,$.  Then due to the
definition of $a_\nu (u,v)\,$, we have necessarily
\[
 w_u^- = -u_-''+ixu_-+\nu u_-~~~\textrm{ and }~~~w_u^+ = -u_+''+ixu_++\nu u_+
\]
in the sense of distributions respectively in $\mathbb R_-$ and
$\mathbb R_+$, and $u$ satisfies the conditions
(\ref{eq:transmission}).  Consequently, the domain of $\mathcal A^+_1$
can be rewritten as
\begin{eqnarray*}
\CD(\mathcal A^+_1) &=& \Big\{u\in \mathcal V : (-u_-'' +i  xu_-,-u_+''+ixu_+)\in L_-^2\times L_+^2\\
& & \textrm{ and }~u\textrm{ satisfies conditions }(\ref{eq:transmission})\big\}\,.
\end{eqnarray*}
We now prove that $\CD(\mathcal A^+_1)  =\widehat D$ where
\begin{eqnarray*}
\widehat D &=& \Big\{u\in \mathcal V:   (u_-,u_+) \in H^2_-\times H^2_+\,,\, (xu_-,xu_+) \in L_-^2\times L_+^2 \\
& &\qquad\qquad \qquad \qquad  \textrm{ and }~u\textrm{ satisfies conditions }(\ref{eq:transmission})\big\}\,.
\end{eqnarray*}
It remains to check that this implies $(u_-,u_+)\in H^2_-\times H^2_+$.
The only problem is at $+\infty$.  Let $u_+$ be as above and let
$\chi$ be a nonnegative function equal to $1$ on $[1,+\infty)$ and
with support in $(\frac 12, +\infty)$.  It is clear that the natural
extension by $0$ of $\chi u_+$ to $\mathbb R$ belongs to $L^2(\mathbb
R)$ and satisfies
\begin{equation*}
\left (-\frac{d^2}{dx^2} +i  x\right) (\chi u_+)\in L^2(\mathbb R)\,. 
\end{equation*}
One can apply for $\chi u_+$ a standard result for the
domain of the accretive maximal extension of the complex Airy operator
on $\mathbb R$ (see for example \cite{Hel1}).\\

Finally, let us notice that that the continuous embedding
\[
 \mathcal V\hookrightarrow L^2(\mathbb{R} ; |x|dx)\cap \big(H_-^1\times H_+^1\big)
\]
implies that $\mathcal A _1^+$  has a compact resolvent; hence its spectrum is
discrete.\\

Moreover, from the characterization of the domain and its inclusion in
$\widehat D$, we deduce the stronger
\begin{proposition}\label{propSchatten}
 There exists $\lambda_0$ ($\lambda_0=0$ for $\kappa >0$) such that
 $(\mathcal A_1^+ -\lambda_0)^{-1}$ belongs to the Schatten class
 $\mathcal C^p$ for any $p > \frac 32$.
\end{proposition}
Note that if it is true for some $\lambda_0$ it is true for any
$\lambda$ in the resolvent set.

\begin{remark}
The adjoint of $\mathcal A_1^+$ is the operator associated by the same
construction with $D_x^2-i  x$.  $\mathcal A_1 ^- +\lambda$ being
injective, this implies by a general criterion \cite{Hel1} that
$\mathcal A _1^++\lambda$ is maximal accretive, hence generates a
contraction semigroup.
\end{remark}

The following statement summarizes the previous discussion.
\begin{proposition}
 The operator $\mathcal A^+_1$ acting as
  \[
 u\mapsto   \mathcal A_1^+ u  = \left(-\frac{d^2}{dx^2}u_-+ixu_-,\, -\frac{d^2}{dx^2}u_+ +i xu_+\right)
 \]
on the domain
\begin{eqnarray}\label{defdomA1}
 \CD(\mathcal A^+_1)& = &\big\{u\in H_-^2\times H_+^2 : xu\in L_-^2\times L_+^2 \nonumber \\
 &&~~~~~~~~~~~~~~~ \textrm{ and }u\textrm{ satisfies conditions }(\ref{eq:transmission})\big\}
\end{eqnarray}
is a closed operator with compact resolvent.\\
There exists some positive $\lambda$ such that the operator $\mathcal A^+_1+\lambda$ is maximal accretive.
\end{proposition}

\begin{remark}\label{remsym}
We have
\begin{equation}
\Gamma  \mathcal A^+_1 =  \mathcal A _1 ^-\,,
\end{equation}
where $\Gamma$ denotes the complex conjugation:
\begin{equation*}
\Gamma (u_-\, ,\, u_+) = ( \bar u_-\, ,\, \bar u_+)\,.
\end{equation*}
This implies that the distribution kernel of the resolvent satisfies:
\begin{equation}\label{sym}
\mathcal G (x,y\,;\lambda) = \mathcal G  (y,x; \lambda)\,,
\end{equation}
for any $\lambda$ in the resolvent set.
\end{remark}

\begin{remark}[PT-Symmetry]
If $(\lambda,u)$ is an eigenpair, then $(\bar
\lambda, \bar u(-x))$ is also an eigenpair.  Let indeed $v(x) =
\bar u (-x)$.  This means $v_-(x) = \bar u_+(-x)$ and $v_+(x) = \bar
u_- (-x)$.  Hence we get that $v$ satisfies \eqref{Condbis} if $u$
satisfies the same condition:
\begin{equation*}
v'_ - (0) = - \bar u_+'(0) =  \kappa (\bar u_- (0) - \bar u_+(0)) = + \kappa \left(v_+(0)-v_-(0)\right)\,.    
\end{equation*}
Similarly one can verify that 
\begin{equation*}
\begin{array}{ll}
 \left( -\frac{d^2}{dx^2} + i  x \right) v_+ (x)& = \overline{  \left( -\frac{d^2}{dx^2} - i  x \right) u_- (- x)}\\
 & =  \overline{  \left( \left(-\frac{d^2}{dx^2} + i  x \right) u_-\right) (- x)}\\
 &  =\bar \lambda \, v_+(x)\,.
 \end{array}
\end{equation*}

\end{remark}

\section{Exponential decay of the associated semi-group.}\label{s5}

In order to control the decay of the associated semi-group, we follow
what has been done for the Neumann or Dirichlet realization of the
complex Airy operator on the semi-axis (see for example \cite{Hel1} or
\cite{Hen,Hen1}).
\begin{theorem}\label{thmGP} 
Assume $\kappa >0$, then for any $\omega <
\inf\{\Re\sigma(\mathcal A^+_1)\}$, there exists $M_\omega$ such that
\begin{equation*}
|| \exp (- t \mathcal A_1^+)||_{\mathcal L (L^2_-\times L^2_+)} \leq  M_\omega \exp (- \omega t)\, ,
\end{equation*}
where $\sigma(\mathcal A^+_1)$ is the spectrum of $\mathcal A^+_1$.
\end{theorem}
To apply the quantitative Gearhart-Pr\"uss theorem (see \cite{Hel1})
to the operator $\mathcal A_1^+$, we should prove that
\[
 \sup_{\Re z \leq\omega}\|(\mathcal A^+_1-z)^{-1}\|\leq C_\omega\,,
\]
for all $\omega<\inf\Re\sigma(\mathcal A^+_1):=\omega_1\,$. \\ First
we have by accretivity (remember that $\kappa >0$), for $ \Re \lambda
< 0\,$,
\begin{equation}\label{b10} 
||(\mathcal A_1^+ -\lambda)^{-1}|| \leq \frac{1}{|\Re \lambda|}\,.
\end{equation}
So it remains to analyze the resolvent in
the set
\begin{equation*}
 0 \leq \Re \lambda\leq \omega_1-\epsilon\,,\quad |\Im \lambda| \geq C_\epsilon>0\,,
\end{equation*}
where $C_\epsilon >0$ is sufficiently large.  Let us show the
following lemma.
\begin{lemma}\label{Lemma4.2}~\\
For any $\alpha >0$, there exist $C_\alpha >0$ and $D_\alpha >0$ such
that for any $ \lambda \in \{ \omega \in \mathbb C\,:\, \Re \omega \in
[-\alpha, +\alpha] \mbox{ and } | \Im \omega| >D_\alpha\}\,,\, $
\beq\label{7.5a}
 ||(\mathcal A_1^\pm -\lambda)^{-1}|| \leq C_\alpha\,.
\eeq
\end{lemma}
{\bf Proof.}\\ Without loss of generality, we treat the case when $\Im
\lambda >0\,$.  As in \cite{BHHR}, the main idea of the proof is to
approximate $(\mathcal A_1^+-\lambda)^{-1}$ by a sum of two operators:
one of them is a good approximation when applied to functions
supported near the transmission point, while the other one takes care
of functions whose support lies far away from this point.

The first operator $\check{\mathcal A}$ is associated with the
sesquilinear form $\check{a}$ defined for $u=(u_-,u_+)$ and
$v=(v_-,v_+)$ by
\begin{eqnarray}
\check a(u,v) &=& \int_{-\Im \lambda/2}^0\Big(u_-'(x)\bar v_-'(x) +i \,xu_-(x)\bar v_-(x)+\lambda\, u_-(x)\bar v_-(x)\Big)\,dx \nonumber \\
& & + \int_0^{\Im \lambda/2}\Big(u_+'(x)\bar v_+'(x) +i \,xu_+(x)\bar v_+(x)+\lambda\, u_+(x)\bar v_+(x)\Big)\,dx \nonumber \\
& & \, + \,  \kappa\,  \big(u_+(0)-u_-(0)\big)
 \big(\barr{v_+(0)-v_-(0)}\big)\,, \label{defForm1dchapeau}
\end{eqnarray}
where  $u$ and $v$ belong to  the following space:
\begin{equation*}
\mathbb H_0^1 (\CS_\lambda,\mathbb C): = \left( H^1(S_\lambda^-) \times H^1(S_\lambda^+) \right)  \cap \{ u_-(-\Im \lambda/2)= 0\,,\, u_+ (\Im \lambda /2)=0\}\,,
\end{equation*}
with  $\mathcal S^-_\lambda :=(-\Im \lambda/2,0)$ and $\mathcal
S^+_\lambda:=(0,+\Im \lambda/2)$.\\
The domain $D(\check {\mathcal A}) $ of $\check {\mathcal A}$ is the
set of $u\in H^2 (\mathcal S^-_\lambda) \times H^2 (\mathcal
S^+_\lambda)$ such that $ u_-(-\Im \lambda/2)= 0$\,,\,$ u_+ (\Im
\lambda /2)=0$ and $u$ satisfies conditions (\ref{eq:transmission}).
Denote the resolvent of $\check{\mathcal A}$ by $R_1(\lambda)$ in
$\mathcal L(L^2(\mathcal S_\lambda^-,\mathbb C) \times L^2(\mathcal
S_\lambda^+,\mathbb C) )$ and observe also that $R_1(\lambda) \in
\mathcal L(L^2(\mathcal S_\lambda^-,\mathbb C) \times L^2(\mathcal
S_\lambda^+,\mathbb C ), {\mathbb H}_0^1(\CS_\lambda,\mathbb C))$.

We easily obtain (looking at the imaginary part of the sesquilinear
form) that
\begin{equation}
\label{7.8a}
 \| R_1(\lambda)\| \leq \frac{2}{\Im \lambda}\,.
\end{equation}
Furthermore, we have, for $u= R_1(\lambda)\, f$ (with $u=(u_-,u_+)$,
$f=(f_-,f_+)$)
\begin{displaymath}\aligned
\|D_x  R_1(\lambda)f\|^2 & 
=\|D_x u\|^2\\ 
& \leq \|(\mathcal A^+_1-\lambda)u\|\|u\|+\Re\lambda\|u\|^2\\
& \leq \|f\|\| R_1(\lambda)f\|+|\alpha|\|\mathcal R_1(\lambda)f\|^2\\& 
\leq \left(\frac{2}{| \Im \lambda|}
+\frac{4|\alpha|}{|\Im \lambda|^2}\right)\|f\|^2.
\endaligned
\end{displaymath}
Hence there exists $C_0(\alpha)$ such that, for $\Im \lambda \geq 1$
and $\Re \lambda \in [-\alpha,+\alpha]$,
\begin{equation}
  \label{7.9a}
\|D_x  R_1(\lambda)\|  \leq  C_0(\alpha)\,  |\Im \lambda|^{-\frac 12}  \,.
\end{equation}
Far from the transmission point $0$\,, we approximate by the resolvent
$\mathcal G_0^+$ of the complex Airy operator  $\mathcal A^+$ on the
line.  Denote this resolvent by $R_2(\lambda)$ when considered as an
operator in $\mathcal L (L^2_-\times L^2_+)$.  We recall from Section
\ref{s2a} that
\begin{equation}\label{eq:BMa}
  \| R_2(\lambda)\|_{\mathcal L(L^2)} \sim \sqrt{\frac \pi 2}  \, (\Re \lambda)^{- \frac 14}  \exp \left(\frac 43  (\Re \lambda)^\frac 32\right)\,.
\end{equation}
Recall also that for the same reason the norm $\| R_2(\lambda)\|$ is
independent of $\Im\lambda$.  Since $ R_2(\lambda)$ is an entire
function in $\lambda$, we easily obtain a uniform bound on $\|
R_2(\lambda)\|$ for $\Re \lambda \in [-\alpha,+\alpha]$.  Hence,
\begin{equation}
  \label{7.10a}
\| R_{2} (\lambda)\|\leq C_1(\alpha)\,.
\end{equation}
 As for the proof of \eqref{7.9a}, we then show
\begin{equation}
\label{eq:1}
  \|D_x R_2(\lambda)\|  \leq C(\alpha) \,.
\end{equation}

We now use a partition of unity in the $x$ variable in order to
construct an approximate inverse $R^{\rm app}(\lambda)$ for $\mathcal
A_1^+ -\lambda$.  We shall then prove that the difference between the
approximation and the exact resolvent is well controlled as
$\Im\lambda\to+\infty$.  For this purpose, we define the following
triple $(\phi_-,\psi, \phi_+)$ of cutoff functions in
$C^\infty(\mathbb R,[0,1])$ satisfying
\begin{displaymath}\aligned
&\phi_-(t)=1\; \mbox{ on } (-\infty,- 1/2]\,,\quad    \phi_-(t)= 0\; \mbox{ on } [-1/4,+\infty)  \\
&\psi (t) =1\; \mbox{ on } [-1/4, 1/4] \,,\quad  \psi(t) =0 \mbox{ on }(-\infty,- 1/2]\cup  [ 1/2, +\infty) \,, \\
&\phi_+(t)=1\; \mbox{ on } [1/2,+\infty) \,,\quad    \phi_+(t)= 0\; \mbox{ on } (-\infty, 1/4]  ,\\
&\phi_-(t)^2+\psi(t)^2 + \phi_+ (t)^2 =  1 \text{ on } \mathbb R \,,
\endaligned
\end{displaymath}
and then set
\begin{displaymath}
\phi_{\pm, \lambda }(x)= \phi_\pm \Big(\frac{x}{\Im \lambda}\Big)\,, \quad \psi_\lambda(x)
=\psi\Big(\frac{x}{\Im \lambda}\Big)\,.
\end{displaymath}

The approximate inverse $R^{\rm app}(\lambda)$ is then constructed as
\begin{equation}\label{7.11a}
R^{\rm app}(\lambda) = \phi_{-,\lambda} R_2 (\lambda)\phi_{-,\lambda} +  \psi_\lambda R_1(\lambda) \psi_\lambda
 + \phi_{+,\lambda } R_2 (\lambda)\phi_{+,\lambda}\,,
\end{equation}
where $\phi_{\pm,\lambda}$ and $\psi_\lambda$ denote the operators of
multiplication by the functions $\phi_{\pm,\lambda} $ and
$\psi_\lambda $.  Note that $\psi_\lambda$ maps $L^2_-\times L^2_+$
into $L^2(\mathcal S_\lambda^-) \times L^2(\CS_\lambda^+)$.  In
addition,
\begin{displaymath}\aligned
& \psi_\lambda:~D(\check {\mathcal A} ) \to D(\mathcal A^+_1),\\
&\phi_\lambda:~  D(\mathcal A^+) \to D(\mathcal A^+_1)\,.
\endaligned
\end{displaymath}
Here we can define $\phi_\lambda (u_-,u_+)$ as $(\phi_{-,\lambda}\, u_-\,,\,
\phi_{+,\lambda}\,u_+)$.\\ From \eqref{7.8a} and \eqref{7.10a} we get,
for sufficiently large $\Im\lambda$,
\begin{equation}\label{7.12a}
\| R^{\rm app}(\lambda)\|\leq C_3(\alpha).
\end{equation}
Note that
\begin{equation}\label{7.14a}
|\phi^\prime_\lambda (x)| + |\psi^\prime_\lambda (x)| \leq \frac{C}{|\Im \lambda|},\quad 
|\phi^{\prime\prime}_\lambda (x)| + |\psi^{\prime\prime}_\lambda (x)|  \leq \frac{C}{|\Im \lambda|^2}\,.
\end{equation}
Next, we apply $\mathcal A_1^+-\lambda$ to $R^{\rm app}$ to obtain
that
\begin{equation}\label{7.15a}
(\mathcal A_1^+- \lambda) R^{\rm app}(\lambda) = I + [\mathcal A_1^+\,, \,\psi_\lambda] R_1 (\lambda)\psi_\lambda +  [\mathcal A_1^+\,,\,
\phi_\lambda] R_2 (\lambda)\phi_\lambda \,,
\end{equation}
where $I$ is the identity operator on $L^2_-\times L^2_+$, and
\begin{eqnarray}
[\mathcal A_1^+\,,\,\phi_\lambda]& := & \mathcal A_1^+ \phi_\lambda-\phi_\lambda\mathcal A_1^+ \nonumber \\  
& = & [D_x^2\,,\,\phi_\lambda] \nonumber \\
& = & -\frac{2i}{\Im \lambda} \phi^\prime\Big(\frac{x}{\Im \lambda}\Big)
 D_x \; - \; \frac{1}{(\Im \lambda)^2}\phi^{\prime \prime}\Big(\frac{x}{\Im
\lambda}\Big)\,. \label{7.16a}
\end{eqnarray}

A similar relation holds for $[\mathcal A_1^+,\psi_\lambda]$.  Here we
have used \eqref{7.11a}, and the fact that
\begin{displaymath}
(\mathcal A_1^+-\lambda) R_1(\lambda)\psi_\lambda u=\psi_\lambda u\,,\quad (\mathcal A_1^+-\lambda) R_2(\lambda)\phi_\lambda
u=\phi_\lambda  u \,,\quad  \forall u\in L^2_-\times L^2_+\,.
\end{displaymath}
Using \eqref{7.8a}, \eqref{7.9a}, \eqref{eq:1}, and \eqref{7.16a} we
then easily obtain, for sufficiently large $\Im \lambda$,
\begin{equation}\label{7.17a}
\|[\mathcal A_1^+, \psi_\lambda]\, R_1 (\lambda)\|  + \|[\mathcal A_1^+, \phi_\lambda]\, R_2 (\lambda) \| \leq \frac{C_4(\alpha)}{|\Im
  \lambda|}\,.
\end{equation}
Hence, if $|\Im\lambda|$ is large enough then $ I + [\mathcal A_1^+,
\psi_\lambda] R_1 (\lambda)\psi_\lambda + [\mathcal A_1^+,
\phi_\lambda] R_2 (\lambda)\phi_\lambda $ is invertible in $\mathcal
L(L^2_-\times L^2_+)$, and
\begin{equation}\label{7.18a}
\Big\| \Big(I + [\mathcal A_D, \psi_\lambda] R_1 (\lambda)\psi_\lambda +  [\mathcal A_D,
\phi_\lambda] R_2 (\lambda)\phi_\lambda \Big)^{-1}\Big\|\leq C_5(\alpha)\,.
\end{equation}
Finally, since
\begin{displaymath}
(\mathcal A^+_1-\lambda)^{-1}= R^{\rm app}(\lambda) \circ \left( I + [\mathcal A^+_1, \psi_\lambda] R_1 (\lambda)\psi_\lambda +  [\mathcal A^+_1,
\phi_\lambda] R_2 (\lambda)\phi_\lambda \right)^{-1}\,,
\end{displaymath}
we have 
\begin{displaymath}
  \|(\mathcal A^+_1-\lambda)^{-1}\| \leq  \| R^{\rm app}(\lambda)\|  \big\|\big( I + 
[\mathcal A^+_1, \psi_\lambda] R_1 (\lambda)\psi_\lambda +  [\mathcal A^+_1,
\phi_\lambda] R_2 (\lambda)\phi_\lambda \big)^{-1}\big\| \,.
\end{displaymath}
Using \eqref{7.12a} and \eqref{7.18a} we conclude that \eqref{7.5a} is
true.
\hfill $\square$

\begin{remark}\label{remarkGP}
One could also use more directly the expression of the kernel
$\mathcal G^+(x,y\,;\lambda)$ of $(\mathcal A^+_1-\lambda)^{-1}$ in
terms of $\Ai$ and $\Ai'\,$, together with the asymptotic expansions
of the Airy function, see Appendix \ref{AppA} and the discussion at
the beginning of Section~\ref{s7}.
\end{remark}

\section{Integral kernel of the resolvent and its poles}\label{s6}

Here we revisit some of the computations of \cite{Grebenkov14b,Gr2}
with the aim to complete some formal proofs. We are looking for the
distribution kernel $\mathcal G ^-(x,y ; \lambda)$ of the resolvent
$(\mathcal A_1 ^- -\lambda)^{-1}$ which satisfies in the sense of
distribution
\begin{equation}
\label{eq:BT_g}
\biggl(-\lambda  - ix - \frac{\partial^2}{\partial x^2}\biggr) \mathcal G ^-(x,y\,; \lambda) = \delta(x-y),
\end{equation}
as well as the boundary conditions
\begin{equation}
\label{eq:cond_perm2}
\begin{split}
\biggl[\frac{\partial}{\partial x} \mathcal G ^-(x,y\,;\lambda)\biggr]_{x=0^+} & 
= \biggl[\frac{\partial}{\partial x} \mathcal G ^-(x,y\,;\lambda)\biggr]_{x=0^-} \\
& = \kappa\,  \bigl[\mathcal G ^-(0^+,y;\lambda) - \mathcal G ^-(0^-,y;\lambda)\bigr] \,. \\
\end{split}
\end{equation}
Sometimes, we will write $\mathcal G ^-(x,y\,;\lambda,\kappa)$, in
order to stress the dependence on $\kappa$.\\ 
Note that one can easily come back to the kernel of the resolvent of
$\mathcal A_1^+$ by using
\begin{equation}
\mathcal G^+ (x,y\,;\lambda) = \overline{ \mathcal G ^-(y,x; \bar \lambda)}\,.
\end{equation}
Using \eqref{sym}, we also get
\begin{equation}
\mathcal G^+ (x,y\,;\lambda) = \overline{ \mathcal G ^- (x,y\,; \bar \lambda)}\,.
\end{equation}
We search for the solution $\mathcal G ^- (x,y\,;\lambda)$ in three
subdomains: the negative semi-axis $(-\infty,0)$, the interval
$(0,y)$, and the positive semi-axis $(y,+\infty)$ (here we assumed
that $y > 0$; the opposite case is similar).  For each subdomain, the
solution is a linear combination of two Airy functions:
\begin{equation}
\mathcal G ^-(x,y\,;\lambda) = \begin{cases} 
A^- \Ai(e^{-i\alpha}w_x) + B^- \Ai(e^{i\alpha}w_x) \qquad (x < 0)\, , \cr
A^+ \Ai(e^{-i\alpha}w_x) + B^+ \Ai(e^{i\alpha}w_x) \quad (0 < x < y)\, , \cr
C^+ \Ai(e^{-i\alpha}w_x) + D^+ \Ai(e^{i\alpha}w_x) \qquad (x > y) \,, \end{cases}
\end{equation}
with six unknown coefficients (which are functions of $y >0$).  Here
we have set
\begin{equation*}
\alpha =\frac{2\pi}{3}
\end{equation*}
and 
\[
 w_x = ix+\lambda\,.
\]
The boundary conditions (\ref{eq:cond_perm2}) read as
\begin{equation}\label{trsm}
\begin{split}
& B^- ie^{i\alpha} \Ai'(e^{i\alpha} w_0) \\
& = A^+ ie^{-i\alpha} \Ai'(e^{-i\alpha}w_0) + B^+ ie^{i\alpha} \Ai'(e^{i\alpha}w_0)   \\
& = \kappa\, \bigl[A^+ \Ai(e^{-i\alpha}w_0) + B^+ \Ai(e^{i\alpha} w_0) - B^- \Ai(e^{i\alpha} w_0)\bigr] , \\
\end{split}
\end{equation}
where $w_0 = \lambda$ and we set $A^- = 0$ and $D^+ = 0$ to ensure the
decay of $\mathcal G ^-(x,y\,;\lambda)$ as $x\to -\infty$ and as $x\to +
\infty\,$, respectively.\\ 
We now look at the condition at $x=y$ in order to have \eqref{eq:BT_g}
satisfied in the distribution sense.  We write the continuity
condition,
\begin{equation*}
A^+ \Ai(e^{-i\alpha}w_y) + B^+ \Ai(e^{i\alpha}w_y) =
C^+ \Ai(e^{-i\alpha}w_y)  \,,
\end{equation*}
and the discontinuity jump of the derivative,
\begin{equation*}
A^+ i e^{-i\alpha}   \Ai'(e^{-i\alpha}w_y) + B^+i e^{i\alpha} \Ai'(e^{i\alpha}w_y) =
C^+i e^{-i\alpha} \Ai'(e^{-i\alpha}w_y)  + 1  \,.
\end{equation*}
This can be considered as a linear system for $A^+$ and $B^+\,$.
Using the Wronskian (\ref{eq:Wronskian}), one expresses $A^+$ and
$B^+$ in terms of $C^+$:
\begin{equation}
\label{eq:AB_C}
A^+  = C^+  -  2\pi \Ai(e^{i\alpha} w_y) \,,\quad 
B^+  =   2\pi \Ai(e^{-i\alpha} w_y) \,.
\end{equation}
We can rewrite \eqref{trsm} in the form
\begin{equation}\label{trsm1a}
B^- = e^{-2i\alpha} \frac{\Ai'(e^{-i\alpha} w_0)}{\Ai'(e^{i\alpha} w_0)} A^+ + B^+\,,
\end{equation}
and
\begin{equation}\label{trsm1b}
\begin{array}{l}
 A^+ ie^{-i\alpha} \Ai'(e^{-i\alpha}w_0) + B^+ ie^{i\alpha} \Ai'(e^{i\alpha}w_0) \\
 ~~~~~~~~~~~ = \kappa \,  A^+ \big[\Ai(e^{-i\alpha}w_0) 
 - e^{-2i \alpha} \Ai(e^{i\alpha}w_0) \frac{\Ai'(e^{-i\alpha} w_0)}{\Ai'(e^{i\alpha} w_0)} \bigr].
 \end{array}
\end{equation}
Using again the property of the Wronskian \eqref{eq:Wronskian}, we
obtain
\begin{equation*}\label{trsm2b}
 A^+  \Ai'(e^{-i\alpha}w_0) + B^+ e^{2i\alpha} \Ai'(e^{i\alpha}w_0)  =- \kappa  A^+ 
 \frac{1}{2\pi \Ai'(e^{i\alpha}w_0)} \,,
\end{equation*}
that is
\begin{equation*}
A^+ (f(\lambda)+ \kappa ) + B^+ (2\pi) e^{2i\alpha} \left( \Ai'(e^{i\alpha}w_0) \right)^2 =0 \,,
\end{equation*}
where
\begin{equation}
\label{eq:Airy_h}
f(\lambda) :=2\pi \Ai'(e^{-i\alpha} \lambda) \Ai'(e^{i\alpha} \lambda)\, .
\end{equation}
So we now get
\begin{equation}
A^+= -\frac{1}{f(\lambda) + \kappa } (2\pi)^2 e^{2i\alpha} \left( \Ai'(e^{i\alpha}w_0) \right)^2 \Ai(e^{-i\alpha} w_y)\,,
\end{equation}
\begin{equation}
B^-  = 2\pi \Ai(e^{-i\alpha} w_y) - 2\pi  \frac{f(\lambda)}{f(\lambda)+\kappa } \Ai(e^{-i\alpha} w_y)\,,~~~~~~
\end{equation}
and
\begin{equation}
C^+  = 2\pi \Ai(e^{i\alpha} w_y) - 4\pi^2 \frac{e^{2i\alpha} [\Ai'(e^{i\alpha} \lambda)]^2}{f(\lambda) + \kappa }~ \Ai(e^{-i\alpha} w_y)\,.
\end{equation}
Combining these expressions, one finally gets
\begin{equation}\label{eq:Gtilde_full}
\mathcal G ^-(x,y\,;\lambda,\kappa) =  \mathcal G _0 ^-(x,y\,;\lambda) + \mathcal G _1(x,y\,;\lambda,\kappa) \,,
\end{equation}
where $ \mathcal G _0 ^-(x,y\,;\lambda)$ is the distribution kernel of
the resolvent of the operator $\mathcal A_0^*:= -\frac{d^2}{dx^2} -i
x$ on the line (given by Eq. (\ref{eq:G0_free})), whereas $\mathcal G
_1(x,y\,;\lambda,\kappa)$ is given by the following expressions
\begin{equation}\label{eq:G1tilde}
\mathcal G _1(x,y\,;\lambda,\kappa) = \left\{
  \begin{array}{ll}
      - 4\pi^2 \frac{e^{2i\alpha} [\Ai'(e^{i\alpha} \lambda)]^2}{f(\lambda) 
+ \kappa } \Ai(e^{-i\alpha} w_x) \Ai(e^{-i\alpha} w_y)\,, & (x>0) \,, \\
      - 2\pi  \frac{f(\lambda)}{f(\lambda)+\kappa } \Ai(e^{i\alpha} w_x) \Ai(e^{-i\alpha} w_y)\,, &(x<0) \,, \end{array} \right.
\end{equation}
for $y>0$, and
\begin{equation}\label{yneg}
 \mathcal G _1(x,y\,;\lambda,\kappa) = \left\{
 \begin{array}{ll}
  -2\pi\frac{f(\lambda)}{f(\lambda)+\kappa }\Ai(e^{-i\alpha}w_x)\Ai(e^{i\alpha} w_y)\,, & (x>0)\,, \\
  -4\pi^2\frac{e^{-2i\alpha}[\Ai'(e^{-i\alpha}\lambda)]^2}{f(\lambda)+\kappa}\Ai(e^{i\alpha}w_x)\Ai(e^{i\alpha}w_y)\,, & (x<0)\,,
 \end{array}
\right.
\end{equation}
for $y<0$.
Hence the poles are determined by the equation
\begin{equation}\label{maineq}
f(\lambda) =- \kappa \,,
\end{equation} 
with $f$ defined in \eqref{eq:Airy_h}.
\begin{remark}\label{remN}
For $\kappa =0$, one recovers the conjugated pairs associated with the
zeros $a'_n$ of $\Ai'$.  We have indeed as poles
\begin{equation}
\lambda_n^+ = e^{i\alpha} a'_n \,,\quad \lambda_n^{-} = e^{-i \alpha} a'_n \,,
\end{equation}
where $a'_n $ is the $n$-th zero (starting from the right) of $\Ai'$.
 Note that $a'_n < 0$ so that $\Re \lambda^\pm_n > 0$, as
expected.\\
In this case, the restriction to $\mathbb R_+^2$ of
$\mathcal G _1(x,y\,;\lambda,0)$ is the kernel of the resolvent of the
Neumann problem in $\mathbb R_+$.
\end{remark}

We also know that the eigenvalues for the Neumann problem are simple.
Hence by the local inversion theorem we get the existence of a
solution close to each $\lambda_n^\pm$ for $\kappa $ small enough
(possibly depending on $n$) if we show that $f'(\lambda_n^\pm) \neq
0$.  For $\lambda_n^+$, we have, using the Wronskian relation
\eqref{eq:Wronskian} and $\Ai'(e^{-i\alpha}\lambda^+_n) =0\,$,
\begin{equation}
  \begin{array}{ll}
  f'(\lambda_n^+)& =  2\pi \, e^{- i\alpha}\,  \,\Ai''(e^{-i\alpha}\lambda_n^+ ) \Ai'(e^{i\alpha}\lambda_n^+)\\
  &  =  2\pi e^{-2 i\alpha} \lambda_n^+   \Ai(e^{-i\alpha}\lambda_n^+) \Ai'(e^{i\alpha} \lambda_n^+)\\
  & =-  i \lambda_n^+\,.
  \end{array}
\end{equation}
Similar computations hold for $\lambda_n^-$.  We recall that 
\begin{equation*}
\lambda_n^+= \overline{\lambda_n^-}\,.
\end{equation*}

The above argument shows that $f'(\lambda_n)\neq 0$, with $\lambda_n
=\lambda_n^+$ or $\lambda_n=\lambda_n^-$.  Hence by the holomorphic
inversion theorem we get that, for any $n\in \mathbb N^*$, and any
$\epsilon$, there exists $h_n(\epsilon)$ such that for $|\kappa|\leq
h_n(\epsilon)$, we have a unique solution $\lambda_n(\kappa)$ of
\eqref{maineq} such that $|\lambda_n(\kappa) - \lambda_n| \leq
\epsilon$.\\

We would like to have a control of $h_n(\epsilon)$ with respect to
$n$.  What we should do is inspired by the Taylor expansion given in
\cite{Gr2} (Formula (33)) of $\lambda_n^\pm(\kappa)$ for fixed $n$\,:
\begin{equation}\label{eq:lambda_kappa}
\lambda_n^\pm (\kappa) = \lambda_n^\pm  + e^{\pm i\frac \pi 6}\frac{1}{a'_n } \kappa + \mathcal O_n (\kappa^2)\,.
\end{equation}
Since $|\lambda_n|$ behaves as $n^\frac 23$ (see Appendix \ref{AppA}),
the guess is that $\lambda_{n+1} ^\pm(\kappa) -
\lambda_n^\pm (\kappa) $ behaves as $n^{-\frac 13}$.\\
To justify this guess, one needs to control the derivative in a
suitable neighborhood of $\lambda_n$.
\begin{proposition}
\label{sec:prop1}
There exists $\eta >0$ and $h_\infty >0$, such that, for all $n\in
\mathbb N^*$, for any $\kappa$ such that $|\kappa|\leq h_\infty$ there
exists a unique solution of \eqref{maineq} in $B(\lambda_n, \eta
|\lambda_n|^{-1})$ with $\lambda_n = \lambda_n^\pm$.
\end{proposition}
{\bf Proof of the proposition}\\ Using the previous arguments, it is
enough to establish the proposition for $n$ large enough.  Hence it
remains to establish a local inversion theorem uniform with respect to
$n$ for $n\geq N$.  For this purpose, we consider the holomorphic
function
\begin{equation*}
 B(0,\eta) \ni  t \mapsto \phi_n (t) = f(\lambda_n + t \lambda_n^{-1})\,.
\end{equation*}
To have a  local inversion theorem uniform with respect to $n$, we need to control
$|\phi'_n(t)|$ from below.\\
\begin{lemma}
For any  $\eta >0$, there exists $N$ such that, $\forall n \geq N$,
\begin{equation}
|\phi'_n (t)| \geq \frac 12 \,,\quad \forall t \in B (0,\eta)\,.
\end{equation}
\end{lemma}
{\bf Proof of the lemma.}\\
We have 
\begin{equation*} 
 \phi'_n (t) = \lambda_n^{-1} f'(\lambda_n + t \lambda_n^{-1}) \,,
\end{equation*}
and
\begin{equation*}
 \phi'_n (0) =-i
 \,.
\end{equation*}
Hence it remains to control $ \phi'_n (t)-\phi'_n(0)$ in
$B(0,\eta)$. We treat the case $\lambda_n=\lambda_n^+$. \\ 
We recall that
\begin{equation}
\label{eq:Airy_hprime}
\begin{array}{ll}
f'(\lambda) &=2\pi e^{-i\alpha} \Ai''(e^{-i\alpha} \lambda) \Ai'(e^{i\alpha} \lambda) 
 + 2\pi e^{i\alpha} \Ai(e^{-i\alpha} \lambda) \Ai''(e^{i\alpha} \lambda)\, \\
 & =2\pi \lambda \left (e^{- 2 i\alpha}   \Ai(e^{-i\alpha} \lambda) \Ai'(e^{i\alpha} \lambda) 
 +  e^{2 i\alpha} \Ai'(e^{-i\alpha} \lambda) \Ai(e^{i\alpha} \lambda)\right)\\
 &=- i \lambda +  4 \pi \lambda e^{2i \alpha} \Ai'(e^{-i\alpha} \lambda) \Ai(e^{i\alpha} \lambda) .
 \end{array}
\end{equation}
Hence we have
\begin{equation}
\phi'_n(t) - \phi'_n (0) = 4 \pi \lambda \lambda_n^{-1} e^{2i\alpha}\,   \Ai'(e^{-i\alpha} \lambda) \Ai(e^{i\alpha} \lambda),
\end{equation}
with $\lambda = \lambda_n + t \lambda_n^{-1}$.\\

We will control $ \Ai'(e^{-i\alpha} \lambda) \Ai(e^{i\alpha} \lambda)$
in $B(\lambda_n, \eta |\lambda_n|^{-1})$ and show that this expression
tends to zero as $n\ar +\infty$.\\ We have
\begin{equation*}
\Ai'(e^{-i\alpha} \lambda)  = e^{-i\alpha} (\lambda -\lambda_n) \Ai''(e^{-i\alpha} \tilde \lambda)
= e^{-2 i \alpha} (\lambda-\lambda_n) \,\tilde \lambda \,   \Ai(e^{-i\alpha} \tilde \lambda)\,,
\end{equation*}
with $\tilde \lambda \in B(\lambda_n, \eta |\lambda_n|^{-1})$.

Hence it remains to show that the product $ \, |\Ai(e^{-i\alpha}
\tilde \lambda) \Ai(e^{i\alpha} \lambda)|$ for $\lambda$ and $\tilde
\lambda$ in $B(\lambda_n, \eta |\lambda_n|^{-1})$ tends to $0$.\\
Here we use the known expansion for the Airy function recalled in
Appendix~\ref{AppA} in the balls $B(e^{-i \alpha} \lambda_n, \eta
|\lambda_n|^{-1})$ and $B(e^{i \alpha} \lambda_n, \eta
|\lambda_n|^{-1})$. \\~\\
(i) For the first one, we need the expansion of $\Ai$ in the
neighborhood of $a'_n $.  Using the asymptotic relation (\ref{6}), we
observe that
\begin{equation*}
\exp \left(\pm  i \frac 23 z^{\frac 32} \right) = \exp \left(\pm  i \left(\frac 23 (-a'_n )^{\frac 32} (1+ \mathcal O (1/|a'_n| ^2))\right)\right)
 = \mathcal O (1)\,.
\end{equation*}
Hence we get
\begin{equation*} 
 | \Ai (e^{-i \alpha} \lambda)| \leq C\,  |a'_n |^{-\frac 14}  \quad \forall ~ \lambda\in B(\lambda_n,\eta|\lambda_n|^{-1})\, .
\end{equation*}
(ii) For the second one, we use (\ref{5}) to observe that
\begin{equation*}
\exp\left(- \frac 23 (e^{i \alpha} \lambda)^{\frac 32}\right) = \exp\left( -i \frac 23 (-a'_n )^{\frac 32} 
 (1 + \mathcal O ((-a'_n )^{-2}) ) \right)\,,
\end{equation*}
and we get, for $\lambda  \in  B(\lambda_n, \eta |\lambda_n|^{-1})$
\beq
|\Ai(e^{i\alpha} \lambda)| \leq C \,  |a'_n |^{-\frac 14}\,.
\eeq
This completes the proof of the lemma and of the proposition.\\
Actually, we have proved on the way the more precise
\begin{proposition} 
For all $\eta >0$ and $0\leq \kappa < \frac \eta 2 $, there exists $N$
such that, for all $n \geq N$, there exists a unique solution of
\eqref{maineq} in $B(\lambda_n, \eta \, |\lambda_n|^{-1})$.
\end{proposition}

\begin{figure}
\begin{center}
\includegraphics[scale=0.6]{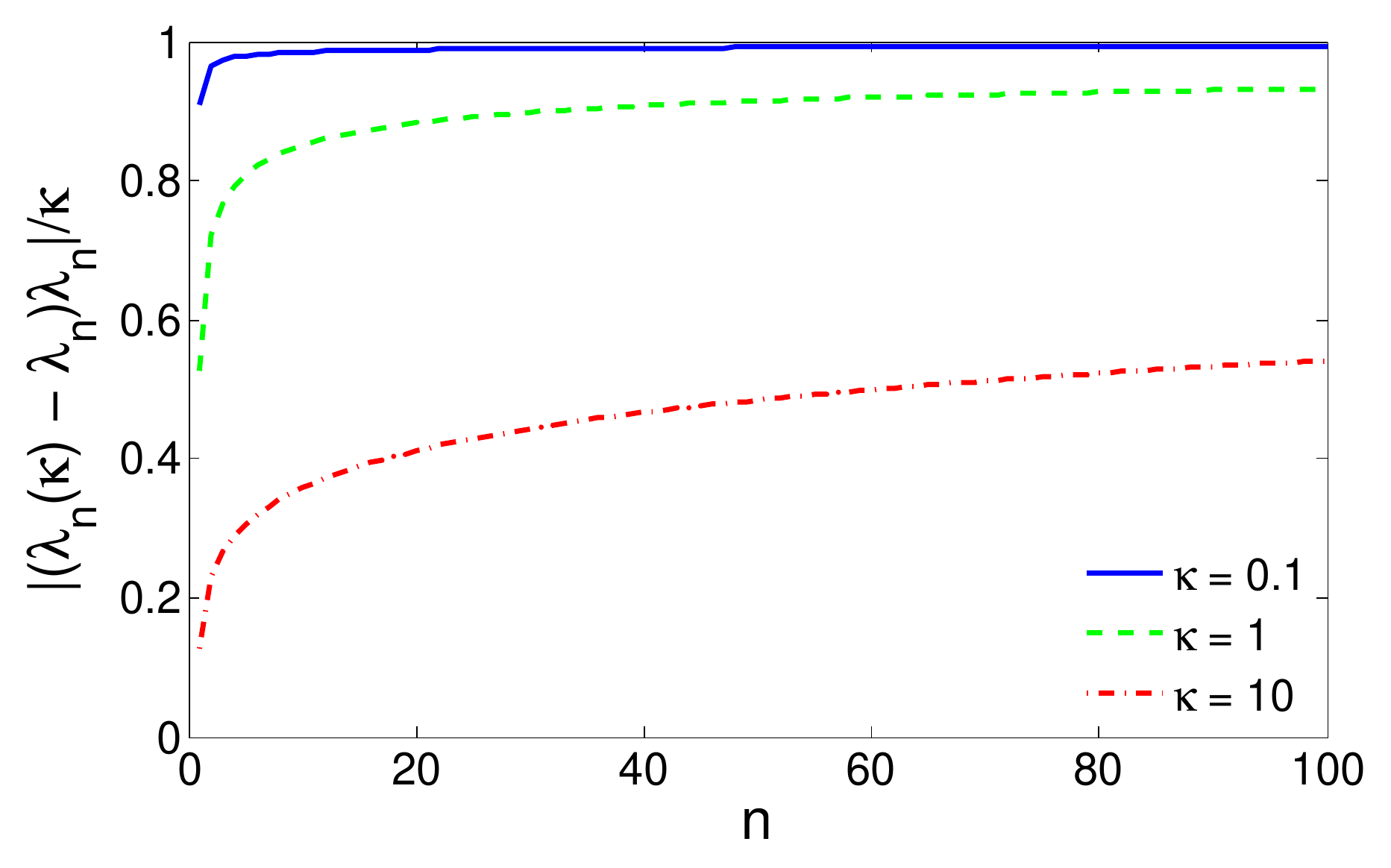}  
\includegraphics[scale=0.6]{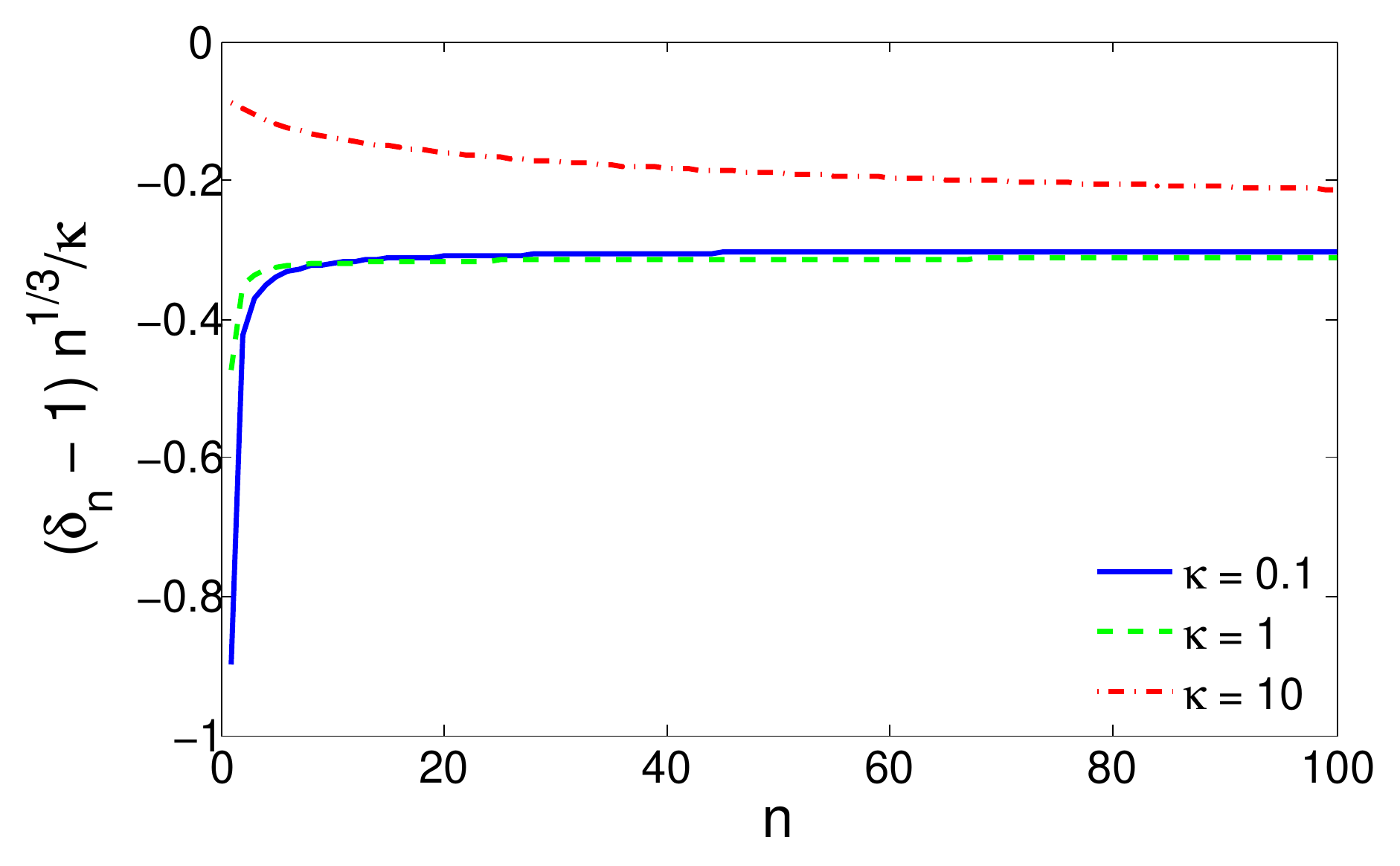}  
\end{center}
\caption{
Illustration of Proposition \ref{sec:prop1} by the numerical
computation of the first 100 zeros $ \lambda_n^+(\kappa)$ of
(\ref{maineq}).  At the top, the rescaled distance $\delta_n(\kappa)$
from (\ref{eq:delta}) between $ \lambda_n^+(\kappa)$ and $ \lambda_n^+
= \lambda_n^+(0)$.  At the bottom, the asymptotic behavior of this
distance.}
\label{fig:lambdan}
\end{figure}

Figure \ref{fig:lambdan} illustrates Proposition \ref{sec:prop1}.
Solving Eq. (\ref{maineq}) numerically, we find the first 100 zeros
$\lambda_n(\kappa)$ with $\Im \lambda_n(\kappa) > 0$.  According to
Proposition~\ref{sec:prop1}, these zeros are within distance
$1/|\lambda_n|$ from the zeros $\lambda_n = \lambda_n(0) = e^{i\alpha}
a'_n$ which are given explicitly through the zeros $a'_n$.  Moreover,
the second order term in (\ref{eq:lambda_kappa}) that was computed
in \cite{Gr2}, suggests that the rescaled distance
\begin{equation}
\label{eq:delta}
\delta_n(\kappa) = |\lambda_n(\kappa) - \lambda_n| |\lambda_n|/\kappa,
\end{equation}
behaves as
\begin{equation}
\label{eq:delta_asympt}
\delta_n(\kappa) = 1 - c \kappa n^{-1/3} + o(n^{-1/3}) ,
\end{equation}
with a nonzero constant $c$.  Figure \ref{fig:lambdan}(top) shows that
the distance $\delta_n(\kappa)$ remains below $1$ for three values of
$\kappa $: $ 0.1$, $1$, and $10$.  The expected asymptotic behavior given in 
(\ref{eq:delta_asympt}) is confirmed by Figure
\ref{fig:lambdan}(bottom), from which the constant $c$ is estimated to
be around $0.31$.

\begin{remark}
The local inversion theorem with control with respect to $n$ permits
to have the asymptotic behavior of the $\lambda_n(\kappa)$ uniformly
for $\kappa$ small:
\begin{equation}\label{eq:lambda_kappaunif}
\lambda_n^\pm (\kappa) = \lambda_n^\pm  + e^{\pm i\frac \pi 6}\frac{1}{a'_n } \kappa + \frac{1}{a'_n }  \mathcal O (\kappa^2)\,.
\end{equation}
An improvment of \eqref{eq:lambda_kappaunif} (as formulated by
\eqref{eq:delta_asympt}) results from a good estimate on $\phi''_n(t)$.
Observing that $|\phi''_n(t)| \leq C |a'_n|^{-\frac 12}$ in the ball
$B(0,\eta)$, we obtain
\begin{equation}\label{eq:lambda_kappaunifimp}
\lambda_n^\pm (\kappa) = \lambda_n^\pm  + e^{\pm i\frac \pi 6}\frac{1}{a'_n } \kappa + \frac{1}{(a'_n)^{\frac 32} }  \mathcal O (\kappa^2)\,.
\end{equation}
If one asks for finer estimates, one should compute $\phi''_n(0)$ and
estimate $\phi'''_n$, and so on.

It would also be interesting to analyze the case $\kappa \ar +
\infty$.  See \cite{Gr2} for a preliminary non rigorous analysis. The
limiting problem in this case is the realization of the complex Airy
operator on the line which has empty spectrum.
\end{remark}

In the remaining part of this section, we describe the distribution
kernel of the projector $ \Pi_n^\pm $ associated with $\lambda_n^ \pm
(\kappa)$.
\begin{proposition}\label{propjordan}
There exists $\kappa_0 >0$ such that, for any $\kappa \in
[0,\kappa_0]$ and any $n\in \mathbb N^*$, the rank of $\Pi_n^\pm$ is
equal to one. Moreover, if $\psi_n^\pm$ is an eigenfunction, then
\begin{equation} \label{integ}
 \int_{-\infty}^{+\infty} \psi_n^\pm (x)^2 \,dx \neq 0\,.
\end{equation}
\end{proposition}
{\bf Proof}\\
To write the projector $ \Pi_n^\pm $ associated with an eigenvalue
$\lambda_n^\pm$ we integrate the resolvent along a small contour
$\gamma_n^\pm$ around $\lambda_n^\pm$.
\begin{equation}
  \Pi_n^\pm  =\frac{1}{2i\pi}\int_{\gamma_n^\pm} ( \mathcal A_1 ^\pm -  \lambda)^{-1} d\lambda\,.
\end{equation}
If we consider the associated kernels, we get, using
\eqref{eq:Gtilde_full} and the fact that $ \mathcal G _0 ^-$ is
holomorphic in $\lambda$:
\begin{equation}
 \Pi_n^\pm (x,y\,;\kappa) = \frac{1}{2i\pi}\int_{\gamma_n^\pm} \mathcal G_1 (x,y\,;\lambda,\kappa)\,  d\lambda\,.
\end{equation} 
 
The projector is given by the following expression (with $w^{\pm,n}_x
= ix +\lambda_n^\pm$) for $y>0\,$
\begin{equation}\label{eq:Pi1}
\Pi_n^\pm(x,y\,;\kappa) = \left\{ 
  \begin{array}{ll} - 4\pi^2 \frac{e^{2i\alpha} [\Ai'(e^{i\alpha} \lambda_n^\pm )]^2}{f'(\lambda_n^\pm)  } 
\Ai(e^{-i\alpha} w^{\pm,n}_x ) \Ai(e^{-i\alpha} w^{\pm,n}_y)\,  & (x>0) \,, \\
  2\pi  \frac{\kappa}{f'(\lambda_n^\pm) } \Ai(e^{i\alpha} w^{\pm,n}_x ) \Ai(e^{-i\alpha} w^{\pm,n}_y)\, &  (x< 0) \,, \\  \end{array}  \right.
\end{equation}
and for $y<0\,$
\begin{equation}\label{eq:Pi2}
 \Pi_n^\pm(x,y\,;\kappa)= \left\{
 \begin{array}{ll}
  2\pi\frac{\kappa}{f'(\lambda_n^\pm)}\Ai(e^{-i\alpha}w^{\pm,n}_x )\Ai(e^{i\alpha} w^{\pm,n}_y)\, & (x>0)\,, \\
  -4\pi^2\frac{e^{-2i\alpha}[\Ai'(e^{-i\alpha}\lambda_n^\pm )]^2}{f'(\lambda_n^\pm)}\Ai(e^{i\alpha}w^{\pm,n}_x )\Ai(e^{i\alpha}w^{\pm,n}_y)\, & (x<0)\,.
 \end{array}
\right.
\end{equation}
Here we recall that we have established that for $|\kappa|$ small
enough $f'(\lambda_n^\pm) \neq 0$. It remains to show that the rank of
$\Pi_n^\pm$ is one and we will get at the same time an expression for
the eigenfunction.  It is clear that the rank of $\Pi_n^\pm$ is at
most two and that every function in the range of $\Pi_n^\pm$ has the
form $(c_- \Ai(e^{i\alpha} w^{\pm,n}_x )\,,\, c_+ \Ai(e^{-i\alpha}
w^{\pm,n}_x ))\,$, where $c_-,c_+\in\mathbb{R}\,$.  It remains to
establish the existence of a relation between $c_-$ and $c_+$.  This
is a consequence of $f(\lambda_n^\pm)=-\kappa$.  If $\kappa \neq 0$,
the functions in the range have the form
\begin{equation*}
c_n \left( Ai '(e^{-i\alpha}\lambda_n^\pm) \Ai(e^{ i\alpha} w^{\pm,n}_x ),   e^{2i\alpha} \Ai'(e^{i\alpha} \lambda_n^\pm ) \Ai(e^{-i\alpha} w^{\pm,n}_x )\right)\,.
\end{equation*}
Inequality \eqref{integ} results from an abstract lemma in \cite{AD}
once we have proved that the rank of the projector is one.  We have
indeed
\begin{equation}
 || \Pi_n^\pm || = \frac{1}{|\int_{-\infty}^{+\infty}\psi_n^\pm (x)^2\, dx|}\,.
\end{equation}

More generally, the proof of the proposition can be formulated in this
way:
\begin{proposition}
If $f(\lambda) + \kappa =0$ and $f'(\lambda)\neq 0\,$, then the
associated projector has rank $1$ (no Jordan block).
\end{proposition}
The condition of $\kappa$ being small in Proposition \ref{propjordan}
is only used for proving the property $f'(\lambda) \neq 0$.  For the
case of the Dirichlet or Neumann realization of the complex Airy
operator in $\mathbb R_+$, we refer to Section \ref{s2a}.  The
nonemptiness was obtained directly by using the properties of the Airy
function.  Note that our numerical solutions did not reveal projectors
of rank higher than~$1$.  We conjecture that the rank of these
projectors is $1$ for any $0 \leq \kappa < +\infty\,$ but  can only
prove the weaker
\begin{proposition} \label{prop:finite_number}
For any $\kappa \geq 0$, there is at most a finite number of
eigenvalues with nontrivial Jordan blocks.
\end{proposition}
{\bf Proof}\\
We start from 
\begin{equation*}
f(\lambda):= 2\pi \Ai'(e^{2\pi i/3} \lambda) \Ai'(e^{-2\pi i/3} \lambda) \,,
\end{equation*}
and get by derivation
\begin{equation}
\frac{1}{2\pi} f'(\lambda)= e^{i\alpha}  \Ai''(e^{i\alpha} \lambda) \Ai'(e^{-i\alpha} \lambda)  + e^{-i\alpha} \Ai'(e^{i\alpha} \lambda) 
\Ai''(e^{-i\alpha} \lambda)\,.
\end{equation}
What we have to prove is that $f'(\lambda)$ is different from $0$ for
a large solution $\lambda$ of $f(\lambda)=-\kappa$.  We know already
that $\Re \lambda \geq 0$.  We note that $f(0) >0$.  Hence $0$ is not
a pole for $\kappa \geq 0$.  More generally $f$ is real and strictly
positive on the real axis.  Hence $f(\lambda) +\kappa >0$ on the real
axis. \\
We can assume that $\Im \lambda >0$ (the other case can be treated
similarly).  Using the equation satisfied by the Airy function, we get
\begin{equation}
\frac{1}{2\pi \lambda } f'(\lambda)= e^{-i\alpha}  \Ai (e^{i\alpha} \lambda) \Ai'(e^{-i\alpha} \lambda)  
+ e^{i\alpha} \Ai'(e^{i\alpha} \lambda) \Ai(e^{-i\alpha} \lambda)\,,
\end{equation}
and by the Wronskian relation (\ref{eq:Wronskian}):
\begin{equation}
e^{-i\alpha} \Ai'(e^{-i\alpha} \lambda ) \Ai(e^{i\alpha} \lambda) - e^{i\alpha} \Ai'(e^{i\alpha} \lambda) \Ai(e^{-i\alpha} \lambda) = \frac{i}{2\pi} \,.
\end{equation}
Suppose that $f(\lambda)=-\kappa$ and that $f'(\lambda)=0\,$.\\
We have 
\begin{equation*}
- e^{i\alpha} \Ai'(e^{i\alpha} \lambda) \Ai(e^{-i\alpha} \lambda) = e^{-i\alpha} \Ai'(e^{-i\alpha} \lambda ) \Ai(e^{i\alpha} \lambda) = \frac{i}{4\pi}  .
\end{equation*}
and  get
\begin{equation*}
\kappa =  \frac{i}{2} \Ai'(e^{2\pi i/3} \lambda) / \Ai (e^{2\pi i/3} \lambda) = - \frac{i}{2} \Ai'(e^{-2\pi i/3} \lambda) / \Ai (e^{-2\pi i/3} \lambda) 
\end{equation*}
Using the last equality and the asymptotics \eqref{5}, \eqref{5a} for
$\Ai$ and $\Ai'$, we get as $|\lambda| \ar +\infty$ satisfying the
previous condition
\begin{equation*}
\kappa \sim \frac 12  | \lambda|^\frac 12
\end{equation*}
which cannot be true for $\lambda$ large.  This achieves the proof of
the proposition.

\section{Resolvent estimates as $|\Im \lambda| \rightarrow + \infty$}\label{s7}

\begin{figure}
\begin{center}
\includegraphics[scale=0.6]{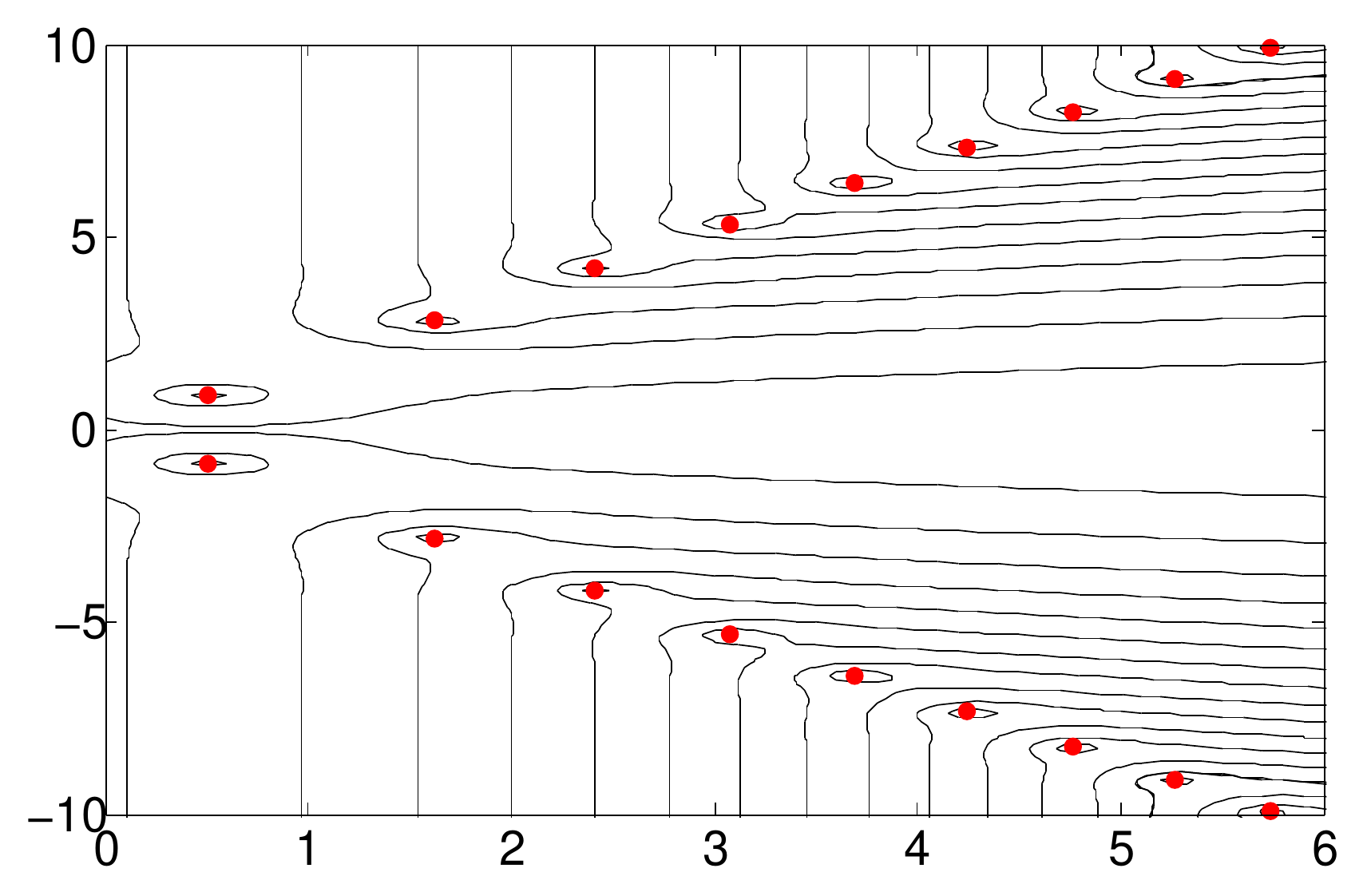}  
\includegraphics[scale=0.6]{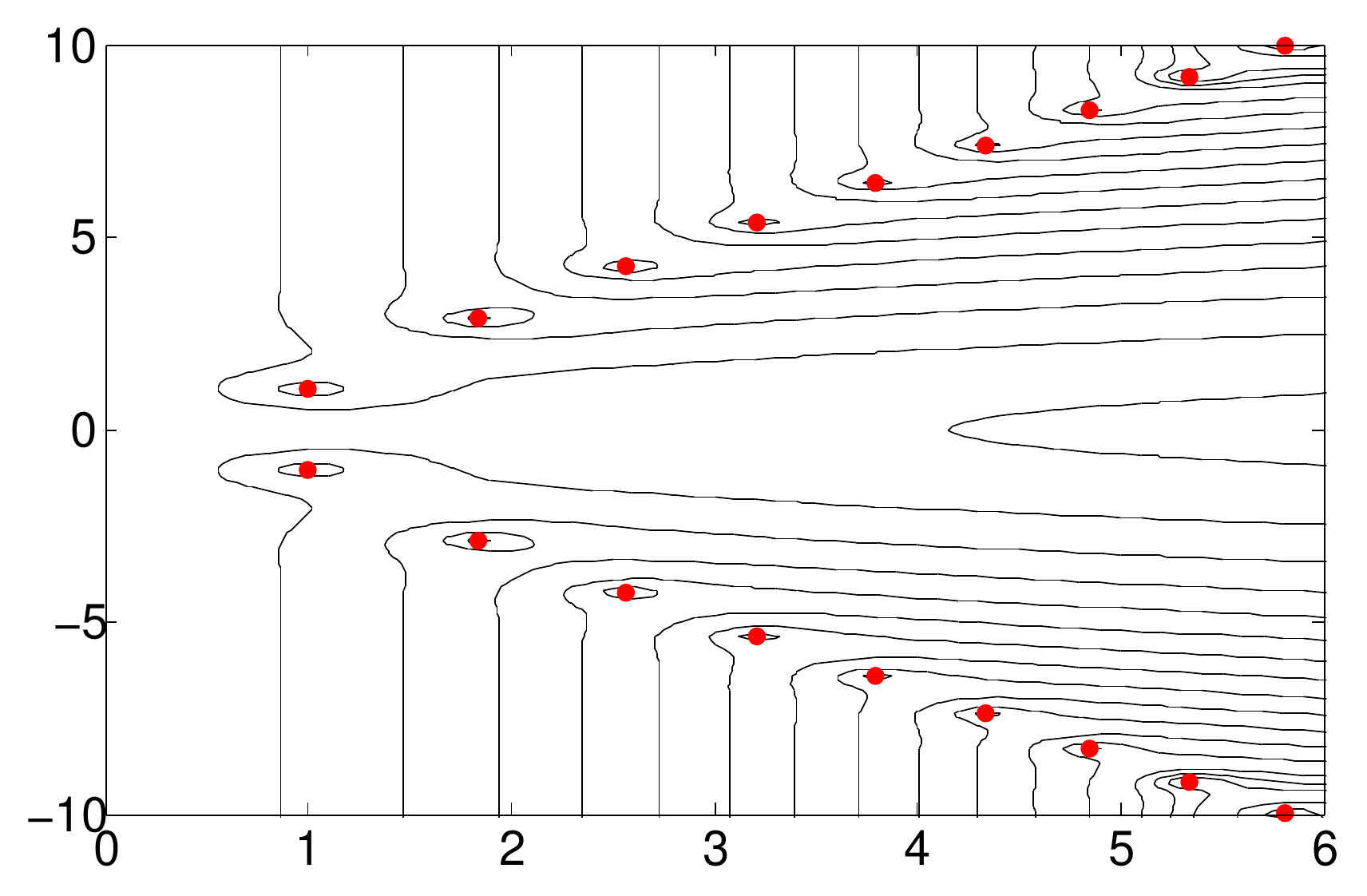}  
\end{center}
\caption{
Numerically computed pseudospectrum in the complex plane of the
complex Airy operator with Neumann boundary conditions (top) and with
the transmission boundary condition at the origin with $\kappa = 1$
(bottom).  The red points show the poles $\lambda_n^\pm(\kappa)$ found
by solving numerically Eq. (\ref{maineq}) that corresponds to the
original problem on $\R$.  The presented picture corresponds to a zoom
(eliminating numerical artefacts) in a computation done for a large
interval $[-L,+L]$ with Dirichlet boundary conditions at $\pm L$.  The
pseudospectrum was computed with $L^3= 10^4$ by projecting the complex
Airy operator onto the orthogonal basis of eigenfunctions of the
corresponding Laplace operator and then diagonalizing the obtained
truncated matrix representation (see Appendix~\ref{sec:numerics} for
details).  We only keep a few lines of pseudospectra for the clarity
of the picture.  As predicted by the theory, the vertical lines are
related to the pseudospectrum of the free complex Airy operator on the
line.}
\label{fig:pseudo}
\end{figure}

The resolvent estimates have been already proved in Section \ref{s5}
and were used in the analysis of the decay of the associated
semigroup.  We propose here another approach which leads to more
precise results.  We keep in mind \eqref{eq:Gtilde_full} and the
discussion in Section \ref{s5}.\\ For $\lambda = \lambda_0+ i\eta\,$,
we have
\[
 \| \mathcal G _0 ^-(\,\cdot\,,\,\cdot \, ; \lambda)\|_{L^2(\mathbb{R}^2)} = \| \mathcal G _0 ^-(\,\cdot\,,\,\cdot\, ; \lambda_0)\|_{L^2(\mathbb{R}^2)}\,.
\]
Hence the Hilbert-Schmidt norm of the resolvent $(\mathcal A^+ -
\lambda)^{-1}$ does not depend on the imaginary part of
$\lambda\,$. \\ 
As a consequence, to recover Lemma \ref{Lemma4.2} by this approach, it
only remains to check the following lemma
\begin{lemma}
For any $\lambda_0$, there exist $C >0$ and $\eta_0 >0$ such that
\begin{equation}
 \sup_{|\eta | > \eta_0}\|\mathcal G _1(\cdot\,,\,\cdot ; \lambda_0+i\eta)\|_{L^2(\mathbb{R}^2)} \leq C\,.
\end{equation}
\end{lemma}
The proof is included in the proof of the following improvement which
is the main result of this section and is confirmed by the numerical
computations.  One indeed observes that the lines of the
pseudospectrum are asymptotically vertical as $\Im \lambda \ar \pm
\infty$ when $\Re \lambda >0\,$,  see Figure \ref{fig:pseudo}.

\begin{proposition}
For any $\lambda_0 >0$, 
\begin{equation*}
\lim_{\eta \ar \pm \infty} \|\mathcal G _1(\cdot\,,\,\cdot ; \lambda_0+i\eta)\|_{L^2(\mathbb{R}^2)} =0\,.
\end{equation*}
Moreover, this convergence is uniform for $\lambda_0$ in a compact
set.
\end{proposition}
{\bf Proof}\\
We have
\[
 e^{i\alpha} \lambda = e^{i\alpha} \lambda_0 - e^{i\pi/6}\eta
\]
and
\[
 e^{-i\alpha} \lambda  = e^{-i\alpha}\lambda_0 + e^{-i\pi/6}\eta\,.
\]
Then according  to \eqref{5a}, one can easily check that the term
$\Ai'(e^{\pm i\alpha}\lambda)$ decays exponentially as
$\eta\rightarrow \mp\infty$ and grows exponentially as
$\eta\rightarrow\pm\infty\,$.  On the other hand, the term
$\Ai'(e^{i\alpha}\lambda )$ decays exponentially as
$\eta\rightarrow\pm\infty$\,.\\ More precisely, we have
\begin{equation}\label{asya}
\begin{array}{ll}
|\Ai'( e^{i\alpha} (\lambda_0+ i \eta) )|^2 & \sim |c|^2 \eta^\frac 12 \exp\left( \frac{2 \sqrt{2}}{3}\eta^{3/2}\right)\,,\, \mbox{ as } \eta \ar + \infty \,;\\
& \sim |c|^2 (-\eta)^\frac 12 \exp\left(- \frac{2\sqrt{2}}{3}\eta^{3/2}\right)\,,\, \mbox{ as } \eta \ar - \infty\,;\\
|\Ai'( e^{-i\alpha} (\lambda_0+ i \eta) )|^2 & \sim |c|^2 \eta^\frac 12 \exp\left(-\frac{2\sqrt{2}}{3}\eta^{3/2}\right)\,,\, \mbox{ as } \eta \ar + \infty \,;\\
& \sim |c|^2 (-\eta)^\frac 12 \exp\left(\frac{2\sqrt{2}}{3}\eta^{3/2}\right)\,,\, \mbox{ as } \eta \ar - \infty\,.
\end{array}
\end{equation}
As a consequence, the function $f(\lambda)$, which was defined in
\eqref{eq:Airy_h} by
\begin{equation*}
f(\lambda) :=2\pi \Ai'(e^{-i\alpha} \lambda) \Ai'(e^{i\alpha} \lambda)\, ,
\end{equation*}
has the following asymptotic behavior as $\eta\rightarrow\mp
\infty\,$:
\begin{equation} \label{asf}
 f(\lambda_0+i\eta) = 2 \pi  |c|^2\, |\eta|^{1/2}\big(1+o(1)\big)\,.
\end{equation}

{\bf We treat the case $\eta >0$} (the other case can be deduced by
considering the complex conjugate).\\ Coming back to the two formulas
giving $\mathcal G _1$ in \eqref{eq:G1tilde} and \eqref{yneg} and
starting with the first one, we have to analyze the $L^2$ norm over
$\mathbb R_+\times \mathbb R_+$ of
\begin{equation*}
(x,y)\mapsto - 4\pi^2 \frac{e^{2i\alpha} [\Ai'(e^{i\alpha} \lambda)]^2}{f(\lambda) 
+ \kappa } \Ai(e^{-i\alpha} w_x) \Ai(e^{-i\alpha} w_y)\,
\end{equation*}
This norm $N_1$ is given by
\begin{equation*}
N_1:= 4\pi^2 |\Ai'(e^{i\alpha} \lambda)|^2 \, | f(\lambda) 
+ \kappa | ^{-1}  || \Ai(e^{-i\alpha} w_x)||_{L^2(\mathbb R_+)}^2 \,.
\end{equation*}
Hence we have to estimate $ \int_0^{+\infty} | \Ai(e^{-i \alpha }w_x) |^2 dx$.
We observe that
\begin{equation*}
e^{-i \alpha} w_x = e^{-i \frac \pi 6} (x+ \eta) + e^{-i \alpha} \lambda_0\,.
\end{equation*}
This is rather simple for $\eta >0$ because $x$ and $\eta$ have the
same sign.  We can use the asymptotics  \eqref{5} 
in order to get
\beq \label{upb}
 \int_0^{+\infty} | \Ai(e^{-i \alpha }w_x) |^2 dx \leq C \, (|\eta|^2 +1)^{-\frac 12}  \,\exp \left(- \frac {2 \sqrt{2}}{3}  |\eta|^{\frac 32}\right)
 \,.
 \eeq
Here we have used that, for $\beta > 0$,
\begin{equation*}
 \int_\eta^{+ \infty} \exp \left(- \beta y^{\frac 32}\right) dy = \frac{2}{3 \beta} \exp \left(- \beta \eta ^{\frac 32} (1 + \mathcal O (|\eta|^{-\frac 12}))\right)
 \,.
\end{equation*}
We use the control of $|\Ai'(e^{i\alpha} (\lambda_0 + i \eta))|^2$
given in \eqref{asya} and \eqref{asf} to finally obtain
\begin{equation}
 N_1  \lesssim (|\eta|^2 +1)^{-\frac 12}\,.
\end{equation}
By the notation $\lesssim\,$, we mean that there exists a constant $C$
such that
\begin{equation*}
 N_1  \leq C  (|\eta|^2 +1)^{-\frac 12}\,.
\end{equation*}
For the $L^2$-norm of the second term (see \eqref{yneg}),
\begin{equation*}
N_2 := \left|\left| - 2\pi  \frac{f(\lambda)}{f(\lambda)+\kappa } \Ai(e^{i\alpha} w_x) \Ai(e^{-i\alpha} w_y)
\right|\right|_{L^2 ( \mathbb R_x^-\times \mathbb R_y^+)}\,,
\end{equation*}
we observe that
\begin{equation*}
N_2 \lesssim || \Ai(e^{i\alpha} w_x) ||_{L^2(\mathbb R_-)}\, || \Ai(e^{-i\alpha} w_x)||_{L^2 ( \mathbb R_+)}\,,
\end{equation*}
and having in mind \eqref{upb}, we have only to bound 
 $
 \int_{-\infty}^{0} | \Ai(e^{i \alpha }w_x) |^2 dx \,.
 $
We can no more use the asymptotic for the Airy function as $(x+\eta)$
is small.  We have indeed
\begin{equation*}
e^{i \alpha} w_x = -  e^{i \frac \pi 6} (x+ \eta) + e^{i \alpha} \lambda_0\,.
\end{equation*}  
We rewrite the integral as the sum
\begin{equation*}
 \begin{array}{ll}
 \int_{-\infty}^{0} | \Ai(e^{i \alpha }w_x) |^2 dx  &=   \int_{-\infty}^{- \eta - C } | \Ai(e^{i \alpha }w_x) |^2 dx \\ &\quad  +
   \int_{-\eta - C }^{-\eta + C } | \Ai(e^{i \alpha }w_x) |^2 dx +  \int_{-\eta + C }^{0 } | \Ai(e^{i \alpha }w_x) |^2 dx\,.
  \end{array}
\end{equation*}
The integral in the middle of the r.h.s. is bounded.  The first one is
also bounded according to the behavior of the Airy function.  So the
dominant term is the third one
\begin{equation*}
\begin{array}{ll}
    \int_{-\eta + C }^{0 } | \Ai(e^{i \alpha }w_x) |^2 dx & = \int_{C}^{ \eta} | \Ai(  -  e^{i \frac \pi 6} x 
    + e^{i \alpha} \lambda_0   ) |^2 dx \\ &  \leq \tilde C     (|\eta|^2 +1)^{\frac 14}\, \,\exp \left(+ \frac {2 \sqrt{2}}{3}  |\eta|^{\frac 32}\right)
 \,.
 \end{array}
\end{equation*}
Combining with \eqref{upb}, the $L^2$-norm of the second term decays
as $\eta \ar +\infty$:
\begin{equation}\label{estn2}
   N_2 \lesssim   (|\eta|^2 +1)^{-\frac 18}\,.
\end{equation}
This achieves the proof of the proposition, the uniformity for
$\lambda_0$ in a compact being controlled at each step of the proof.

\section{Proof of the completeness}\label{s8}

We have already recalled or established in Section \ref{s2a}
(Propositions \ref{CompD}, \ref{CompN}, and \ref{CompR}) the results
for the Dirichlet, Neumann or Robin realization of the complex Airy
operator in $\mathbb R_+$.  The aim of this section is to establish
the same result in the case with transmission.  The new difficulty is
that the operator is no more sectorial.\\

\subsection{Reduction to the case $\kappa=0$} 

We first reduce the analysis to the case $\kappa=0$ by comparison of
the two kernels.  We have indeed
\begin{equation}\label{ref8.1}
\begin{array}{ll}
\mathcal G ^- (x,y\,;\lambda,\kappa) - \mathcal G ^- (x,y\,;\lambda,0)
& = \mathcal G _1 (x,y\,;\lambda,\kappa) - \mathcal G  _1 (x,y\,;\lambda,0)\\
 &  = - \kappa  (f(\lambda) +\kappa)^{-1}  \mathcal G _1 (x,y\,;\lambda,0)\,,
  \end{array}
\end{equation}
where $\mathcal G ^- (x,y\,;\lambda,\kappa)$ denotes the kernel of the
resolvent for the transmission problem associated to $\kappa \geq 0$
and $D_x^2-ix$.\\ We will also use the alternative equivalent
relation: 
\begin{equation}\label{nw}
\mathcal G ^- (x,y\,;\lambda,\kappa)=   \mathcal G ^-  (x,y\,;\lambda,0) f(\lambda) \, ( f(\lambda) +\kappa)^{-1} 
 + \kappa  (f(\lambda) +\kappa)^{-1} \,   \mathcal G _0 ^-(x,y\,;\lambda,0)\,.
\end{equation}

\begin{remark}
This formula gives another way for proving that the operator with
kernel $\mathcal G^\pm (x,y\,;\lambda,\kappa) $ is in a suitable
Schatten class (see Proposition~\ref{propSchatten}).  It is indeed enough to
have the result for $\kappa=0$, that is to treat the Neumann case on
the half line.
\end{remark}

Another application of this formula is
\begin{proposition}\label{conj}
There exists $M>0$ such that for all $\lambda>0\,$,
\begin{equation}
  \|(\mathcal A^\pm_1-\lambda)^{-1}\|_{HS} \leq M(1+|\lambda|)^{-\frac 14} (\log \lambda)^\frac 12 \,.
\end{equation}
\end{proposition}

{\bf Proof } Proposition \ref{conj} is a consequence of Proposition
\ref{conjb}, and Formula \eqref{ref8.1}.

\begin{remark}
Similar estimates are obtained in the case without boundary (typically
for a model like the Davies operator $D_x^2 + i x^2$) by
Dencker-Sj\"ostrand-Zworski \cite{DSZ} or more recently by Sj\"ostrand
\cite{Sj}.  
\end{remark}

\subsection{Estimate for $f(\lambda)$} 

We recall that $f(\lambda)$ was defined in \eqref{eq:Airy_h} by
\begin{equation*}
f(\lambda) :=2\pi \Ai'(e^{-i\alpha} \lambda) \Ai'(e^{i\alpha} \lambda)\, .
\end{equation*}

Recalling the asymptotic expansions (\ref{5a}) and (\ref{6b}) of $\Ai'$, it
is immediate to get
\begin{lemma}\label{lemmatype32}
The function $\lambda \mapsto f(\lambda)$ is an entire function of
type $\frac 32$, i.e. there exists $D >0$ such that
\begin{equation}
|f(\lambda) | \leq D \exp \bigl(D |\lambda|^\frac 32\bigr)\,,\quad \forall \lambda \in \mathbb C\,.
\end{equation}
\end{lemma}

Focusing now on the main purpose of this section, we get from
\eqref{5a} the existence of $\lambda_1>0$ such that, for
$\lambda\geq \lambda_1$,
\begin{equation}
|\Ai'(e^{i\alpha}\lambda)|^2= |\Ai'(e^{-i\alpha}\lambda)|^2 \geq \frac{1}{4\pi} \lambda^{1/2} \exp\left(\frac 43\lambda^{3/2}\right),
\end{equation}
where $c_1>0\,$.\\
Thus there exists  $C_1 >0$ such that, for $\lambda \geq 1$, 
\begin{equation}\label{eq:est_f}
 \frac{1}{|f(\lambda) |} \leq \frac{C_1}{\lambda^\frac 12} \, \exp\left(-\frac 43\lambda^{3/2}\right)\,.
\end{equation}

\subsection{Estimate of the $L^2$ norm of $\mathcal G _1(\cdot\,,\,\cdot\,;\lambda,0)$} 

Having in mind \eqref{eq:G1tilde}-\eqref{yneg} and noting that
$\frac{ [\Ai'(e^{i\alpha} \lambda)]^2}{|f(\lambda)| } 
=\frac{1}{2\pi}$, it is enough to estimate
\begin{equation}
\label{eq:I0def}
\int_{0}^{+\infty} |\Ai (e^{-i \alpha }(ix + \lambda))|^2 dx = I_0 (\lambda) =\int_{-\infty}^{0} |\Ai (e^{i \alpha }(ix + \lambda))|^2 dx   \,.
\end{equation}
It is enough to observe from \eqref{eq:G0_free} that
\begin{equation}
2 I_0(\lambda)^2 \leq  || \mathcal G_0^-(\cdot\,,\,\cdot\,; \lambda) ||^2\,.
\end{equation}
Applying \eqref{BMM}, we get
\begin{equation}\label{estI0}
I_0(\lambda)\lesssim \lambda^{-\frac 14} \exp \left(\frac 43 \lambda^\frac 32 \right)\,.
\end{equation}
Hence, coming back to \eqref{ref8.1}, we have obtained
\begin{proposition}
There exist $\kappa_0$, $C$ and $\lambda_0 >0$
such that, for all $\kappa \in [0,\kappa_0]$, for all $\lambda \geq
\lambda_0$,
\begin{equation}
|| \mathcal G ^- (\cdot\,,\,\cdot\,; \lambda,\kappa)\, - \,  \mathcal G ^- (\cdot\,,\,\cdot\,; \lambda,0) ||_{L^2(\mathbb R^2)}
\leq C \kappa \, |\lambda|^{-\frac 34  }\,.
\end{equation}
\end{proposition}
Hence we are reduced to the case $\kappa=0$ which can be decoupled
(see Remark~\ref{remN}) in two Neumann problems on $\mathbb R_-$ and
$\mathbb R_+$.\\

Using \eqref{nw} and the estimates established for $\mathcal G_0^-
(\cdot\,,\,\cdot\,;\lambda,0)$ (which depends only on $\Re
\lambda$) (see \eqref{BMM} or \eqref{eq:BM}), we have
\begin{proposition}
For all $\kappa_0$, there exist a constant $C$ and $\lambda_0 >0$ such
that, for all $\kappa \in [0,\kappa_0]$, for all real $ \lambda \geq
\lambda_0$, one has
\begin{equation}
|| \mathcal G ^- (\cdot\,,\,\cdot\,; \lambda,\kappa)\, - \, (f(\lambda) (f(\lambda) +\kappa)^{-1} ) \mathcal G ^- (\cdot\,,\,\cdot\,; \lambda,0) ||_{L^2(\mathbb R^2)}
\leq C \kappa \, |\lambda|^{-\frac 34  }\,.
\end{equation}
\end{proposition}
This immediately implies
\begin{proposition}\label{prop:newnew}
 For any $g= (g_-,g_+)$, $h= (h_-, h_+)$ in $L^2_-\times L^2_+$, we have
\begin{equation}
| \langle \mathcal G ^- (\lambda,\kappa) g\,,\, h \rangle\, - \, (f(\lambda) (f(\lambda) +\kappa)^{-1} ) \langle \mathcal G ^- (\lambda,0) g\,,\, h\rangle |
\leq C(g,h)  \kappa \, |\lambda|^{-\frac 34  }\,,
\end{equation}
where $\langle \cdot , \cdot\rangle$ denotes the scalar product
in the Hilbert space  $L^2_-\times L^2_+$.
\end{proposition}

We now adapt the proof of the completeness from \cite{Agm}.\\
If we denote by $E$ the closed space generated by the generalized
eigenfunctions of $\mathcal A_1^-$, the proof of \cite{Agm} in the
presentation of \cite{Hen} consists in introducing
\begin{equation*}
F(\lambda) = \langle \mathcal G ^- (\lambda,\kappa) g\,,\, h \rangle ,
\end{equation*}
where 
\begin{equation}
\label{eq:cond_auxil1}
h\in E^\perp \quad \textrm{and} \quad g \in L^2_-\times L^2_+. 
\end{equation}
As a consequence of the assumption on $h$, one observes that
$F(\lambda)$ is an entire function and the problem is to show that $F$
is identically $0$.  The completeness is obtained if we prove this for
any $g$ and $h$ satisfying the condition (\ref{eq:cond_auxil1}).\\
Outside the numerical range of $\mathcal A_1^-$, i.e. in the negative
half-plane, it is immediate to see that $F(\lambda)$ tends to zero as
$\Re \lambda \to -\infty$.  If we show that $|F(\lambda)| \leq C
(1+|\lambda|)^M$ for some $M>0$ in the whole complex plane, we will
get by Liouville's theorem that $F$ is a polynomial and, with the
control in the left half-plane, we should get that $F$ is identically
$0$.\\ Hence it remains to control $F(\lambda)$ in a neighborhood of
the positive half-plane $\{\lambda\,,\, \Re \lambda \geq0\}$.\\
   
As in \cite{Agm}, we apply Phragmen-Lindel\"of principle (See Appendix
\ref{AppD}).  The natural idea (suggested by the numerical picture) is
to control the resolvent on the positive real axis.  We first recall
some additional material present in Chapter 16 in
\cite{Agm}.
\begin{theorem}\label{thabove}
Let $\phi(\lambda)$ be an entire complex valued function of finite
order $\rho$.  Then for any $\epsilon >0$ there exists a sequence $r_1
< r_2 < \cdots < r_k$ such that
\begin{equation*}
\min_{|\lambda| = r_k}| \phi(\lambda)  | > \exp (- r_k^{\rho +\epsilon})\,.
\end{equation*}
\end{theorem}
For this theorem (Theorem 6.2 in \cite{Agm}), S. Agmon refers to the
book of Titchmarsh \cite{Ti} (p. 273).

This theorem is used for proving an inequality of the type $\rho$ with
$\rho=2$ in the Hilbert-Schmidt case.  We avoid an abstract lemma
\cite{Agm} (Lemma 16.3) but follow the scheme of its proof for
controlling directly the Hilbert-Schmidt norm of the resolvent along
an increasing sequence of circles.
\begin{proposition}
For $\epsilon >0$, there exists a sequence $r_1 < r_2 < \cdots < r_k$
such that
\begin{equation*}
\max_{|\lambda| = r_k} ||\mathcal G^\pm(\cdot\,,\,\cdot\,; \lambda,\kappa)||_{HS} \leq  \exp \bigl(r_k^{\frac 32 +\epsilon} \bigr)\,.
\end{equation*}
\end{proposition}
{\bf Proof.}\\
We start from  
\begin{equation}\label{nww}
\mathcal G ^- (x,y\,;\lambda,\kappa)=   \mathcal G ^- (x,y\,;\lambda,0) f(\lambda) \, ( f(\lambda) +\kappa)^{-1} 
 + \kappa  (f(\lambda) +\kappa)^{-1} \,   \mathcal G _0 ^-(x,y\,;\lambda,0)\,.
\end{equation}
We apply Theorem \ref{thabove} with $\phi(\lambda) = f(\lambda) +
\kappa$.  It is proven in Lemma~\ref{lemmatype32} that $f$ is of type
$\frac 32$.  Hence we get for $\epsilon >0$ (arbitrary small) the
existence of a sequence $r_1 < r_2 < \cdots < r_k$ such that
\begin{equation*}
\max_{|\lambda| = r_k}  \left|\frac{1}{f(\lambda) +\kappa} \right|  \leq  \exp \bigl(r_k^{\frac 32 +\epsilon}\bigr)\,.
\end{equation*}
In view of \eqref{nww}, it remains to control the Hilbert-Schmidt norm of 
\begin{equation*}
\mathcal G ^- (x,y\,;\lambda,0) f(\lambda) \,  + \kappa   \,   \mathcal G _0 ^-(x,y\,;\lambda,0)\,.
\end{equation*}
Hence the remaining needed estimates only concern the case
$\kappa=0$. The estimate on the Hilbert-Schmidt norm of $ \mathcal G
_0 ^-$ is recalled in \eqref{BMM}.  It remains to get an estimate
for the entire function $\mathcal G^- (x,y\,;\lambda,0) f(\lambda)$.\\
Because $\kappa=0\,$, this can be reduced to a question for the
Neumann problem on the half-line for the complex Airy operator $D_x^2
- ix$.  For $y>0\,$ and $x>0\,$, $f(\lambda) \,\mathcal G
^N_1(x,y\,;\lambda)$ is given by the following expression
\begin{equation}\label{eq:G1tildeneumann}
f(\lambda) \, \mathcal G ^N_1(x,y\,;\lambda) =  - 4\pi^2 [e^{2i\alpha} \Ai'(e^{i\alpha} \lambda)]^2 \Ai(e^{-i\alpha} w_x) \Ai(e^{-i\alpha} w_y)\, .
\end{equation}
We only need the estimate for $\lambda$ in a sector containing
$\mathbb R_+\times \mathbb R_+$.  This is done in \cite{Hen} but we
will give a direct proof below.  In the other region, we can first
control the resolvent in $\mathcal L (L^2)$ and then use the resolvent
identity
\begin{equation*}
\mathcal G ^{\pm, N}(\lambda) - \mathcal G ^{\pm,N}(\lambda_0) = (\lambda-\lambda_0) \mathcal G ^{\pm, N}(\lambda) \mathcal G ^{\pm,N}(\lambda_0)\,.
\end{equation*}
This shows that in order to control the Hilbert-Schmidt norm of
$\mathcal G ^{\pm,N}(\lambda)$ for any $\lambda$, it is enough to
control the Hilbert-Schmidt norm of $\mathcal G ^{\pm,N}(\lambda_0)$
for some $\lambda_0$, as well as the $\mathcal L (L^2)$ norm of
$\mathcal G ^{\pm, N}(\lambda)$, the latter being easier to
estimate.\\

More directly the control of the Hilbert-Schmidt norm is reduced to
the existence of a constant $C>0$ such that 
\begin{equation*}
\int_0^{+\infty} | \Ai(e^{-i\alpha} (ix+\lambda))|^2 dx \leq C \, \exp (C |\lambda|^\frac 32) \,.
\end{equation*}

In this case, we have to control the resolvent in a neighborhood of
the sector $\Im \lambda \leq 0\,,\, \Re\lambda \geq 0$, which corresponds to 
the numerical range of the operator.\\ 
As $x\ar +\infty$, the dominant term in the argument of the Airy
function is $e^{i (-\alpha+\frac \pi 2)} x = e^{-i \frac \pi 6} x$.
As expected we arrive in a zone of the complex plane where the Airy
function is exponentially decreasing.  It remains to estimate for
which $x$ we enter in this zone.  We claim that there exists $C > 0$
such that if $x \geq C |\lambda|$ and $|\lambda| \geq 1$, then $$|\Ai(
e^{-i\alpha} (ix+\lambda)) | \leq C \exp (- C( x + |\lambda|)^{\frac
32})\,.$$
In the remaining zone, we obtain easily an upper bound of the integral
by $\mathcal O \bigl(\exp (C |\lambda|^\frac 32)\bigr)$.\\
 
We will then use the Phragmen-Lindel\"of principle (Theorem
\ref{thPL}).  For this purpose, it remains to control the resolvent on
the positive real line.  It is enough to prove the theorem for
$g^+=(0,g_+)$ and $g^-= (g_-,0)$.  In other words, it is enough to
consider $F_+$ (resp. $F_-$) associated with $g^+$ (resp. $g^-$).\\
Let us treat the case of $F_+$ and use Formula \eqref{nw} and
Proposition \ref{prop:newnew}:
\begin{equation}\label{compa}
|\langle \mathcal G ^- (\lambda,\kappa) g^+\,,\, h \rangle\, - \, (f(\lambda) (f(\lambda) +\kappa)^{-1} ) \langle \mathcal G ^- (\lambda,0) g_+\,,\, h_+\rangle |
\leq C(g,h)  \kappa \, |\lambda|^{-\frac 34  }\,.
\end{equation}
This estimate is true on the positive real axis.  It remains to
control the term $|\langle \mathcal G ^- (\lambda,0) g^+\,,\, h
\rangle |$.  Along this positive real axis, we have by Proposition
\ref{conjb} the decay of $F_+(\lambda)$.  Using Phragmen-Lindel\"of principle 
completes the proof.
 
Note that for $F_-(\lambda)$, we have to use the symmetric (with
respect to the real axis) curve in $\Im \lambda >0$.
 
In summary, we have obtained the following proposition
\begin{proposition}\label{CompT}
For any $\kappa \geq 0$, the space generated by the generalized
eigenfunctions of the complex Airy operator with transmission is dense
in $L^2_-\times L^2_+$.
\end{proposition}

\newpage 
\appendix 
\section*{Appendices}

\section{Basic properties of the Airy function}  \label{AppA}

In this Appendix, we summarize the basic properties of the Airy
function $\Ai(z)$ and its derivative $\Ai'(z)$ that we used (see
\cite{AS} for details).\\
We recall that the Airy function is the unique solution  of
\begin{equation*}
(D_x^2 + x) u =0\,,
\end{equation*}
on the line such that $u(x)$ tends to $0$ as $x \rightarrow +\infty$ and 
\begin{equation*}
\Ai (0) = 1/ \left( 3^{\frac 23}\, \Gamma \left(\frac 23\right)\right).
\end{equation*}
This Airy function extends into a holomorphic function in $\mathbb C\,$.

The Airy function is positive decreasing on $\mathbb R_+$ but has an
infinite number of zeros in $\mathbb R_-$.  We denote by $a_n$ ($n\in
\mathbb N$) the decreasing sequence of zeros of $\Ai$.  Similarly we
denote by $a'_n$ the sequence of zeros of $\Ai'$.  They have the
following asymptotics (see for example \cite{AS}), as $n\ar +\infty$,
\begin{equation}\label{zeros}
a_n \underset{\tiny{n\rightarrow+\infty}}{\sim} - \left(\frac{3\pi}{2}(n-1/4)\right)^{2/3}\,,
\end{equation}
and 
\begin{equation}\label{critiques}
a'_n \underset{\tiny{n\rightarrow+\infty}}{\sim} - \left(\frac{3\pi}{2}(n-3/4)\right)^{2/3}\,.
\end{equation}

The functions $\Ai(e^{i\alpha} z)$ and $\Ai(e^{-i\alpha } z )$ (with
$\alpha = 2\pi/3$) are two independent solutions of the differential
equation
\begin{equation*}
\left(-\frac{d^2}{dz^2} - i z\right) w(z) = 0\,  .
\end{equation*}
Considering their Wronskian, one gets
\begin{equation}
\label{eq:Wronskian}
e^{-i\alpha} \Ai'(e^{-i\alpha} z) \Ai(e^{i\alpha} z) - e^{i\alpha} \Ai'(e^{i\alpha} z) \Ai(e^{-i\alpha} z) = \frac{i}{2\pi} \,  \quad \forall~ z\in \mathbb C\, .
\end{equation}
Note that these two functions are related to $\Ai(z)$ by the identity
\begin{equation}\label{ide}
\Ai (z) + e^{-i \alpha} \Ai (e^{-i \alpha} z) + e^{i\alpha} \Ai (e^{i\alpha}z) =0\,   \quad \forall~ z\in \mathbb C\, .
\end{equation}

The Airy function and its derivative satisfy different asymptotic
expansions depending on their argument: \\ (i) For $|\arg z|<\pi$,
\begin{eqnarray} \label{5} 
\Ai(z) &=& \frac 12 \pi^{-\frac 12}z^{-1/4} \,  \exp\left(-\frac{2}{3}z^{3/2}\right) \bigl(1 + \mathcal O  (|z|^{-\frac 32}) \bigr), \\
\label{5a}
\Ai'(z) &=& - \frac 12 \pi^{-\frac 12} \, z^{1/4}\, \exp\left(-\frac{2}{3}z^{3/2}\right) \bigl(1+ \mathcal O (|z|^{-\frac 32}) \bigr)  
\end{eqnarray}
(moreover $\mathcal O$ is, for any $\epsilon >0$, uniform when $|\arg
z| \leq \pi -\epsilon$)\,.  \\

(ii) For $|\arg z|<\frac{2}{3}\pi\,$,
\begin{eqnarray}
\label{6}
\Ai(-z) &=& \pi^{-\frac 12}z^{-1/4} \left(  \sin\left(\frac{2}{3}z^{3/2}+\frac{\pi}{4}\right) (1 + \mathcal O (|z|^{-\frac 32})\right. \\
\nonumber
& & \left. -  \frac{5}{72} \left(\frac 23 z^\frac 32\right)^{-1}  \cos\left(\frac{2}{3}z^{3/2}+\frac{\pi}{4}\right) (1 + \mathcal O (|z|^{-\frac 32}) \right) \\
\label{6b}
\Ai'(-z) &=& - \pi^{-\frac 12} z^{1/4} 
\left( \cos \left(\frac{2}{3}z^{3/2}+\frac{\pi}{4}\right) (1+ \mathcal O (|z|^{-\frac 32}))\right.  \\
\nonumber
& &  \left. + \frac{7}{72}  \left(\frac{2}{3}z^{3/2}\right)^{-1} \sin \left(\frac{2}{3}z^{3/2}+\frac{\pi}{4}\right) (1+ \mathcal O (|z|^{-\frac 32}))\right)
\end{eqnarray}
(moreover $\mathcal O$ is for any $\epsilon >0$, uniform in the sector
$\{ |\arg z| \leq \frac{2\pi}{3} -\epsilon\}$).

\section{Analysis of the resolvent of $\mathcal A^+$ on the line for $\lambda>0$ (after \cite{Mar})}\label{AppB}

On the line $\R$, $\mathcal A^+$ is the closure of the operator
$\mathcal A_0^ +$ defined on $C_0^\infty(\mathbb R)$ by $\mathcal
A_0^+ =D_x^2 +ix$.  A detailed description of its properties can be
found in \cite{Hel1}.  In this appendix, we give the asymptotic
control of the resolvent $(\mathcal A^+-\lambda)^{-1}$ as $\lambda \ar
+\infty$.  We successively discuss the control in $\mathcal
L(L^2(\mathbb R))$ and in the Hilbert-Schmidt space $\mathcal
C^2(L^2(\mathbb R))$.  These two spaces are equipped with their
canonical norms.

\subsection{Control in $\mathcal L (L^2(\mathbb R))$.}

Here we follow an idea present in the book of Davies \cite{Dav2} and
used in Martinet's PHD \cite{Mar} (see also \cite{Hel1}).
\begin{proposition}\label{propmajoration}~\\
For all $\lambda>\lambda_0\,$,
   \begin{eqnarray}
     \|(\mathcal A^+-\lambda)^{-1}\|_{\mathcal{L}(L^2(\mathbb{R}))}\le
     \sqrt{2\pi}~ \lambda^{-\frac{1}{4}}\exp\left(\frac{4}{3}\lambda^{\frac{3}{2}}\right) \bigl(1+o(1)\bigr)\,.
   \end{eqnarray}
\end{proposition}
{\bf Proof } The proof is obtained by considering $\mathcal A^+$ in
the Fourier space, i.e.
\begin{equation}
\widehat {\mathcal A}^+=\xi^2 + \frac{d}{d\xi}\,.
\end{equation}
The associated semi-group $T_t:= \exp (-\widehat{\mathcal A}^+ t)$ is
given by
\begin{eqnarray}
 {T}_tu(\xi)=\exp \left(-\xi^2t- \xi
  t^2-\frac{t^3}{3}\right) u(\xi-t)\,,\quad \forall\,u\in \mathcal S(\mathbb{R})\,.
\end{eqnarray}
$T_t$ appears as the composition of a multiplication by $\exp(-\xi^2t-
\xi t^2-\frac{t^3}{3})$ and a translation by $t$.  Computing $\sup_\xi
\exp(-\xi^2t- \xi t^2-\frac{t^3}{3})$ leads to
\begin{eqnarray}
  \|{T}_t\|_{\mathcal{L}(L^2(\mathbb{R}))}\leq \exp\left(-\frac{t^3}{12}\right)\,.
\end{eqnarray}
It is then easy to get an upper bound for the resolvent. For $\lambda >0$, we have
\begin{eqnarray}
   \|(\mathcal A^+-\lambda)^{-1}\|_{\mathcal{L}(L^2(\mathbb{R}))} & =  & \|(\widehat{ \mathcal A}^+-\lambda)^{-1}\|_{\mathcal{L}(L^2(\mathbb{R}))} \\ 
    &\le&
     \int_0^{+\infty}\exp(t\lambda)\|{T}_t\|_{\mathcal{L}(L^2(\mathbb{R}))}dt\\
     &\le&\int_0^{+\infty}\exp\left(t\lambda-\frac{t^3}{12}\right)dt\,.
\end{eqnarray}
The right hand side can be estimated by using the Laplace integral
method.  Setting $t=\lambda^{\frac{1}{2}}s$, we have
\begin{eqnarray}
  \int_0^{+\infty}\exp\left(t\lambda-\frac{t^3}{12}\right)dt = 
\lambda^{\frac{1}{2}}  \int_0^{+\infty} \exp\left(\lambda^{\frac{3}{2}}\left(s-\frac{s^3}{12}\right)\right)ds\,.
\end{eqnarray}
We observe that $\hat \phi(s)=s-\frac{s^3}{12}$ admits a global
non-degenerate maximum on $[0,+\infty)$ at $s=2$ with $\hat
\phi(2)=\frac{4}{3}$ and $\hat \phi''(2)=- 1 $.  The Laplace integral
method gives the following equivalent as $\lambda \ar +\infty$~:
\begin{eqnarray}
  \int_0^{+\infty}\exp\left(\lambda^{\frac{3}{2}}(s-\frac{s^3}{12})\right) ds \sim 
\sqrt{2\pi}~ \lambda^{-\frac{3}{4}}\exp\left(\frac{4}{3}\lambda^{\frac{3}{2}}\right) \,.
\end{eqnarray}
This completes the proof of the proposition.  We note that this upper
bound is not optimal in comparison with Bordeaux-Montrieux's formula
\eqref{eq:BM}.

\subsection{Control in Hilbert-Schmidt norm}

In this part, we give a proof of Proposition \ref{PropMart}.  As in
the previous subsection, we use the Fourier representation and analyze
$\widehat {\mathcal A}^+$.  Note that
\begin{equation}\label{invHS}
 \|(\widehat{\mathcal A}^+-\lambda)^{-1}\|_{HS}^2 = \| ({\mathcal A}^+-\lambda)^{-1}\|_{HS}^2 
\end{equation}
We have then an explicit description of the resolvent by
\begin{equation*}
  (\widehat{\mathcal A}^+-\lambda)^{-1}u(\xi)=\int_{-\infty}^\xi u(\eta)\exp\bigg(\frac{1}{3}(\eta^3-\xi^3)+\lambda(\xi- \eta)\bigg)d\eta\,.
\end{equation*}
Hence, we have to compute
\begin{equation*}
  \|(\widehat{\mathcal A}^+-\lambda)^{-1}\|_{HS}^2 =\int\int_{\eta<\xi}\exp\bigg(\frac{2}{3}(\eta^3-\xi^3)+2\lambda(\xi-\eta)\bigg)d\eta d\xi\,.
\end{equation*}
After the change of variable
$(\xi_1,\eta_1)=(\lambda^{-\frac{1}{2}}\xi,\lambda^{-\frac{1}{2}}\eta)$,
we get
\begin{equation*}
 \|(\widehat{\mathcal A}^+-\lambda)^{-1}\|_{HS}^2 =\lambda \int\int_{\eta_1<\xi_1}\exp\left(\lambda^{\frac{3}{2}}
\left[\frac{2}{3}(\eta_1^3-\xi_1^3)+2(\xi_1-\eta_1)\right]\right)\, d\xi_1d\eta_1\,.
\end{equation*}
With
\begin{equation}\label{deft}
h=\lambda^{-\frac 32}\,,
\end{equation}
we can write
\begin{equation}\label{forHS}
  \|(\widehat{\mathcal A}^+-\lambda)^{-1}\|_{HS}^2 = h^{-\frac 23} \Phi(h) ,
  \end{equation}
where 
\begin{equation}
 {\Phi}(h)=\int_{y<x}\exp\left(\frac{2}{h}[\phi(x)-\phi(y)]\right)dxdy\,,\label{int1}
\end{equation}
with
\begin{eqnarray}
  \phi(x)=-\frac{x^3}{3}+x\,.
\end{eqnarray}
$\Phi(h)$ can now be split in three terms
\begin{eqnarray}
\Phi(h) = I_1(h)+I_2(h)+I_3(h)\,,
\end{eqnarray}
with
\begin{eqnarray}
  &&I_1(h)=\int_{\substack{y<x\\y>0}}\exp\left(\frac{2}{h}[\phi(x)-\phi(y)]\right)\,dxdy\,,\nonumber\\
  &&I_2(h)=\int_{\substack{y<x\\x<0}}\exp\left(\frac{2}{h}[\phi(x)-\phi(y)]\right)\,dxdy\,,\nonumber\\
  &&I_3(h)=\int_{\substack{x\in\mathbb{R}^+\\y\in\mathbb{R}^-}}\exp\left(\frac{2}{h}[\phi(x)-\phi(y)]\right)\, dxdy\nonumber\,.
\end{eqnarray}
We observe now that by the change of variable $(x,y)\mapsto (-y,-x)$,
we get
\begin{equation*}
  I_1(h)=I_2(h)\,,
\end{equation*}
and that
\begin{equation*}
  I_3(h)=I_4(h)^2\,,
\end{equation*}
with
\begin{equation*}
  I_4(h)=\int_{\mathbb{R}^+}\exp\left(\frac{2}{h}\phi(x)\right)dx\,.
\end{equation*}
Hence, it remains to estimate, as $h\ar 0$, the integrals $I_1(h)$ and
$I_4(h)$.\\
\paragraph{Control of $I_1(h)$}~\\
The function $\phi(x)$ is positive on $ (0,\sqrt{3})$ and negative
decreasing on $(\sqrt{3},+\infty)$ ($\phi(0)=\phi(\sqrt{3})=0$).  It
admits a unique (non degenerate) maximum at $x=1$ with $\phi(1)=\frac
23$.\\ Using the trivial estimates
\begin{equation*}
  \exp\left(-\frac{2}{h}\phi(y) \right)\le 1\,,\quad \forall\,y\in[0,\sqrt{3}]\,,
\end{equation*}
\begin{equation*}
\begin{array} {ll}
  \exp\left(-\frac{2}{h}\phi(y)\right)&=-\frac{h}{2}\frac{1}{1-y^2}\frac{d}{dy}[\exp(-\frac{2}{h}\phi(y))]\nonumber\\
  &\le-\frac{h}{2}\frac{1}{1-x^2}\frac{d}{dy}[\exp(-\frac{2}{h}\phi(y))]\,,\quad \mbox{ if } \sqrt{3} < x < y\,,
\end{array}
\end{equation*}
and
\begin{equation*}
  \exp\left(\frac{2}{h}\phi(x)\right)\le1\,,\quad \forall\,x\in[\sqrt{3},+\infty[\,,
\end{equation*}
we can estimate $I_1$ from above in the following way
\begin{eqnarray}
  I_1(h)&=&\int_0^{\sqrt{3}}\exp\left(\frac{2}{h}\phi(x)\right)\left(\int_0^x\exp\left(-\frac{2}{h}\phi(y)\right)dy\right)dx\nonumber\\
  &&+\int_{\sqrt{3}}^{+\infty}\exp\left(\frac{2}{h}\phi(x)\right)\left( \int_0^{\sqrt{3}}\exp\left(-\frac{2}{h}\phi(y)\right)dy \right) dx \nonumber\\
  &&+\int_{\sqrt{3}}^{+\infty}\exp\left(\frac{2}{h}\phi(x)\right )\left(\int_{\sqrt{3}}^x\exp\left(-\frac{2}{h}\phi(y)\right)dy\right)dx \nonumber\\
  &\le&\int_0^{\sqrt{3}}\exp\left(\frac{2}{h}\phi(x)\right)\left( \int_0^{\sqrt{3}}\exp\left(-\frac{2}{h}\phi(y)\right)dy\right)dx\nonumber\\
  &&+\sqrt{3}\int_{\sqrt{3}}^{+\infty}\exp\left(\frac{2}{h}\phi(x)\right)dx\nonumber\\
  &&-\frac{h}{2}\int_{\sqrt{3}}^{+\infty}\frac{1}{1-x^2}\exp\left(\frac{2}{h}\phi(x)\right)
 \left( \int_{\sqrt{3}}^x\frac{d}{dy} \exp\left(-\frac{2}{h}\phi(y)\right)dy\right) dx\nonumber\\
  &\le&3\sup_{[0,\sqrt{3}]}\left\{\exp\left(\frac{2}{h}\phi(x)\right)\right\}\nonumber\\
  &&+\frac{\sqrt{3} h}{2}\int_{\sqrt{3}}^{+\infty}\frac{1}{1-x^2}\frac{d}{dx} \exp\left(\frac{2}{h}\phi(x)\right) dx\nonumber\\
  &&-\frac{h}{2}\int_{\sqrt{3}}^{+\infty}\frac{1-\exp(\frac{2}{h}\phi(x))}{1-x^2}dx\nonumber \\
  &\le & 3 \exp \left(\frac {4}{3h}\right) + \frac{\sqrt{3}h}{4} +\frac h2 \int_{\sqrt{3}}^{+\infty} \frac{1}{x^2-1} dx\nonumber  \,.
\end{eqnarray}
Hence we have shown the existence of $C>0$ and $h_0>0$ such that, for
$h \in (0,h_0)$,
\begin{equation}
  I_1(h)\le C\exp\left(\frac{4}{3h}\right)\,.
\end{equation}
Hence $I_1(h)$ and $I_2(h)$ appear as remainder terms.\\
\paragraph{Asymptotic of  $I_4(h)$}
Here, using the properties of $\phi$, we get by the standard Laplace
integral method
\begin{eqnarray}\label{devasymp}
  I_4(h) \sim \sqrt{\pi /2} \sqrt{h}\exp\left(\frac{4}{3h}\right)\,.
\end{eqnarray}
Hence, putting altogether the estimates, we get, as $h \ar 0$,
\begin{equation}
 {\Phi}(h)\sim \frac{\pi h}{2} \exp\left(\frac{8}{3h}\right)
\end{equation}
Coming back to \eqref{invHS}, \eqref{deft} and \eqref{forHS}, this
achieves the proof of Proposition \ref{PropMart}.

\section{Analysis of the resolvent for the Dirichlet realization in the half-line.}\label{AppC}

\subsection{Main statement}

The aim of this appendix is to give the proof of Proposition
\ref{conja}.  Although it is not used in our main text, it is
interesting to get the main asymptotic for the Hilbert-Schmidt norm of
the resolvent in Proposition \ref{conja}.
\begin{proposition}
As $\lambda \ar +\infty$, we have:
\begin{equation} 
|| \mathcal G^{-,D}(\lambda) ||_{HS} \sim \frac{\sqrt{3}}{2\sqrt{2}}     \,  \lambda^{-\frac 14} (\log \lambda)^\frac 12\,.
\end{equation}
\end{proposition}

\subsection{The Hilbert-Schmidt norm of the resolvent for real $\lambda$ }

The Hilbert-Schmidt norm of the resolvent can be written as
\begin{equation}
\label{eq:GnormD}
|| \mathcal G^{-,D} ||^2_{HS} = \int\limits_{\R_+^2} |\mathcal G ^{-,D} (x,y\,;\lambda)|^2 dx dy = 8\pi^2 \int\limits_0^\infty Q(x;\lambda) dx ,
\end{equation}
where
\begin{equation}
\label{eq:Qdef}
\begin{split}
Q(x;\lambda) & = \frac{|\Ai(e^{-i\alpha}(ix+\lambda))|^2}{|\Ai(e^{-i\alpha}\lambda)|^2} \times \\
&\times  \int\limits_0^x \left|\Ai(e^{i\alpha}(iy+\lambda)) \Ai(e^{-i\alpha}\lambda) - \Ai(e^{-i\alpha}(iy+\lambda)) \Ai(e^{i\alpha}\lambda)\right|^2 dy .\\
\end{split}
\end{equation}
Using the identity (\ref{ide}), we observe that
\begin{equation}\label{cancel}
\begin{array}{l}
\Ai(e^{i\alpha} (iy+\lambda)) \Ai (e^{-i \alpha} \lambda) - \Ai (e^{-i\alpha} (iy+\lambda)) \Ai (e^{i \alpha} \lambda) 
\\ \qquad =  e^{-i \alpha} \left( \Ai (e^{-i \alpha} (iy+\lambda)) \Ai(\lambda) - \Ai (iy+\lambda) \Ai (e^{-i \alpha} \lambda)\right)\,.
\end{array}
\end{equation}
Hence we get
\begin{equation}
\label{eq:Qdefa}
Q(x;\lambda) = |\Ai(e^{-i\alpha}(ix+\lambda))|^2  
\int\limits_0^x  \left| \Ai (e^{-i \alpha}(iy+\lambda)) \frac{\Ai(\lambda)}{\Ai(e^{-i\alpha}\lambda)} - \Ai(iy+\lambda)\right|^2 dy \, .
\end{equation}

\subsection{More facts on Airy expansions}
\begin{figure}
\begin{center}
\includegraphics[width=60mm]{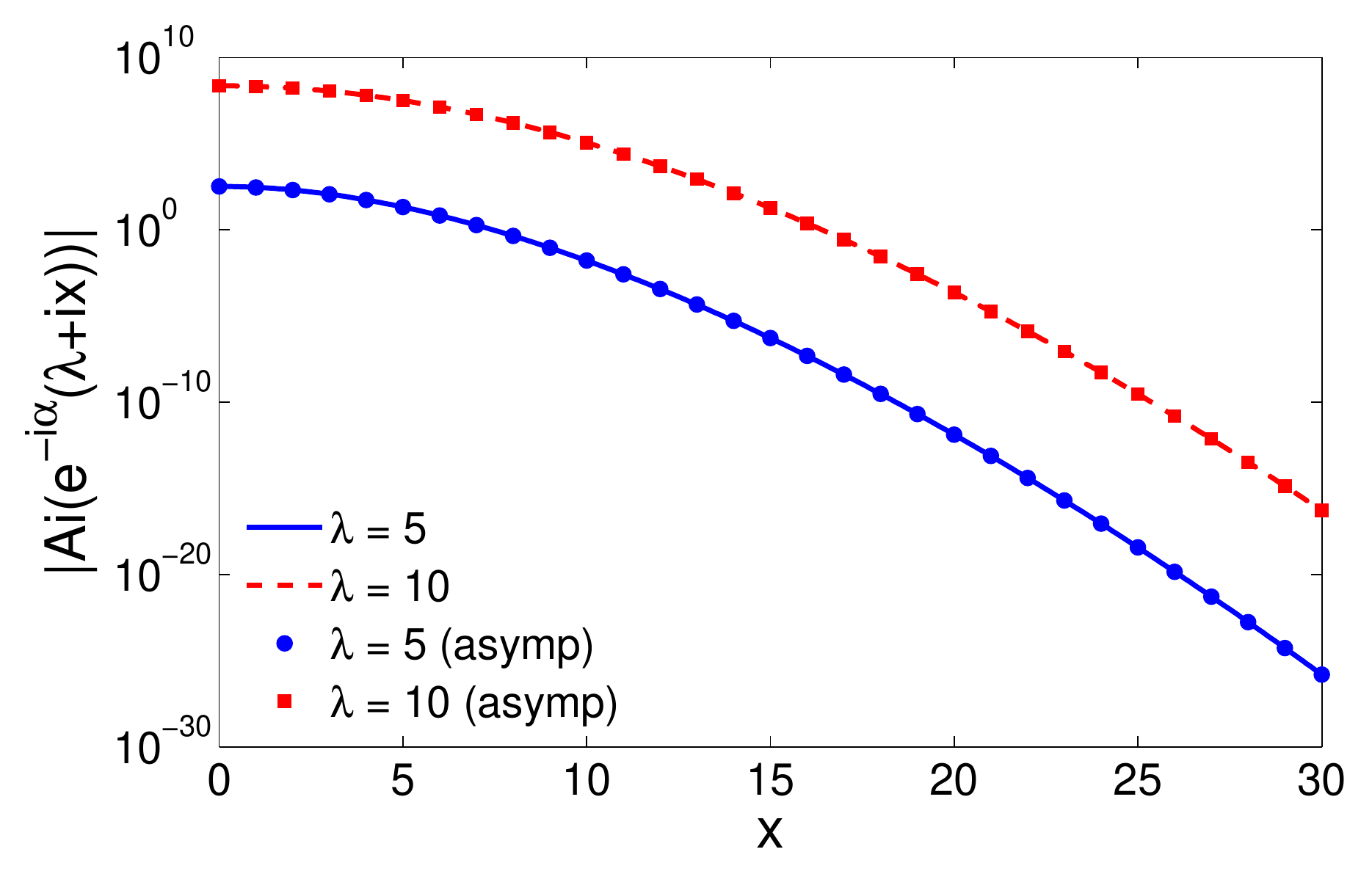}  
\includegraphics[width=60mm]{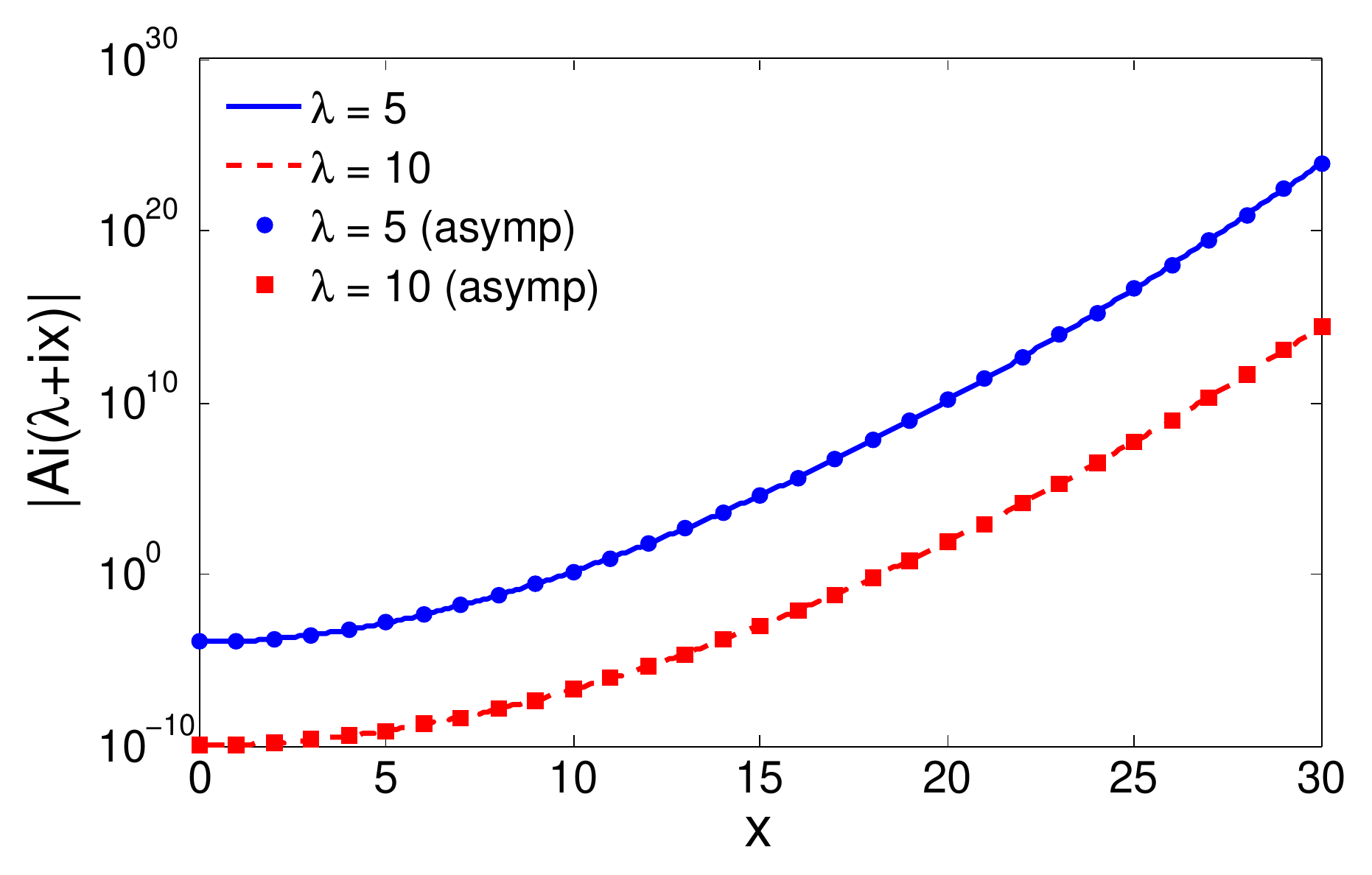}  
\includegraphics[width=60mm]{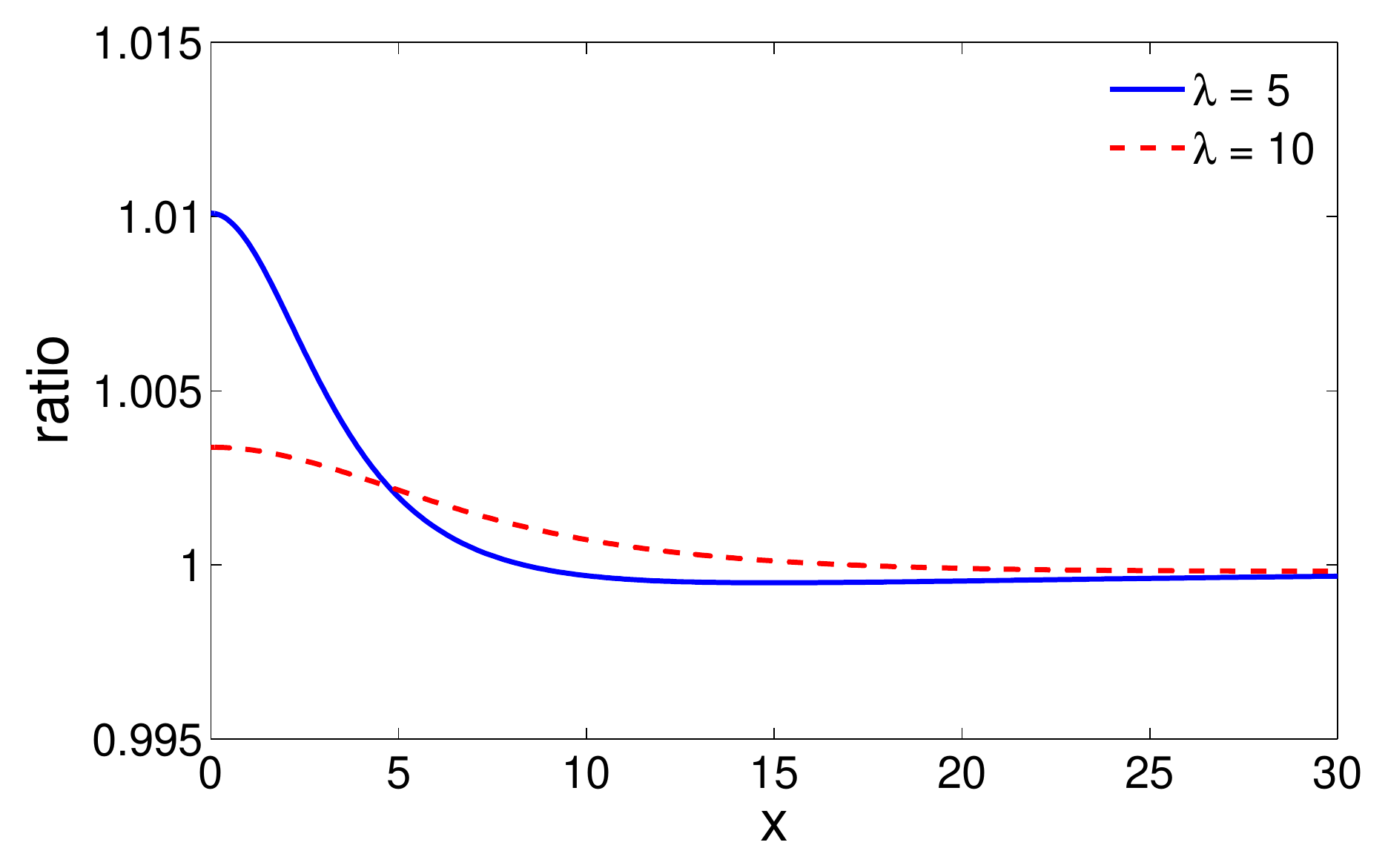}  
\includegraphics[width=60mm]{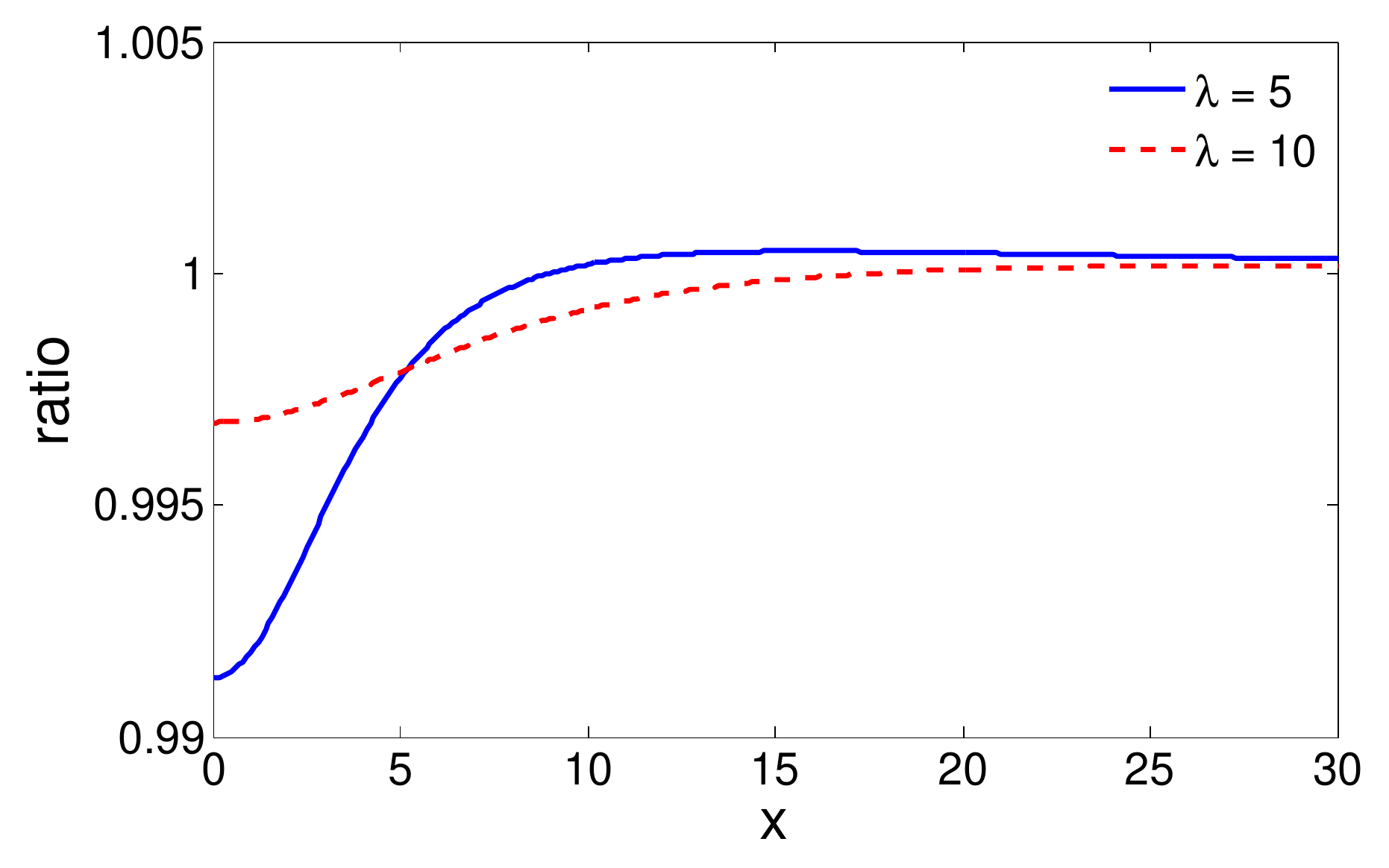}  
\end{center}
\caption{
(Top) Asymptotic behavior of $|\Ai(e^{-i\alpha}(ix+\lambda))|$ (left)
and \break $|\Ai(ix+\lambda)|$ (right) for large $\lambda$.  (Bottom)
The ratio between these functions and their asymptotics given by
\eqref{eq:airywm_asympt} and \eqref{eq:airy_asympt}. }
\label{fig:airy_asympt}
\end{figure}

As a consequence of \eqref{5}, we can write
\begin{equation}
\label{eq:airywm_asympt}
|\Ai(e^{-i\alpha}(ix+\lambda))| = \frac{\exp\bigl(- \frac23 \lambda^{3/2} u(x/\lambda) \bigr)}{2\sqrt{\pi} (\lambda^2 + x^2)^{1/8}}  
(1 + \mathcal O (\lambda^{-\frac 32})),
\end{equation}
where
\begin{equation}
\label{eq:u}
\begin{split}
u(s) & = - (1 + s^2)^{3/4} \cos\left(\frac32 \tan^{-1}(s)\right) \\
& = \frac{\sqrt{\sqrt{1+s^2}+1} ~(\sqrt{1+s^2}-2)}{\sqrt{2}}. \\
\end{split}
\end{equation}
We note indeed that $|e^{-i\alpha}(ix+\lambda)|=\sqrt{x^2 +
\lambda^2}\geq \lambda \geq \lambda_0$ and that we have a control of
the argument $\arg (e^{-i\alpha}(ix+\lambda))\in [-\frac { 2\pi}{ 3},
-\frac \pi 6]$ which permits to apply \eqref{5}.\\ 
Similarly, we obtain
\begin{equation}
\label{eq:airy_asympt}
|\Ai(ix+\lambda)| = \frac{\exp\bigl(\frac23 \lambda^{3/2} u(x/\lambda) \bigr)}{2\sqrt{\pi} (\lambda^2 + x^2)^{1/8}}  (1 + \mathcal O (\lambda^{-\frac 32}))\,.
\end{equation}
We note indeed that $|ix+\lambda|=\sqrt{x^2 + \lambda^2}$ and $\arg
((ix+\lambda))\in [0, + \frac \pi 2]$ so that one can then again apply
\eqref{5}.  In particular the function $|\Ai(ix+\lambda)|$ grows
super-exponentially as $x\rightarrow +\infty$. \\
Figure \ref{fig:airy_asympt} illustrates that, for large $\lambda$,
both equations (\ref{eq:airywm_asympt}) and (\ref{eq:airy_asympt}) are
very accurate approximations for $|\Ai(e^{-i\alpha}(ix+\lambda))|$ and
$|\Ai(ix+\lambda)|$, respectively. \\
The control of the next order term (as given in \eqref{5}) implies
that there exist $C> 0\,$ and $\epsilon_0 >0$,  such that, for any $\epsilon \in (0,\epsilon_0]$, any $\lambda>\varepsilon^{-\frac23}$
and any $x \geq 0$, one has
\begin{eqnarray}
 |\Ai(e^{-i\alpha}(ix+\lambda))| &\leq& (1+ C \epsilon) \frac {1}{ 2\sqrt{\pi} }  
\frac{\exp\bigl(- \frac23 \lambda^{3/2} u(x/\lambda) \bigr)}{(\lambda^2 + x^2)^{1/8}} , \\
|\Ai(ix+\lambda)| &\leq& (1+ C \epsilon)  \frac {1}{ 2\sqrt{\pi} } \frac{\exp\bigl(\frac23 \lambda^{3/2} u(x/\lambda) \bigr)}{(\lambda^2 + x^2)^{1/8}}\, ,
\end{eqnarray}
and
\begin{eqnarray}
(1-C \epsilon) \frac {1}{ 2\sqrt{\pi} }  \frac{\exp\bigl(- \frac23 \lambda^{3/2} u(x/\lambda) \bigr)}{(\lambda^2 + x^2)^{1/8}} & 
  \leq& |\Ai(e^{-i\alpha}(ix+\lambda))| \,, \\
(1- C \epsilon)  \frac {1}{ 2\sqrt{\pi} } \frac{\exp\bigl(\frac23 \lambda^{3/2} u(x/\lambda) \bigr)}{(\lambda^2 + x^2)^{1/8}} &\leq &|\Ai(ix+\lambda)| \,,
\end{eqnarray}
where the function $u$ is explicitly defined in Eq. (\ref{eq:u}).\\

{\bf Basic properties of $u$.} \\ Note that
\begin{equation}
u'(s) = \frac{3}{2\sqrt{2}} ~ \frac{s}{\sqrt{1 + \sqrt{1+s^2}}} \geq 0  \qquad (s\geq 0),
\end{equation}
and  $u$ has the following expansion at the origin
\begin{equation}
u(s) = -1 + \frac 38 s^2  + \mathcal O (s^4)\,.
\end{equation}
For large $s$, one has 
\begin{equation}
u(s) \sim
\frac{s^{3/2}}{\sqrt{2}}\,, \qquad  u'(s) \sim \frac{3 s^{1/2}}{2\sqrt{2}}  \,.
\end{equation}
One concludes that the function $u$ is monotonously increasing from
$-1$ to infinity.

\subsection{Upper bound}

We start from the simple upper bound (for any $\epsilon >0$)
\begin{equation}\label{ub}
Q(x,\lambda) \leq \left(1+\frac 1 \epsilon\right) Q_1(x,\lambda) +  (1+\epsilon) Q_2(x,\lambda)\,,
\end{equation}
with
\begin{equation*}
Q_1(x, \lambda):=  |\Ai(e^{-i\alpha}(ix+\lambda))|^2 \, \frac{|\Ai(\lambda)|^2}{|\Ai(e^{-i\alpha}\lambda)|^2}\, 
\int\limits_0^x  | \Ai (e^{-i \alpha} (iy+\lambda)) |^2\, dy 
\end{equation*}
and
\begin{equation*}
Q_2(x,\lambda):= |\Ai(e^{-i\alpha}(ix+\lambda))|^2 \int\limits_0^x  |\Ai(iy+\lambda)|^2\, dy \,.
\end{equation*}
We then write
\begin{equation*}
Q_1(x, \lambda)\leq  |\Ai(e^{-i\alpha}(ix+\lambda))|^2 \,\frac{|\Ai(\lambda)|^2}{|\Ai(e^{-i\alpha}\lambda)|^2} \, 
\int\limits_0^{+\infty}  | \Ai (e^{-i \alpha} (iy+\lambda))  |^2\, dy 
\end{equation*}
and integrating over $x$
\begin{equation*}
\int_0^{+\infty} Q_1(x, \lambda) dx \leq   I_0(\lambda)^2  \, \frac{|\Ai(\lambda)|^2}{|\Ai(e^{-i\alpha}\lambda)|^2} ,
\end{equation*}
where $I_0(\lambda)$ is given by (\ref{eq:I0def}).\\

Using \eqref{estI0} and \eqref{5}, we obtain
\begin{equation}\label{q1}
\int_0^{+\infty} Q_1(x, \lambda) dx \leq  C \lambda^{-\frac 12} \,.
\end{equation}

Hence at this stage, we have proven the existence of $C >0$,
$\epsilon_0>0$ and $\lambda_0$ such that such that for any $\epsilon
\in (0,\epsilon_0]$ and any $\lambda \geq \lambda_0$:
\begin{equation} \label{eq:GnormDstep1}
|| \mathcal G^{-,D} ||^2_{HS} \leq (1+\epsilon)   \left( 8\pi^2\,  \int\limits_0^\infty Q_2 (x;\lambda) dx\right)   + C \lambda^{-1} \epsilon^{-1}\,.
\end{equation}

It remains to estimate
\begin{equation}\label{intq2}
\int_0^{+\infty} Q_2(x,\lambda)dx =  \int_0^{+\infty} dx  \, \int\limits_0^x  | \Ai(e^{-i\alpha}(ix+\lambda) ) \Ai (iy +\lambda) |^2\, dy \,.
\end{equation}

Using the estimates \eqref{eq:airywm_asympt} and
\eqref{eq:airy_asympt}, we obtain
\begin{lemma}  
There exist $C$ and $\epsilon_0$, such that, for any $\epsilon \in
(0,\epsilon_0)$, for $\lambda > \epsilon^{-\frac 23}$, the integral of
$Q_2(x;\lambda)$ can be bounded as
\begin{equation}\label{lulb}
\frac 12  (1- C \epsilon)\, I(\lambda)\ \leq 8 \pi^2  \int_0^{+\infty} Q_2(x,\lambda) dx \leq \frac 12  (1+ C \epsilon)\, I(\lambda)\,,
\end{equation}
where
\begin{equation}
I(\lambda) = \int\limits_0^{\infty} dx ~ \frac{\exp\bigl(- \frac43 \lambda^{3/2} u(x/\lambda) \bigr)}{(\lambda^2 + x^2)^{1/4}}
\int\limits_0^x dy~\frac{\exp\bigl(\frac43 \lambda^{3/2} u(y/\lambda) \bigr)}{(\lambda^2 + y^2)^{1/4}} .
\end{equation}
\end{lemma}

{\bf Control of $I(\lambda)$.}\\
It remains to control $I(\lambda)$ as $\lambda \rightarrow
+\infty\,$. Using a change of variables, we get
\begin{equation}
I(\lambda) = \lambda  \int\limits_0^{\infty} dx ~ \frac{\exp\bigl(- \frac43 \lambda^{3/2} u(x) \bigr)}{(1 + x^2)^{1/4}}
\int\limits_0^x dy~ \frac{\exp\bigl(\frac43 \lambda^{3/2} u(y) \bigr)}{(1+ y^2)^{1/4}} \,.
\end{equation}
Hence, introducing
\begin{equation}\label{tlambda}
t= \frac43 \lambda^\frac 32\,,
\end{equation}
we reduce the analysis to $\hat I (t)$ defined for $t \geq t_0$ by
\begin{equation}
\label{eq:hatIt}
\hat I (t):=  \int\limits_0^{\infty} dx ~ \frac{1}{(1 + x^2)^{1/4}}
\int\limits_0^x dy~ \frac{\exp\bigl(t (u(y)-u(x))  \bigr)}{(1+ y^2)^{1/4}}\, ,
\end{equation}
with
\begin{equation} \label{chapeauI}
I(\lambda) = \lambda\, \widehat I (t)\,.
\end{equation}
The following analysis is close to that of the asymptotic behavior of
a Laplace integral.\\

{\bf Asymptotic upper bound of $\hat I (t)$.}\\ Although $u(y) \leq u
(x)$ in the domain of integration in Eq. (\ref{eq:hatIt}), a direct
use of this upper bound will lead to an upper bound by $+\infty$.

Let us start by a heuristic discussion.  The maximum of $u(y) - u (x)$
should be on $x=y$.  For $x-y$ small, we have $u(y)-u(x) \sim (y-x)
u'(x)$.  This suggests a concentration near $x=y=0$, whereas a
contribution for large $x$ is of smaller order.\\ More rigorously, we
write
\begin{equation}\label{decompositionI}
\widehat I(t) = \widehat I_1(t,\epsilon) + \widehat I_2(t,\epsilon,\xi) + \widehat I_3 (t,\epsilon),
\end{equation}
with, for $0< \epsilon < \xi$, 
\begin{equation}
\label{eq:auxil_6}
\begin{split}
\widehat I_1 (t,\epsilon) &= \int_0^\epsilon  \, dx ~ \frac{1}{(1 + x^2)^{1/4}}  \int\limits_0^x dy~ \frac{\exp\bigl(t (u(y)-u(x))  \bigr)}{(1+ y^2)^{1/4}} , \\
\widehat I_2 (t,\epsilon,\xi) &= \int_\epsilon^\xi  \, dx ~ \frac{1}{(1 + x^2)^{1/4}}  \int\limits_0^x dy~ \frac{\exp\bigl(t (u(y)-u(x))  \bigr)}{(1+ y^2)^{1/4}} , \\
\widehat I_3 (t,\xi) &= \int_\xi^{+\infty}  \, dx ~ \frac{1}{(1 + x^2)^{1/4}}  \int\limits_0^x dy~ \frac{\exp\bigl(t (u(y)-u(x))  \bigr)}{(1+ y^2)^{1/4}}\,.
\end{split}
\end{equation}

We now observe that $u(s)$ has the form $u(s) = v(s^2)$ where $v'
>0\,$, so that
\begin{equation}
\begin{array}{l}
 \forall x,y \mbox{ s.t. } 0\leq y \leq x \leq \tau_0 \,, \\
 \quad\quad\left( \sup_{\tau \in [0,\tau_0]}  v' (\tau)\right)  (x^2-y^2)\geq  u(x)- u (y) \geq  
\left( \inf_{\tau \in [0,\tau_0]}  v' (\tau)\right)   (x^2-y^2)\,.\label{ineq}
\end{array}
\end{equation}

{\bf Analysis of $\widehat I_1 (t,\epsilon)$.}\\
Using the right hand side of inequality \eqref{ineq} with
$\tau_0=\epsilon$ , we show the existence of constants $C$ and
$\epsilon_0 >0$, such that, $\forall \epsilon \in (0,\epsilon_0)$
\begin{equation} \label{B.17}
 (1- C\epsilon)  J_\epsilon \left( (1+ C\epsilon) \frac 38 t\right) \leq \widehat I_1(t,\epsilon) 
\leq (1+ C\epsilon)  J_\epsilon \left( (1- C\epsilon) \frac 38 t\right)\,,
\end{equation}
with
\begin{equation*}
J_\epsilon (\sigma):=\int_0^{\epsilon} dx  \int_0^x \exp \bigl(\sigma (y^2-x^2)\bigr) dy\,,
\end{equation*}
which has now to be estimated for large $\sigma$.\\
For $\frac{1}{ \sqrt{\epsilon\sigma}} \leq \epsilon $, we write
\begin{equation*}
J_\epsilon (\sigma) =  J_\epsilon^1 (\sigma) + J_\epsilon^2(\sigma)\,,
\end{equation*}
with
\begin{equation*}
\begin{split}
J_\epsilon^1(\sigma) & := \int_0^{\frac{1}{ \sqrt{\epsilon \sigma}} } dx  \int_0^x \exp \bigl( \sigma (y^2-x^2)\bigr) dy\,, \\
J_\epsilon^2(\sigma ) & := \int_{\frac{1}{ \sqrt{\epsilon \sigma}} }^\epsilon dx  \int_0^x \exp \bigl(\sigma  (y^2-x^2)\bigr) dy\,. \\
\end{split}
\end{equation*}
Using the trivial estimate
\begin{equation*}
 \int_0^x \exp \bigl(\sigma (y^2-x^2) \bigr) dy \leq x ,
\end{equation*}
we get
\begin{equation}
 J_\epsilon^1(\sigma ) \leq \frac{1}{2 \epsilon \sigma} \,.
\end{equation}

We have now to analyze $J_\epsilon^2(\sigma)$.\\
 The formula giving  $J_\epsilon^2(\sigma)$ can be expressed by using
the Dawson function (cf  \cite{AS}, p. 295 and 319)
\begin{equation*}
s \mapsto D(s):=\int_0^s\exp (y^2-s^2)\, dy
\end{equation*}
and its asymptotics as $s \ar +\infty$\,,\\
\begin{equation}
D(s) = \frac{1}{2s} (1 + \delta(s)) \,,
\end{equation}
where the function $\delta(s)$ satisfies $\delta(s)=\mathcal{O}(s^{-1})\,$.\\
We get indeed 
\begin{equation*}
 J_\epsilon^2(\sigma) = \frac{1}{\sigma } \int _{ \epsilon^{-\frac 12}}^{\epsilon\sigma ^\frac 12 } D(s) ds \,.
\end{equation*}
By taking $\epsilon$ small enough to use the asymptotics of
$D(\cdot)\,$,
\begin{equation}\label{j2epsilon}
\begin{array}{lll}
J_\epsilon^2(\sigma) &=&\frac{1}{2 \sigma } \left(\int _{\epsilon^{-\frac 12}}^{ \sigma ^\frac 12 \epsilon} \frac 1s  ds 
+\int _{\epsilon^{-\frac 12}}^{ \sigma ^{\frac 12} \epsilon}\frac{\delta(s)}{s} ds\right) \\
& = &  \frac 14~ \frac{\log \sigma }{\sigma } + \frac{C}{\sigma} (\log \epsilon + \mathcal O(1))\,.
\end{array}
\end{equation}

Hence we have shown the existence of constants $C>0$ and $\epsilon_0$
such that if $t \geq C \epsilon^{-3 }$ and $\epsilon \in
(0,\epsilon_0)$
\begin{equation}\label{estimateI1}
\widehat I_1( t,\epsilon) \leq \frac {2}{3} \frac{\log t}{t} + C \left(\epsilon \frac{\log t}{t} + \frac{1}{\epsilon} \frac 1t \right) \,.
\end{equation}
{\bf Analysis of $\widehat I_3 (t,\xi)$} \\
We start from
\begin{equation*}
\widehat I_3(t,\xi) = \int_\xi^{+\infty}  \, dx ~ \frac{1}{(1 + x^2)^{1/4}}  \int\limits_0^x dy~ \frac{\exp\bigl(t (u(y)-u(x))  \bigr)}{(1+ y^2)^{1/4}} 
\end{equation*}
and will determine the choice of $\xi$ for a good estimate.  Having in
mind the properties of $u $, we can choose $\xi$ large enough in order
to have for some $c_\xi >0$ the property that for $x \geq \xi$ and
$\frac x 2 \leq y \leq x$,
\begin{equation}\label{B.21}
\begin{array}{rl}
u(x) & \geq c_\xi x^\frac 32\,\\
u(x)-u(x/2) &  \geq c_\xi x^\frac 32\,,\\
u(x)-u(y) & \geq c_\xi x^\frac 12\,  (x-y)\,.
\end{array}
\end{equation}
This determines our choice of $\xi$.
Using these inequalities, we rewrite $\widehat I_3 (t,\xi)$ as the sum
\begin{equation*}
\widehat I_3 (t, \xi) =\widehat I_{31} (t) +  \widehat I_{32} (t) \,,
\end{equation*}
with
\begin{equation*}
\begin{split}
\widehat I_{31} (t) & = \int_\xi ^{+\infty}  \, dx ~ \frac{1}{(1 + x^2)^{1/4}}
\int\limits_0^{\frac x2}  dy~ \frac{\exp\bigl(t (u(y)-u(x))  \bigr)}{(1+ y^2)^{1/4}} , \\
\widehat I_{32} (t) & = \int_\xi ^{+\infty}  \, dx ~ \frac{1}{(1 + x^2)^{1/4}}
\int\limits_{\frac x2}^x  dy~ \frac{\exp\bigl(t (u(y)-u(x))  \bigr)}{(1+ y^2)^{1/4}} . \\
\end{split}
\end{equation*}
Using the monotonicity of $u$, we obtain the upper bound
\begin{equation*}
\begin{split}
\widehat I_{31} (t) &  \leq  \int_\xi ^{+\infty}  \, dx ~ \frac{1}{(1 + x^2)^{1/4}} \int\limits_0^{\frac x2}  dy~ \exp\bigl(t (u(y)-u(x))  \bigr) \\
& \leq  \frac 12  \int_\xi ^{+\infty}  \,  x^\frac 12 \,  \exp\bigl(t (u(x/2)-u(x))  \bigr)\, dx\\
& \leq   \frac 12  \int_\xi ^{+\infty}  \,  x^\frac 12 \,  \exp\bigl(- c_\xi \,t \, x^\frac 32\bigr) \, dx\\
& \leq \frac 13 \int_{\xi^\frac 32}^{+\infty} \exp\bigl( - c_\xi t s\bigr) \, ds\\ & 
\leq  \frac{1}{3 c_\xi t } \,  \exp\bigl( - c_\xi {\xi^\frac 32} t\bigr) \,.
\end{split}
\end{equation*}
Hence, there exists $\epsilon_\xi >0$ such that as $t\ar +\infty$\,, 
\begin{equation}\label{B.22}
\widehat I_{31} (t)= \mathcal O \bigl(\exp( -\epsilon_\xi t)\bigr)\,.
\end{equation}
The last term to control is $\widehat I_{32} (t)$.  Using
\eqref{B.21}, we get
\begin{equation}\label{estimateI32}
\begin{array}{ll}
\widehat I_{32} (t) & \leq \sqrt{2}  \int_\xi ^{+\infty}  \, dx ~ \frac{1}{(1 + x^2)^{1/2}}
\int\limits_{\frac x2}^x  dy~ \exp (t (u(y)-u(x)) \\
& \leq \sqrt{2}  \int_\xi ^{+\infty}  \, dx ~ \frac{1}{(1 + x^2)^{1/2}}
\int\limits_{\frac x 2}^x  dy~\exp\bigl( - c_\xi t  x^{\frac 12} (x-y)\bigr)  \\
 &  \leq  \frac{\sqrt{2}}{c_\xi t} \, \int_\xi^{+\infty}   x^{-\frac 3 2} \, dx \, = \frac{ 1}{\sqrt{2\xi} c_\xi t} \, . 
\end{array}
\end{equation}
Hence putting together \eqref{B.22} and \eqref{estimateI32} we have,
for this choice of $\xi$, the existence of $\hat C_\xi >0$ and $t_\xi
>0$ such that
\begin{equation} \label{B.23}
\forall t\geq t_\xi\,,\quad \widehat I_{3} (t)  \leq  \hat C_\xi /t\,.
\end{equation}

{\bf Analysis of $\widehat I_2(t,\epsilon,\xi)$.\\}
We recall that
\begin{equation*}
\widehat I_2 (t,\epsilon,\xi) = \int_\epsilon^\xi  \, dx ~ \frac{1}{(1 + x^2)^{1/4}}  \int\limits_0^x dy~ \frac{\exp\bigl(t (u(y)-u(x))  \bigr)}{(1+ y^2)^{1/4}} \,.
\end{equation*}
We first observe that
\begin{equation*}
\widehat I_2 (t,\epsilon,\xi) \leq  \int_\epsilon^\xi  \, dx   \int\limits_0^x dy \exp\bigl(t (u(y)-u(x))  \bigr) 
\leq   \int_\epsilon^\xi  \, dx   \int\limits_0^x dy \exp\bigl( c_\xi  t (y^2-x^2))  \bigr) \,,
\end{equation*}
with
\begin{equation*}
c_\xi = \inf _{[0,\xi]} v' >0\,.
\end{equation*}

Using now
\begin{equation*}
  \int_0^x \exp \bigl(c_\xi t (y^2-x^2)\bigr) dy \leq  \int_0^x \exp \bigl(c_\xi t x  (y-x)\bigr) dy = 
\frac{1}{c_\xi tx}\, \bigl(1 -  \exp (- c_\xi t x^2)\bigr)   \leq \frac{1}{c_\xi tx}\,,
\end{equation*}
we get
\begin{equation}\label{estimateI2}
\widehat I_2 (t,\epsilon,\xi)   \leq \frac{1}{c_\xi t} \left(\log \xi  -\log \epsilon \right)\,.
\end{equation}
Putting together \eqref{decompositionI}, \eqref{estimateI1},
\eqref{B.23} and \eqref{estimateI2}, we have shown the existence of
$C>0$ and $\epsilon_0$ such that if $t \geq C \epsilon^{-3 }$ and
$\epsilon \in (0,\epsilon_0)$
\begin{equation}
\widehat I ( t) \leq \frac {2}{3} \frac{\log t}{t} + C \left( \epsilon \frac{\log t}{t} + \frac{1}{\epsilon} \frac 1t \right) \,.
\end{equation}
Coming back to \eqref{chapeauI} and using \eqref{eq:GnormDstep1}, we
show the existence of $C>0$ and $\epsilon_0$ such that if $\lambda
\geq C \epsilon^{-2}$:
\begin{equation*}
|| \mathcal G^{-,D}(\lambda) ||_{HS}^2  \leq \frac 38  \lambda^{-\frac 12} \log \lambda + 
C \left(\epsilon  \lambda^{-\frac 12} \log \lambda + \frac 1 \epsilon \lambda^{-\frac 12} \right)\,.
\end{equation*}
Taking $\epsilon = (\log \lambda)^{-\frac 12}$, we obtain
\begin{lemma} There exist $C>0$ and $\lambda_0$ such that for
$\lambda \geq \lambda_0$
\begin{equation*}
||\mathcal G^{-,D}(\lambda) ||_{HS} ^2 \leq \frac 38 \lambda^{-\frac 12} \log \lambda ( 1 + C\,  (\log \lambda)^{-\frac 12})\,.
\end{equation*}
\end{lemma}

\subsection{Lower bound}

Once the upper bounds established, the proof of the lower bound is
easy. We start from the simple lower bound (for any $\epsilon >0$)
\begin{equation}\label{lb} 
Q(x,\lambda) \geq - \frac 1 \epsilon Q_1(x,\lambda) +  (1-\epsilon) Q_2(x,\lambda)\,,
\end{equation} 
and consequently
\begin{equation}\label{lb1}
\int_0^{+\infty} Q(x,\lambda) \,dx \geq  (1-\epsilon)  \int_0^{+\infty}Q_2(x,\lambda)\,dx  - \frac 1 \epsilon \int_0^{+\infty} Q_1(x,\lambda)   \,.
\end{equation}
 Taking $\epsilon = (\log\lambda)^{-\frac 12}$ and using the upper bound \eqref{q1}, 
 it remains to find a lower bound for
$\int_0^{+\infty}Q_2(x,\lambda) dx$, which can be worked out in the same way as for the upper bound. 
We can use \eqref{lulb}, \eqref{B.17}, \eqref{j2epsilon}  and
\begin{equation}
\widehat I (t) \geq \widehat I_1( t,\epsilon) \geq \frac {2}{3} \frac{\log t}{t}  -  
C \left(\epsilon \frac{\log t}{t} + \frac{1}{\epsilon} \frac 1t \right) \,.
\end{equation}
This gives the proof of 
\begin{lemma} There exist $C>0$ and $\lambda_0$ such that for $\lambda \geq \lambda_0$
\begin{equation*}
||\mathcal G^{-,D}(\lambda) ||_{HS} ^2 \geq \frac 3 8  \lambda^{-\frac 12} \log \lambda \, ( 1 - C \, (\log \lambda)^{-\frac 12})\,.
\end{equation*}
\end{lemma}

\section{Phragmen-Lindel\"of theorem}\label{AppD}

The Phragmen-Lindel\"of Theorem (see Theorem $16.1$ in \cite{Agm})
reads
\begin{theorem}{(Phragmen-Lindel\"of)}\label{thPL}
Let us assume that there exist two rays 
\[\mathcal{R}_1=\{re^{i\theta_1} : r\geq0\} \ \textrm{and} \ \mathcal{R}_2=\{re^{i\theta_2} : r\geq0\}\]
with $(\theta_1,\theta_2)$ such that
$|\theta_1-\theta_2|=\frac{\pi}{a}$ and a continuous function $F$ in
the closed sector delimited by the two rays, holomorphic in the open
sector, satisfying the properties
\begin{itemize}
 \item 
\[ \exists C >0, \exists N \in \mathbb R, \mbox{ s. t. }\forall\lambda\in\mathcal{R}_1\cup\mathcal{R}_2, \quad |F(\lambda)|\leq C \langle \lambda \rangle ^N.\]
\item 
There exist an increasing sequence $(r_k)$ tending to $+\infty$, and
$C$ such that
\begin{equation}\label{PL1}
\forall k, \quad \max_{|\lambda|=r_k}|F(\lambda)|\leq \, C\, \exp(r_k^\beta)\, ,
\end{equation}
with $\beta< a$.
\end{itemize}
Then we have
\begin{equation*}
|F(\lambda)|\leq C\,  \langle \lambda \rangle^N\,
\end{equation*}
for all $\lambda$ between the two rays $\mathcal{R}_1$ and
$\mathcal{R}_2$.
\end{theorem}

\section{Numerical computation of eigenvalues} \label{sec:numerics}

In order to compute numerically the eigenvalues of the realization $
\mathcal A^{+,D}_{1,L}$ of the complex Airy operator $\mathcal A^+_0
:= D_x^2 + i x = - \frac{d^2}{dx^2} + i x$ on the real line with a
transmission condition, we impose auxiliary Dirichlet boundary
conditions at $x = \pm L$, i.e., we search for eigenpairs
$\{\lambda_L, u_L(\cdot)\}$ of the following problem:
\begin{equation}
\begin{split}
& \left(- \frac{d^2}{dx^2} + i x\right) u_L(x) = \lambda_L u_L(x),   \qquad (-L < x < L),  \\
& u_L(\pm L) = 0\,, \qquad  u'_L(0_+) = u'_L(0_-) = \kappa \bigl(u_L(0_+) - u_L(0_-)\bigr)\, ,  \\
\end{split}
\end{equation}
with a positive parameter $\kappa$. 

Since the interval $[-L,L]$ is bounded, the spectrum of the above
differential operator is discrete.  To compute its eigenvalues, one
can either discretize the second derivative, or represent this
operator in an appropriate basis in the form of an
infinite-dimensional matrix.  Following \cite{Grebenkov14a}, we choose
the second option and use the basis formed by the eigenfunctions of
the Laplace operator $-\frac{d^2}{dx^2}$ with the above boundary
conditions.  Once the matrix representation is found, it can be
truncated to compute the eigenvalues numerically.  Finally, one
considers the limit $L\to +\infty$ to remove the auxiliary boundary
conditions at $x = \pm L$.\\ 
There are two sets of Laplacian eigenfunctions in this domain:

(i) symmetric eigenfunctions
\begin{equation}
v_{n,1}(x) = \sqrt{1/L}  \cos(\pi (n+1/2) x/L), \quad  \mu_{n,1} = \pi^2 (n+1/2)^2/L^2  ,
\end{equation}
enumerated by the index $n \in \mathbb N$.

(ii) antisymmetric eigenfunctions
\begin{equation}
v_{n,2}(x) = \begin{cases} + (\beta_n/\sqrt{L}) \sin(\alpha_n(1 - x/L)) \quad (x > 0), \cr
- (\beta_n/\sqrt{L}) \sin(\alpha_n(1 + x/L)) \quad (x < 0),  \end{cases}
\end{equation}
with $\mu_{n,2} = \alpha_n^2/L^2$, where $\alpha_n$
($n=0,1,2,\ldots$) satisfy the equation
\begin{equation}
\label{eq:alpha}
\alpha_n \, \ctan (\alpha_n) = - 2\kappa L\, ,
\end{equation}
while the normalization constant $\beta_n$ is 
\begin{equation}
\beta_n =  \left(1 + \frac{2\kappa L}{\alpha_n^2 + 4\kappa^2 L^2}\right)^{-1/2}\, .
\end{equation}
The solutions $\alpha_n$ of Eq. (\ref{eq:alpha}) lie in the intervals
$(\pi n + \pi/2, \pi n + \pi)$, with $n \in \mathbb N$\,. 

In what follows, we use the double index $(n,j)$ to distinguish
symmetric and antisymmetric eigenfunctions and to enumerate
eigenvalues, eigenfunctions, as well as the elements of governing
matrices and vectors.  We introduce two (infinite-dimensional)
matrices $\Lambda$ and $\B$ to represent the Laplace operator and the
position operator in the Laplacian eigenbasis: 
\begin{equation}
\Lambda_{n,j;n',j'} = \delta_{n,n'} \delta_{j,j'} \,\mu_{n,j}\,, 
\end{equation}
and 
\begin{equation}
\B_{n,j;n',j'} = \int\limits_{-L}^L dx ~ v_{n,j}(x)~ x ~ v_{n',j'}(x)\, .
\end{equation}
The symmetry of eigenfunctions $v_{n,j}$ implies $\B_{n,1;n',1} =
\B_{n,2;n',2} = 0$, while
\begin{equation}
\begin{split}
\B_{n,1;n',2} & = \B_{n',2;n,1}  \\
& =  -2 L \beta_{n'} \frac{\sin(\alpha_{n'})(\alpha_{n'}^2 + \pi^2 (n+1/2)^2) - (-1)^n(2n+1)\pi \alpha_{n'}}{(\alpha_{n'}^2 - \pi^2 (n+1/2)^2)^2} . \\
\end{split}
\end{equation}

The infinite-dimensional matrix $\Lambda + i  \B$ represents the
complex Airy operator $\mathcal A^{+,D}_{ 1,L}$  on the interval $[-L,L]$ in the Laplacian
eigenbasis.  As a consequence, the eigenvalues and eigenfunctions can
be numerically obtained by truncating and diagonalizing this matrix.
The obtained eigenvalues are ordered according to their increasing
real part:
\begin{equation*}
\Re \lambda_{1,L} \leq \Re \lambda_{2,L} \leq \ldots
\end{equation*}
Table \ref{tab:convergence} illustrates the rapid convergence of these
eigenvalues to the eigenvalues of the complex Airy operator $\mathcal
A^+_1$ on the whole line with transmission, as $L$ increases.  The
same matrix representation was used for plotting the pseudospectrum of
$\mathcal A^+_1$ (Fig. \ref{fig:pseudo}).

\begin{table}
\begin{center}
\begin{tabular}{| c | c | c | c | c |}   \hline
& $L$      &    $\lambda_{1,L}$ &    $\lambda_{3,L}$ & $\lambda_{5,L}$    \\  \hline
\multirow{5}{3mm}[1mm]{\begin{turn}{90} $\kappa = 0$ \end{turn}}
& 4        & 0.5161 - 0.8918i & 1.2938 - 2.1938i & 3.7675 - 1.9790i \\
& 6        & 0.5094 - 0.8823i & 1.1755 - 3.9759i & 1.6066 - 2.7134i \\
& 8        & 0.5094 - 0.8823i & 1.1691 - 5.9752i & 1.6233 - 2.8122i \\
& 10       & 0.5094 - 0.8823i & 1.1691 - 7.9751i & 1.6241 - 2.8130i \\  \hline
& $\infty$ & 0.5094 - 0.8823i &                  & 1.6241 - 2.8130i \\  \hline  \hline
\multirow{5}{3mm}[1mm]{\begin{turn}{90} $\kappa = 1$ \end{turn}}
& 4        & 1.0516 - 1.0591i & 1.3441 - 2.0460i & 4.1035 - 1.7639i \\
& 6        & 1.0032 - 1.0364i & 1.1725 - 3.9739i & 1.7783 - 2.7043i \\
& 8        & 1.0029 - 1.0363i & 1.1691 - 5.9751i & 1.8364 - 2.8672i \\
& 10       & 1.0029 - 1.0363i & 1.1691 - 7.9751i & 1.8390 - 2.8685i \\  \hline
& $\infty$ & 1.0029 - 1.0363i &                  & 1.8390 - 2.8685i \\  \hline
\end{tabular}
\end{center}
\caption{
The convergence of the eigenvalues $\lambda_{n,L}$ computed by
diagonalization of the matrix $\Lambda + i\B$ truncated to the size
$100\times 100$.  Due to the reflection symmetry of the interval, all
eigenvalues appear in complex conjugate pairs: $\lambda_{2n,L} = \bar
\lambda_{2n-1,L}$.  The last line presents the poles of the resolvent
of the complex Airy operator $\mathcal A_1^+$ obtained by solving
numerically the equation (\ref{maineq}).  The intermediate column
shows the eigenvalue $\lambda_{3,L}$ coming from the auxiliary
boundary conditions at $x = \pm L$ (as a consequence, it does not
depend on the transmission coefficient $\kappa$).  Since the imaginary
part of these eigenvalues diverges as $L\to +\infty$, they can be
easily identified and discarded. }
\label{tab:convergence}
\end{table}

\end{document}